\documentclass[pre,twocolumn,amsmath,amssymb,floatfix,superscriptaddress,nopacs]{revtex4}

\usepackage{graphicx}
\usepackage{latexsym}
\usepackage{amsmath}
\usepackage{amssymb}
\usepackage{amsfonts}
\usepackage{bm}
\usepackage{color}

\newcommand{\la}{\left<}
\newcommand{\ra}{\right>}

\newcommand{\nvecl}{\underline{n}_l}
\newcommand{\pvec}{\ensuremath{\underline{p}}}

\newcommand{\rvec}{\ensuremath{\underline{r}}}
\newcommand{\rvecl}{\ensuremath{\underline{r}_l}}

\newcommand{\ddiff}{\ensuremath{\mathrm{d}}}
\newcommand{\tp}{\ensuremath{t^{\prime}}}

\newcommand{\nlx}{n_{l,x}}
\newcommand{\nly}{n_{l,y}}
\newcommand{\rl}{r_l}

\newcommand{\px}{p_x}
\newcommand{\py}{p_y}
\newcommand{\pix}{p_{i,x}}
\newcommand{\piy}{p_{i,y}}
\newcommand{\rx}{r_x}
\newcommand{\ry}{r_y}

\newcommand{\ahat}{\hat{a}}

\newcommand{\Tver}{\ensuremath{{\cal P}_{\tsamp}}}

\newcommand{\Tglass}{\mbox{$T_{\rm g}$}}

\newcommand{\Vhat}{\hat{V}}
\newcommand{\Vbar}{\overline{V}}
\newcommand{\muA}{\ensuremath{\mu_\mathrm{A}}}
\newcommand{\dmuA}{\ensuremath{\delta \mu_\mathrm{A}}}
\newcommand{\muAbar}{\ensuremath{\overline{\mu}_\mathrm{A}}}
\newcommand{\muAhat}{\ensuremath{\hat{\mu}_\mathrm{A}}}
\newcommand{\muAidhat}{\hat{\mu}_\mathrm{A}^\mathrm{id}}
\newcommand{\muAexhat}{\hat{\mu}_\mathrm{A}^\mathrm{ex}}
\newcommand{\muF}{\ensuremath{\mu_\mathrm{F}}}
\newcommand{\dmuF}{\ensuremath{\delta \mu_\mathrm{F}}}
\newcommand{\muFbar}{\ensuremath{\overline{\mu}_\mathrm{F}}}
\newcommand{\muFone}{\ensuremath{\mu_1}}
\newcommand{\dmuFone}{\ensuremath{\delta \mu_1}}
\newcommand{\muFonebar}{\ensuremath{\overline{\mu}_1}}
\newcommand{\muFtwo}{\ensuremath{\mu_0}}
\newcommand{\dmuFtwo}{\ensuremath{\delta \mu_0}}
\newcommand{\muFtwobar}{\ensuremath{\overline{\mu}_0}}
\newcommand{\Geq}{\mu_\mathrm{eq}}
\newcommand{\GF}{\mu_\mathrm{sf}}
\newcommand{\GFbar}{\overline{\mu}_\mathrm{sf}}
\newcommand{\dGF}{\delta \mu_\mathrm{sf}}
\newcommand{\GFmed}{\mu_\mathrm{sf,med}}
\newcommand{\GFmax}{\mu_\mathrm{sf,max}}
\newcommand{\Hhat}{\hat{\cal H}}
\newcommand{\Hidhat}{\hat{\cal H}^\mathrm{id}}
\newcommand{\Hexhat}{\hat{\cal H}^\mathrm{ex}}

\newcommand{\tauhat}{\ensuremath{\hat{\tau}}}
\newcommand{\tauAhat}{\ensuremath{\hat{\tau}_\mathrm{A}}}
\newcommand{\tauAidhat}{\hat{\tau}_\mathrm{A}^\mathrm{id}}
\newcommand{\tauAexhat}{\hat{\tau}_\mathrm{A}^\mathrm{ex}}

\newcommand{\coeffonetwo}{r_{01}}
\newcommand{\Thalf}{T_{1/2}}

\newcommand{\Gtbar}{\overline{G}(t)}
\newcommand{\dGt}{\delta G(t)}
\newcommand{\ctbar}{\overline{c}(t)}
\newcommand{\dct}{\delta c(t)}
\newcommand{\htbar}{\overline{h}(t)}
\newcommand{\dmu}{\delta \mu}

\newcommand{\mustar}{\mu_{\star}}
\newcommand{\deta}{\delta \eta}

\newcommand{\etainf}{\eta_{\infty}}

\newcommand{\tauinf}{\tau_{\infty}}

\newcommand{\Gloss}{G^{\prime\prime}(\omega)}

\newcommand{\Ubond}{U_\mathrm{bond}}
\newcommand{\rmin}{r_\mathrm{min}}
\newcommand{\rcut}{r_\mathrm{cut}}
\newcommand{\lbond}{l_\mathrm{bond}}
\newcommand{\kbond}{k_\mathrm{bond}}
\newcommand{\dtMD}{\delta t_\mathrm{MD}}
\newcommand{\ttemp}{\Delta t_\mathrm{temp}}
\newcommand{\tsamp}{\Delta t}
\newcommand{\tsampmax}{\Delta t_\mathrm{max}}
\newcommand{\tsamphuge}{\Delta t_\mathrm{huge}}
\newcommand{\an}{\approx 0}

\begin{document}

\title{Shear-stress fluctuations and relaxation in polymer glasses}

\author{I. Kriuchevskyi}
\affiliation{Institut Charles Sadron, Universit\'e de Strasbourg \& CNRS, 23 rue du Loess, 67034 Strasbourg Cedex, France}
\affiliation{LaMCoS, INSA, Av. Jean Capelle, F69621 Villeurbanne Cedex, France}
\author{J.P.~Wittmer}
\email{joachim.wittmer@ics-cnrs.unistra.fr}
\affiliation{Institut Charles Sadron, Universit\'e de Strasbourg \& CNRS, 23 rue du Loess, 67034 Strasbourg Cedex, France}
\author{H. Meyer}
\affiliation{Institut Charles Sadron, Universit\'e de Strasbourg \& CNRS, 23 rue du Loess, 67034 Strasbourg Cedex, France}
\author{O. Benzerara}
\affiliation{Institut Charles Sadron, Universit\'e de Strasbourg \& CNRS, 23 rue du Loess, 67034 Strasbourg Cedex, France}
\author{J. Baschnagel}
\affiliation{Institut Charles Sadron, Universit\'e de Strasbourg \& CNRS, 23 rue du Loess, 67034 Strasbourg Cedex, France}

\begin{abstract}
We investigate by means of molecular dynamics simulation a coarse-grained polymer glass model 
focusing on (quasi-static and dynamical) shear-stress fluctuations as a function of temperature $T$ 
and sampling time $\tsamp$.
The linear response is characterized using (ensemble-averaged) expectation values of the 
contributions (time-averaged for each shear plane) to the stress-fluctuation relation $\GF$ 
for the shear modulus and the shear-stress relaxation modulus $G(t)$.
Using $100$ independent configurations we pay attention to the respective standard deviations.
While the ensemble-averaged modulus $\GF(T)$ decreases continuously with increasing $T$ for all $\tsamp$ sampled,
its standard deviation $\dGF(T)$ is non-monotonous with a striking peak at the glass transition.
The question of whether the shear modulus is continuous or has a jump-singularity 
at the glass transition is thus ill-posed.
Confirming the effective time-translational invariance of our systems, the $\tsamp$-dependence
of $\GF$ and related quantities can be understood using a weighted integral over $G(t)$.
This implies that the shear viscosity $\etainf(T)$ may be readily 
obtained from the $1/\tsamp$-decay of $\GF(\tsamp)$ above the glass transition.
\end{abstract}
\date{\today}
\maketitle

\section{Introduction}
\label{sec_intro}

\paragraph*{Motivation.}
The equilibrium shear modulus $\Geq(T)$ of crystalline solids is known to vanish discontinuously 
at the melting point with increasing temperature $T$ \cite{Barrat88,LXW16}. 
A natural question which arises is that of the behavior of $\Geq(T)$ for 
amorphous solids and glasses in the vicinity of the glass transition temperature $\Tglass$.
(We assume in this paragraph that a thermodynamically properly defined  
shear modulus actually does exist. This is in fact not obvious as discussed below.)
Two qualitatively different theoretical predictions have been put forward suggesting either
a {\em discontinuous jump} at the glass transition 
\cite{GoetzeBook,Szamel11,Ikeda12,Klix12,Klix15,Yoshino14} or 
a {\em continuous} (cusp-like) transition 
\cite{Barrat88,Yoshino12,ZT13,WXP13,LXW16,ivan17c}.
The predicted jump singularity is a result of mean-field theories \cite{GoetzeBook,Biroli09,Yoshino14} which
find the energy barriers for structural relaxation to diverge at $\Tglass$ so that liquid-like
flow stops. However, in experimental or simulated glass formers the barriers do not diverge abruptly.
Such non-mean-field effects are expected to smear out the sharp transition \cite{Yoshino14}.

Another line of recent research has focused on the elastic properties deep in the glass
\cite{Gardner,Biroli16,Procaccia16}.
At $T \ll \Tglass$ a transition in the solid is found, where multiple particle arrangements occur
as different competing glassy states. This so-called ``Gardner transition" is accompanied by
strong fluctuations of $\Geq$ (and of higher order elastic moduli) 
from one glass state to the other \cite{Biroli16,Procaccia16}.
Interestingly, strong fluctuations of the shear modulus were also observed in self-assembled networks 
\cite{WKC16} which is a model for vitrimers \cite{Leibler11,Leibler13}.
The results of \cite{Biroli16,Procaccia16,WKC16} beg the question of whether 
also the glass transition is accompanied by strong fluctuations of shear stresses and moduli. 

\paragraph*{Our approach.}
Corroborating a brief account given in Ref.~\cite{ivan17c}, 
we present here numerical data obtained by means of large-scale molecular dynamics (MD) 
simulation \cite{AllenTildesleyBook,LAMMPS} of a standard coarse-grained bead-spring model.
This model has already been used in earlier work on the polymer glass transition
\cite{SBM11,BaschRev16,ivan17a,ivan17c}.
We characterize the shear rigidity in the canonical ensemble \cite{AllenTildesleyBook} 
\begin{itemize}
\item
following the pioneering work by Barrat {\em et al.} \cite{Barrat88}
using as main diagnostics the well-known stress-fluctuation formula $\GF$ for the shear modulus 
\cite{Hoover69,Lutsko88,Barrat88,WTBL02,Barrat06,SBM11,WXP13,WXB15,WXBB15,WKB15,LXW16,WXB16,WKC16,Procaccia16,ivan17a,ivan17c}
and its various contributions as defined below in Sec.~\ref{sec_algo}.
\item
by direct computation of the shear-stress relaxation modulus $G(t)$
using the general fluctuation-dissipation relation appropriate for solid-like 
systems with strong quenched shear stresses \cite{ivan17a,ivan17c}.
See Appendix~\ref{sm_Gt} for details.
\end{itemize}
Particular attention will be paid to the standard deviations and cross-correlations of the different 
contributions of the two main observables $\GF$ and $G(t)$.
We will characterize in detail the (ensemble averaged) effects of the time pre-averaging performed 
over a finite sampling time $\tsamp$ for each independent configuration and shear plane.
This is of importance since the difference between time and ensemble averages corresponds to
the standard experimental procedure
--- properties are first averaged for each shear plane and only then ensemble-averaged ---
and since the detailed averaging procedure matters for all observables characterizing 
fluctuations \cite{WXP13,WXB16,WKC16}.

\paragraph*{Key results.}
We remind \cite{WXP13,WXB15,WXBB15,WKB15,WXB16}
that if a proper sampling time independent thermodynamic 
equilibrium modulus $\Geq$ characterizing the glass transition existed, this would 
imply $\GF(\tsamp) \to \Geq$ for sufficiently large sampling times $\tsamp \gg \tauinf$ 
and also $G(t) \to \Geq$ for large times $t \gg \tauinf$ with $\tauinf(T)$ 
being the terminal relaxation time of the system.
Our numerical results are in fact qualitatively quite different and much more in line
with our recent study on self-assembled transient networks \cite{WKC16}.
We highlight three key results demonstrated below:
\begin{itemize}
\item[I)]
$\GF(T)$ decreases continuously and monotonously for all temperatures $T$ and 
sampling times $\tsamp$. Being $\tsamp$-dependent $\GF$ is {\em not}
an equilibrium storage modulus.
Albeit the crossover of $\GF(T)$ at $\Tglass$ becomes systematically sharper with 
increasing $\tsamp$, our data are {\em not} consistent with a jump-singularity.
\item[II)]
The standard deviation $\dGF$ is strongly non-monotonous with respect to $T$ with a 
remarkable peak at $\Tglass$. The transition characterized by $\GF(T)$ 
is thus masked by very strong fluctuations \cite{foot_however}.
\item[III)]
We demonstrate that $\GF$ is identical to the weighted moment $\mu(\tsamp)$
over the shear-stress relaxation modulus $G(t)$ defined by
\begin{equation}
\mu(\tsamp) \equiv \frac{2}{\tsamp^2} \int_0^{\tsamp} \ddiff t \ (\tsamp -t) \ G(t).
\label{eq_mu} 
\end{equation}
\end{itemize}
The observed $\tsamp$-dependence of $\GF$ is thus not due to non-equilibrium (``aging")
processes but can be traced back to the finite sampling time (time-averaged) stress fluctuations 
need to explore the phase space. The historically thermodynamically rooted $\GF$ takes
due to Eq.~(\ref{eq_mu}) the meaning of a ``generalized modulus" also containing information 
about dissipation processes associated to the plastic reorganization of the particle contact 
network \cite{Alexander98}.
It thus follows as a corollary from Eq.~(\ref{eq_mu}) that
\begin{itemize}
\item[IV)]
the shear viscosity $\etainf(T)$ above the glass transition may
be obtained most readily using
\begin{equation}
\GF(\tsamp) = \frac{2\etainf(T)}{\tsamp} 
\mbox{ for } \tsamp \gg \tauinf(T),
\label{eq_etainf}
\end{equation}
\end{itemize}
which we would like to stress as the forth key result of the presented work \cite{foot_Gloss2eta}.
Due to the inevitable \cite{AllenTildesleyBook} too low precision of the relaxation modulus $G(t)$ for large times, 
especially for supercooled liquids close to the glass transition, this method is shown to be much more precise 
than the commonly used approach using the asymptotic behavior of the generalized shear viscosity 
\begin{equation}
\eta(\tsamp) \equiv \int_{t=0}^{\tsamp} \ddiff t \  G(t)
\mbox{ with } \etainf = \lim_{\tsamp\to \infty} \eta(\tsamp). \label{eq_eta}
\end{equation}
Moreover, the third key result will allow us to express $G(t)$, $\eta(\tsamp)$ and 
the related generalized terminal relaxation time $\tau(\tsamp)$
in terms of the numerically better behaved $\GF(\tsamp)$.

\paragraph*{Outline.}
The present paper is organized as follows.
Our polymer model is defined in Sec.~\ref{sec_algo} where we also explain technical
details concerning the quench protocol, the time series stored and the different
time and ensemble averages computed.
We begin the presentation of our numerical results in Sec.~\ref{sec_GF} where
we focus on the (ensemble-averaged) expectation values of the stress-fluctuation
prediction $\GF$ for the shear modulus and its related contributions.
Standard deviations, fluctuations and cross-correlations of the different contributions
to $\GF$ are discussed in Sec.~\ref{sec_dGF}.
We turn then in Sec.~\ref{sec_Gt} to the shear-stress relaxation function $G(t)$ and 
the associated sampling time dependent moments $\mu(\tsamp)$ and $\eta(\tsamp)$.
We demonstrate in Sec.~\ref{res_GtGF} that $\GF(\tsamp)$ and $\mu(\tsamp)$ are identical.
Various important consequences are discussed in Sec.~\ref{res_GtGF_continued} and Sec.~\ref{res_eta}.
We show especially that Eq.~(\ref{eq_etainf}) must hold.
The standard deviation $\dGt$ of $G(t)$ is considered in Sec.~\ref{res_dGt}.
We verify in Sec.~\ref{res_V} that our results are not due to finite-size effects.
The paper is summarized in Sec.~\ref{conc_summary} and an outlook on on-going
work is given in Sec.~\ref{conc_outlook}.
Appendix~\ref{sm_affine} reminds the connection between canonical affine shear transformations 
and the instantaneous shear stress $\tauhat$ and the affine shear modulus $\muAhat$.
Focusing in Appendix~\ref{sm_highT} on temperatures above the glass transition we determine 
the shear viscosity $\etainf(T)$ and the terminal relaxation time $\tauinf(T)$ 
from the $\tsamp$-dependence of $\GF$.
Additional details concerning the shear-stress relaxation modulus $G(t)$  
are given in Appendix~\ref{sm_Gt}. 
The derivation of Eq.~(\ref{eq_mu}) for systems with time-translational invariance 
is given in Appendix~\ref{sm_GtGF}. 
The generalized terminal relaxation time $\tau(\tsamp)$ is considered in Appendix~\ref{sm_tau}.

\section{Algorithmical details}
\label{sec_algo}

\subsection{Model Hamiltonian}
\label{algo_hamiltonian}
\begin{figure}[t]
\centerline{\resizebox{1.0\columnwidth}{!}{\includegraphics*{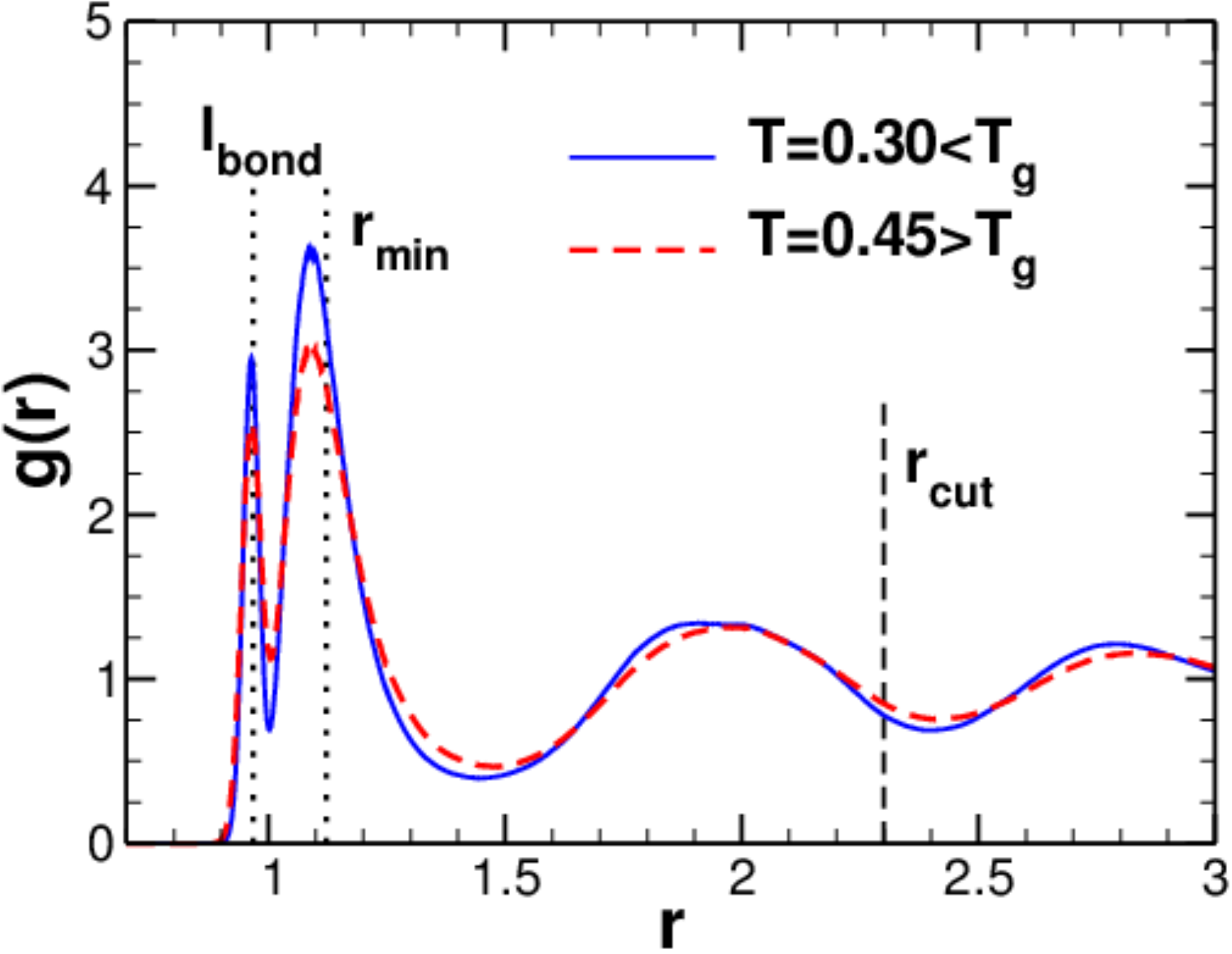}}}
\caption{Radial pair correlation function $g(r)$ for one temperature above and one below 
the glass transition temperature $\Tglass$ showing that our coarse-grained
polymer model does not crystallize. Also indicated are the equilibrium bond length $\lbond=0.967$,
the position of the minimum of the LJ potential $\rmin = 2^{1/6}$ and the potential cut-off distance
$\rcut =  2.3$.
}
\label{fig_pairV}
\end{figure}

Our data have been obtained by MD simulation \cite{AllenTildesleyBook} of a 
bead-spring model already used in earlier work on the polymer glass transition 
\cite{SBM11,Frey15,BaschRev16,ivan17a,ivan17c}. 
In this model all monomers, that are not connected by bonds, interact via a monodisperse 
Lennard-Jones (LJ) potential. LJ units \cite{AllenTildesleyBook} are thus used. 
To increase the numerical efficiency the LJ potential is truncated at $\rcut=2.3 \approx 2\rmin$, 
with $\rmin=2^{1/6}$ being the potential minimum, and shifted at $\rcut$ to make it continuous. 
(See Sec.~\ref{algo_trunc} below.)
The flexible bonds are represented by a harmonic spring potential $\Ubond(r) = (\kbond/2) \ (r- \lbond)^2$
with $r$ being the distance between the beads, $\kbond = 1110$ the spring constant and 
$\lbond=0.967$ the equilibrium bond length as indicated in Fig.~\ref{fig_pairV}.

\subsection{Operational parameters}
\label{algo_para}
As in Refs.~\cite{ivan17a,ivan17c} we focus in this work on data obtained using $m=100$ independent 
configurations containing $M=3072$ chains of length $N=4$. 
As may be seen from Fig.~\ref{fig_pairV}, this chain length is sufficient to avoid the 
crystallization tendency of the monodisperse LJ beads \cite{BaschRev16}. 
It is, however, not large enough to neglect finite-chain size effects, i.e. important properties such as 
the glass transition temperature $\Tglass$ or the affine shear modulus $\muA$ have not yet reached 
their $N$-independent asymptotic values \cite{BaschRev16,foot_otherdata}. 
The total number of monomers $n = N M = 12288$ is sufficient to make continuum mechanics applicable.
See Sec.~\ref{res_V} below for a brief comment on our on-going work on system-size effects.
The large number $m$ of independent systems allows the precise characterization of ensemble averages, 
standard deviations and error bars.
For the numerical integration of the equation of motion we use a
velocity-Verlet scheme with time steps of length $\dtMD = 0.005$.
The temperature $T$ is imposed by means of the Nos\'e-Hoover algorithm 
and the average normal pressure $P$ by the Nos\'e-Hoover-Andersen barostat 
(both provided by LAMMPS \cite{LAMMPS}).
All simulations are carried out at $P=0$.
Standard cubic simulation boxes with periodic boundary conditions are used throughout this work, 
i.e. the shape of the box is imposed and does not fluctuate as was the case in 
recent related studies \cite{WXP13,WXB15,WXBB15,WKB15,WXB16}.

\begin{figure}[t]
\centerline{\resizebox{1.0\columnwidth}{!}{\includegraphics*{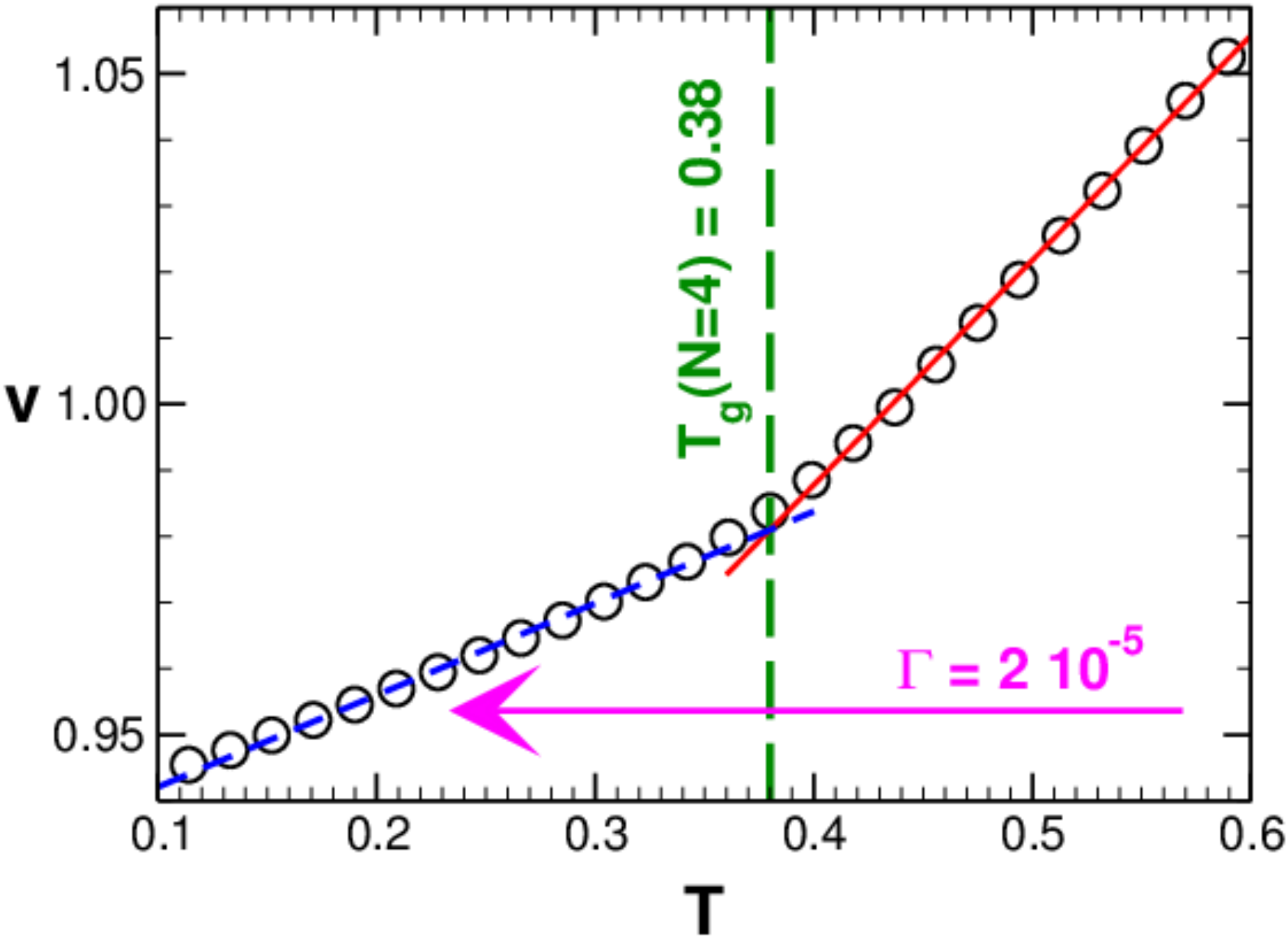}}}
\caption{Average specific volume $v = \langle \Vhat \rangle/n$ 
per monomer as a function of temperature $T$ for $\Gamma = 2 \cdot 10^{-5}$.
The thin lines indicate linear fits to the glass and the liquid branches.
Using this dilatometric criterion one defines a glass transition temperature
$\Tglass \approx 0.38$ from the intersection of both linear asymptotes.
}
\label{fig_quench}
\end{figure}

\subsection{Quench of configuration ensemble}
\label{algo_quench}

We start the quench with $m=100$ independent equilibrated configurations at $T=0.6$ and $P=0$. 
We continuously cool down the configurations with a constant cooling rate $\Gamma = 2 \cdot 10^{-5}$
\cite{foot_otherdata} while keeping constant the average normal pressure $P=0$ letting thus the 
instantaneous volume $\Vhat$ of each configuration fluctuate. 
As may be seen from Fig.~\ref{fig_quench}, the average specific volume $v=\langle \Vhat \rangle/n \approx 1$ 
decreases slightly with decreasing temperature $T$.
Using the intersection of the linear extrapolations of the glass and the liquid branches of $v$
(or of its logarithm) \cite{SBM11,LXW16,BaschRev16,foot_otherdata}, this provides a simple
and experimentally meaningful operational definition of the glass transition temperature $\Tglass$. 
We obtain
\begin{equation}
\Tglass \approx 0.38 \mbox{ for } P=0, N=4 \mbox{ and } \Gamma = 2 \cdot 10^{-5}. 
\label{eq_Tglass}
\end{equation}
(As seen from Fig.~\ref{fig_static} in Sec.~\ref{res_static}, 
a similar value is obtained from the affine shear modulus $\muA(T)$.)
After having reached a specific working temperature $T$,
the configuration is first tempered over $\ttemp=10^5$ at constant pressure
\cite{foot_Vbar}. 
We switch then to the standard canonical ensemble, 
i.e. the volume $\Vbar$ of each configuration is fixed,
and temper the systems again over $\ttemp$. 

\begin{table*}[t]
\begin{center}
\begin{tabular}{|c||c|c|c|c|c|c|c|c|c|c|c||c|c||c|c|}
\hline
$T$ &$v$   & $\muA$&$\dmuA$&$\muFtwo$&$\dmuFtwo$&$\muFone$&$\dmuFone$&$\muF$&$\dmuF$&$\GF$&$\dGF$& $G(t)$&$\dGt$& $\etainf$&$\tauinf$\\ \hline
0.05&0.9382& 85.0  &0.73   &231.7    &246.6     &162.6    &246.6     &68.3  &1.27   &16.6 &1.2   & 17.4  & 1.2  & -      &  -   \\
0.10&0.9439& 84.3  &0.71   &146.9    &110.9     &77.5     &110.8     &68.9  &1.33   &15.3 &1.2   & 16.6  & 1.6  & -      &  - \\
0.15&0.9496& 83.4  &0.71   &115.8    &64.6      &46.3     &64.7      &69.1  &1.44   &14.5 &1.5   & 14.4  & 1.9  & -      &  -\\
0.20&0.9559& 83.1  &0.68   &115.8    &44.7      &32.4     &44.6      &68.7  &1.59   &14.1 &1.5   & 14.2  & 2.5  & -      &  -\\
0.25&0.9624& 81.9  &0.66   &91.8     &28.3      &22.0     &28.2      &68.7  &1.48   &12.4 &1.4   & 14.2  & 3.7  & -      &  -\\
0.30&0.9696& 81.0  &0.63   &84.6     &17.7      &14.5     &20.7      &68.4  &1.49   &11.4 &1.4   & 13.6  & 5.1  & -      &  -\\
0.35&0.9777& 80.2  &0.62   &81.6     &12.4      &8.5      &11.9      &69.4  &3.47   &6.8  &3.5   & 3.9   & 12.6 & -      &  - \\
0.36&0.9797& 80.1  &0.62   &80.9     &9.2       &6.3      &9.0       &69.9  &3.41   &6.3  &3.4   & 5.3   & 11.4 & -      & - \\
0.37&0.9817& 80.1  &0.55   &79.9     &5.8       &3.2      &4.6       &72.0  &3.24   &3.5  &3.3   & -     & 12.3 &275000  & 178000\\
0.38&0.9838& 79.7  &0.55   &79.7     &2.6       &1.0      &1.5       &74.8  &2.19   &1.1  &2.2   & -     & 7.9  & 50000  &  34000 \\
0.39&0.9860& 79.3  &0.46   &79.4     &1.3       &0.26     &0.4       &77.5  &1.22   &0.8  &0.2   &$\an$  &7.3   & 13000  &  8300\\
0.40&0.9888& 78.6  &0.31   &78.9     &0.70      &0.09     &0.1       &78.0  &0.68   &$\an$&0.7   &$\an$  &5.5   &4500    & 4000\\
0.41&0.9915& 78.5  &0.21   &78.5     &0.46      &0.04     &0.1       &78.1  &0.46   &$\an$&0.5   &$\an$  &3.0   &1800    & 2491\\
0.42&0.9949& 78.1  &0.14   &78.1     &0.29      &0.04     &0.05      &77.7  &0.29   &$\an$&0.3   &$\an$  &2.3   &890     & 1641\\
0.43&0.9973& 77.7  &0.12   &77.7     &0.24      &0.01     &0.02      &77.6  &0.23   &$\an$&0.27  &$\an$  &1.7   &460     & 935\\
0.44&1.0002& 77.3  &0.10   &77.3     &0.19      &0.006    &0.008     &77.2  &0.19   &$\an$&0.22  &$\an$  &1.3   &320     & 712\\
0.45&1.0040& 76.7  &0.08   &76.9     &0.17      &0.004    &0.005     &76.9  &0.17   &$\an$&0.19  &$\an$  &1.2   &220     & 525\\
0.50&1.0207& 74.8  &0.04   &74.9     &0.11      &0.001    &0.002     &74.9  &0.11   &$\an$&0.12  &$\an$  &0.8   &59      & 66 \\
0.55&1.0387& 73.1  &0.03   &73.1     &0.10      &0.001    &0.001     &73.0  &0.10   &$\an$&0.10  &$\an$  &$0.7$ &30      & 26 \\
\hline
\end{tabular}
\vspace*{0.5cm}
\caption[]{Some properties as a function of temperature:
volume $v$ per monomer,
affine shear modulus $\muA$ and its standard deviation $\dmuA$,
$\muFtwo$ and $\dmuFtwo$,
$\muFone$ and $\dmuFone$,
shear stress fluctuation $\muF=\muFtwo-\muFone$ and its standard deviation $\dmuF$,
shear stress modulus $\GF=\muA-\muF$ and its standard deviation $\dGF$,
shear stress relaxation modulus $G(t)$ and its standard deviation $\dGt$ taken at $t=10000$,
shear viscosity $\etainf$ obtained in the liquid regime using Eq.~(\ref{eq_etainf})
and terminal relaxation time $\tauinf$ from the $\muFone(\tsamp)$-scaling (Fig.~\ref{fig_mu1_dt}).
The data from the third to the 10th column have been obtained using the largest sampling time $\tsamp=\tsampmax=10^5$.
In agreement with Lutsko \cite{Lutsko88}, $\muF$ does not vanish for small temperatures
and $\muA$ is thus only an upper bound to $\GF$ for all temperatures.
Importantly, $\muFtwo$ deviates from $\muA$ and $\muFone$ from $\GF$ below $T \approx 0.3$.
While $\muA$, $\muFtwo$ and $G(t)$ do not depend on the sampling time $\tsamp$,
this is different for $\muFone$, $\muF$, $\GF$ and the standard deviations
$\dmuA$, $\dmuFtwo$, $\dmuFone$, $\dmuF$, $\dGF$ and $\dGt$.
See Sec.~\ref{res_static} and Sec.~\ref{res_GF_dt} for more details.
\label{tab_T}}
\end{center}
\end{table*}

\subsection{Time averages}
\label{algo_timeaver}
The subsequent production runs are performed over $\tsampmax=10^5$ with entries made every $10 \dtMD$. 
Of importance for the present study are the instantaneous shear stress $\tauhat$ and 
the instantaneous ``affine shear modulus" $\muAhat$ obtained using 
Eqs.~(\ref{eq_tauAidhat}-\ref{eq_muAexhat}) given in Appendix~\ref{sm_affine}.
As reminded there, $\tauhat$ is the first functional derivative of the Hamiltonian with 
respect to an imposed infinitesimal canonical and affine shear transformation and $\muAhat$ 
the corresponding second functional derivative.
(The Born-Lam\'e coefficient $\muAhat$ is elsewhere also called ``affine shear elasticity" or
``high-frequency shear modulus" \cite{WXP13,WXB15,WXB16,WKC16}.)
The stored time-series are used to compute for a given configuration and shear plane
various {\em time averages} (marked by horizontal bars) of instantaneous properties 
computed over a broad range of sampling times $\tsamp \le \tsampmax$. 
Specifically, we shall investigate in Sec.~\ref{sec_GF} the following time averages 
\begin{eqnarray}
\muAbar    & \equiv & \overline{\muAhat} \label{eq_muAbar} \\
\muFtwobar & \equiv & \beta V \ \overline{\tauhat^2} \label{eq_muFtwobar} \\
\muFonebar & \equiv & \beta V \ \overline{\tauhat}^2 \label{eq_muFonebar} \\
\muFbar    & \equiv & \muFtwobar - \muFonebar      \label{eq_muFbar} \\
\GFbar     & \equiv & \muAbar - \muFbar \equiv (\muAbar-\muFtwobar) + \muFonebar  \label{eq_GFbar}
\end{eqnarray}
with $\beta=1/T$ being the inverse temperature and $V = \langle \Vbar \rangle$ the ensemble-averaged volume.
(It would have theoretically been more exact to use here instead the fixed volume $\Vbar$ of each configuration.
But since the volume fluctuations within the ensemble are tiny, this difference is numerically irrelevant.)
The last relation Eq.~(\ref{eq_GFbar}) corresponds to the well-known stress-fluctuation formula
for the shear modulus for one shear plane of a given configuration 
\cite{Hoover69,Lutsko88,Barrat88,WTBL02,Barrat06,SBM11,WXP13,WXB15,WXBB15,WKB15,LXW16,WXB16,WKC16,Procaccia16,ivan17a,ivan17c}.

We also consider dynamical properties related to the stress fluctuations of each shear plane
such as the shear-stress autocorrelation function (ACF) $\ctbar$ and 
the shear-stress mean-square displacement (MSD) $\htbar$ being defined, respectively, by
\begin{eqnarray}
\ctbar & \equiv & \beta V \ \overline{\tauhat(t+t') \tauhat(t')} \label{eq_ctbar} \\
\htbar & \equiv & \frac{\beta V}{2} \ \overline{(\tauhat(t+t')-\tauhat(t'))^2} = \muFtwobar - \ctbar. \label{eq_htbar}
\end{eqnarray}
The bars indicate here that we perform for the time series of each shear plane standard gliding (time) 
averages \cite{AllenTildesleyBook} over all possible pairs of entries $\tauhat(t')$ and $\tauhat(t'+t)$ 
being a time interval $t$ apart. This implies that the number of pairs contributing to the
gliding average decreases linearly with $t$ and the statistics must thus deteriorate for $t \to \tsamp$.
As reminded in Appendix~\ref{sm_Gt}, the time-averaged shear-stress relaxation
modulus of a given configuration and shear plane is then given in general by 
\cite{WXB15,WXBB15,WXB16,WKC16,ivan17a}
\begin{equation}
\Gtbar \equiv \muAbar - \htbar = (\muAbar-\muFtwobar) + \ctbar. 
\label{eq_Gtbar} 
\end{equation}
Note that Eq.~(\ref{eq_Gtbar}) reduces to the commonly assumed $\Gtbar=\ctbar$ 
\cite{HansenBook,AllenTildesleyBook} if and only if $\muAbar=\muFtwobar$ holds.

\subsection{Ensemble averages}
\label{algo_ensembleaver}

By averaging over the $m$ configurations and the three shear planes
($\langle \ldots \rangle$ denotes the corresponding average)
we obtain then the ensemble averages
\begin{eqnarray}
\muA & \equiv & \langle \muAbar \rangle \label{eq_muA}\\
\muFtwo & \equiv & \langle \muFtwobar \rangle \label{eq_muFtwo}\\
\muFone & \equiv & \langle \muFonebar \rangle \label{eq_muFone}\\
\muF & \equiv & \langle \muFbar \rangle \label{eq_muF}\\
\GF  & \equiv & \langle \GFbar \rangle = \muA - \muF = (\muA - \muFtwo) + \muFone \label{eq_GF}\\
c(t) & \equiv & \langle \ctbar \rangle \label{eq_ct} \\
h(t) & \equiv & \langle \htbar \rangle = \muFtwo - c(t) \label{eq_ht}\\
G(t) & \equiv & \langle \Gtbar \rangle = \muA - h(t) = (\muA - \muFtwo) + c(t).   \label{eq_Gt}
\end{eqnarray}
See Table~\ref{tab_T} for some values for different temperatures.
As seen from the Table and as further discussed in Sec.~\ref{res_static}, $\muA=\muFtwo$ for temperatures above $T \approx 0.3$.
As one would expect for liquids, this implies that Eq.~(\ref{eq_GF}) and Eq.~(\ref{eq_Gt}) reduce to 
\begin{equation}
\GF = \muFone \mbox{ and } G(t) = c(t) \mbox{ for } T \ge 0.3
\label{eq_hightT}
\end{equation}
as we have explicitly checked. See Appendix~\ref{sm_highT} for details.
The Table also contains standard deviations of various observables
such as the standard deviation $\dGF$ of the shear modulus $\GF$ given by 
\begin{equation}
\dGF \equiv \sqrt{ \la \GFbar^2 \ra - \la \GFbar \ra^2 }. \label{eq_dGF}
\end{equation}
The error bars are obtained from the indicated standard deviations by dividing by $\sqrt{3m-1}$ 
if one assumes the shear planes to be statistically independent (which is a delicate issue) or 
by $\approx \sqrt{m}$ if one wishes to take a more conservative estimate.

\subsection{Truncation corrections}
\label{algo_trunc}
Albeit the truncated and shifted LJ potential is continuous, it is not continuous with respect to its first derivative.
As can be seen from Eq.~(\ref{eq_muAexhat}), one contribution to $\muA$ depends on 
the second derivative of the potential. Following Ref.~\cite{XWP12}, impulsive truncation corrections 
are thus required for the determination of the Born-Lam\'e coefficient $\muA$. 
These truncation corrections correspond to a shift of about $-0.3$ for all temperatures.
This is taken into account in Table~\ref{tab_T} as elsewhere.
In practice this correction is only relevant for some specific properties at high temperatures.
As shown in Appendix~\ref{sm_highT}, the shear modulus does otherwise not rigorously vanish
as $\GF(\tsamp) \approx \muFone(\tsamp) \sim 1/\tsamp$ to leading order as expected on general grounds related to the
well-known finite-sampling-time corrections of time-preaveraged fluctuations 
\cite{LandauBinderBook,WXP13,WXB15,WXBB15,WKB15,WXB16,WKC16}.

\section{Expectation values}
\label{sec_GF}

\subsection{Shear modulus $\GF$}
\label{res_mean_mu}

\begin{figure}[t]
\centerline{\resizebox{1.0\columnwidth}{!}{\includegraphics*{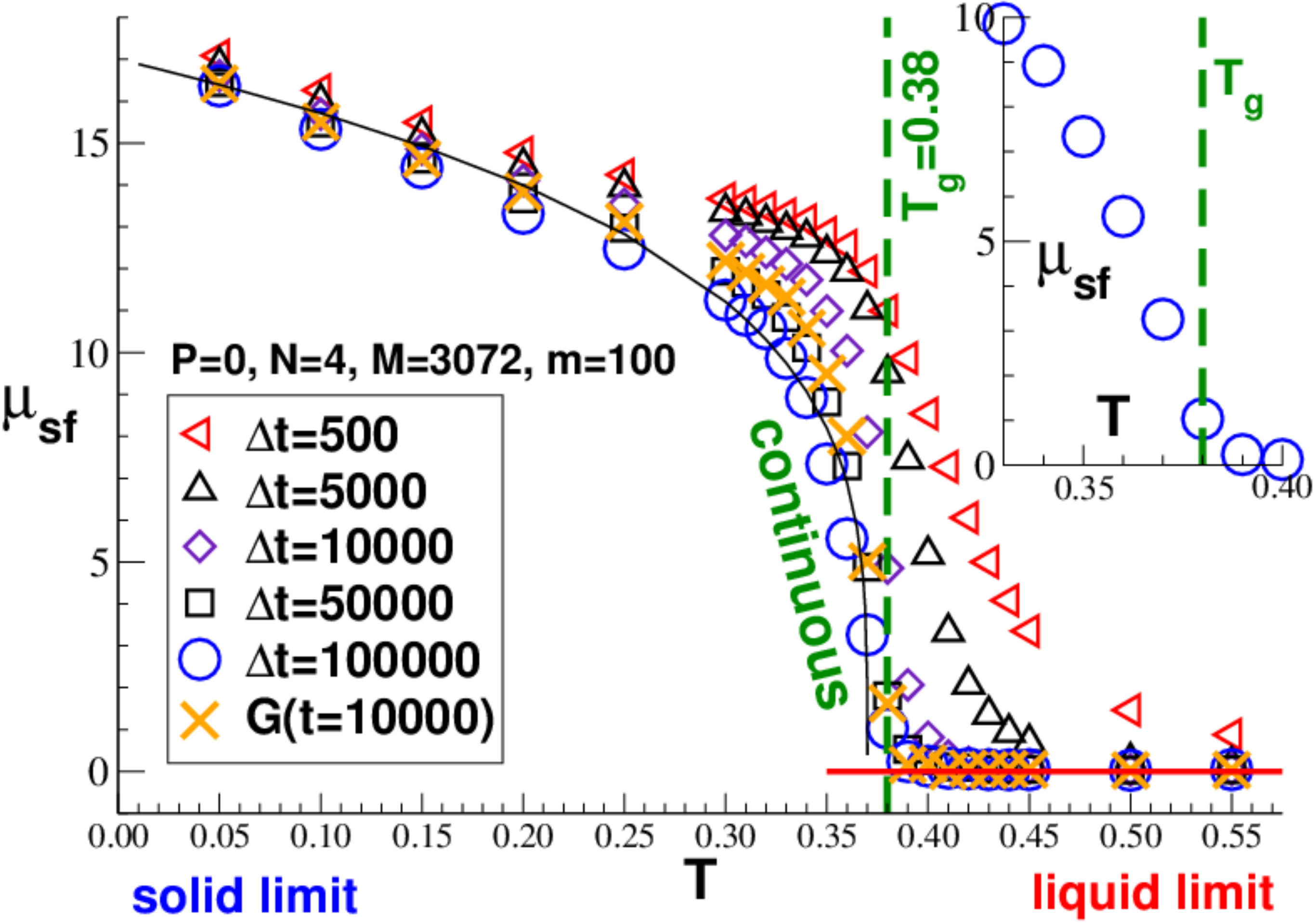}}}
\caption{Shear modulus $\GF(T) = \langle \GFbar(T) \rangle$ for different sampling times $\tsamp$
using a linear representation. The transition becomes more and more step-like with increasing $\tsamp$ 
but remains continuous for all $\tsamp$ sampled.
Also included is the shear stress relaxation modulus $G(t)$ taken at a time $t=10^4$ (crosses).
The vertical dashed line indicates the ($\tsamp$-independent) glass transition temperature, 
Eq.~(\ref{eq_Tglass}), operationally defined using a dilatometric criterion during the 
continuous temperature quench.
The thin solid line corresponds to a cusp singularity with an effective exponent $\alpha \approx 0.2$.
Inset: Zoom for $T$ around $\Tglass$ for $\tsamp=100000$ emphasizing that the transition 
characterized by $\GF(T)$ remains continuous.
\label{fig_mean_mu}
}
\end{figure}

Using a linear representation we present in Fig.~\ref{fig_mean_mu} the shear modulus $\GF(T)$ 
determined by means of the stress-fluctuation relation Eq.~(\ref{eq_GF}). Data for several
sampling times $\tsamp$ are given. There are two points to be emphasized here.
Firstly, $\GF(T;\tsamp)$ depends strongly on $\tsamp$. It is not clear from Fig.~\ref{fig_mean_mu} 
whether this dependence may drop out ultimately in the large-$\tsamp$ limit.
Secondly, while the transition becomes systematically more step-like with increasing $\tsamp$, 
it clearly remains {\em continuous} for all $\tsamp$ available. 
(This is emphasized in the inset of Fig.~\ref{fig_mean_mu}.)
The predicted discontinuous jump \cite{Szamel11,Ikeda12,Klix12,Klix15,Yoshino14} 
must therefore be strongly blurred by relaxation effects.
For large $\tsamp$ our data differ also qualitatively from the parabolic cusp-singularity 
predicted in Ref.~\cite{ZT13} and observed by some of the authors in Monte Carlo 
simulations of two-dimensional polydisperse LJ beads \cite{WXP13}.
In fact, as indicated by the thin solid line, 
a much better phenomenological fit is obtained for $\tsamp=10^5$ using 
\begin{equation}
\GF(T) \approx 17 \ \left(1-1.03 \ T/\Tglass\right)^{\alpha} \mbox{ with } \alpha \approx 0.2.
\label{eq_GFcuspfit}
\end{equation}
This effective power-law exponent corresponds to a much stronger increase below the glass 
transition as predicted by the so-called ``disorder-assisted melting" approach ($\alpha=0.5$) 
put forward in Ref.~\cite{ZT13}. 
Please note that the indicated fit is only shown to describe the data and no physical meaning should 
be attributed to the given constants and the exponent $\alpha$. 
We shall address the observed $\tsamp$-dependence of $\GF$ more systematically in Sec.~\ref{res_GF_dt}.
 
\subsection{Related expectation values}
\label{res_mean}
\begin{figure}[t]
\centerline{\resizebox{1.0\columnwidth}{!}{\includegraphics*{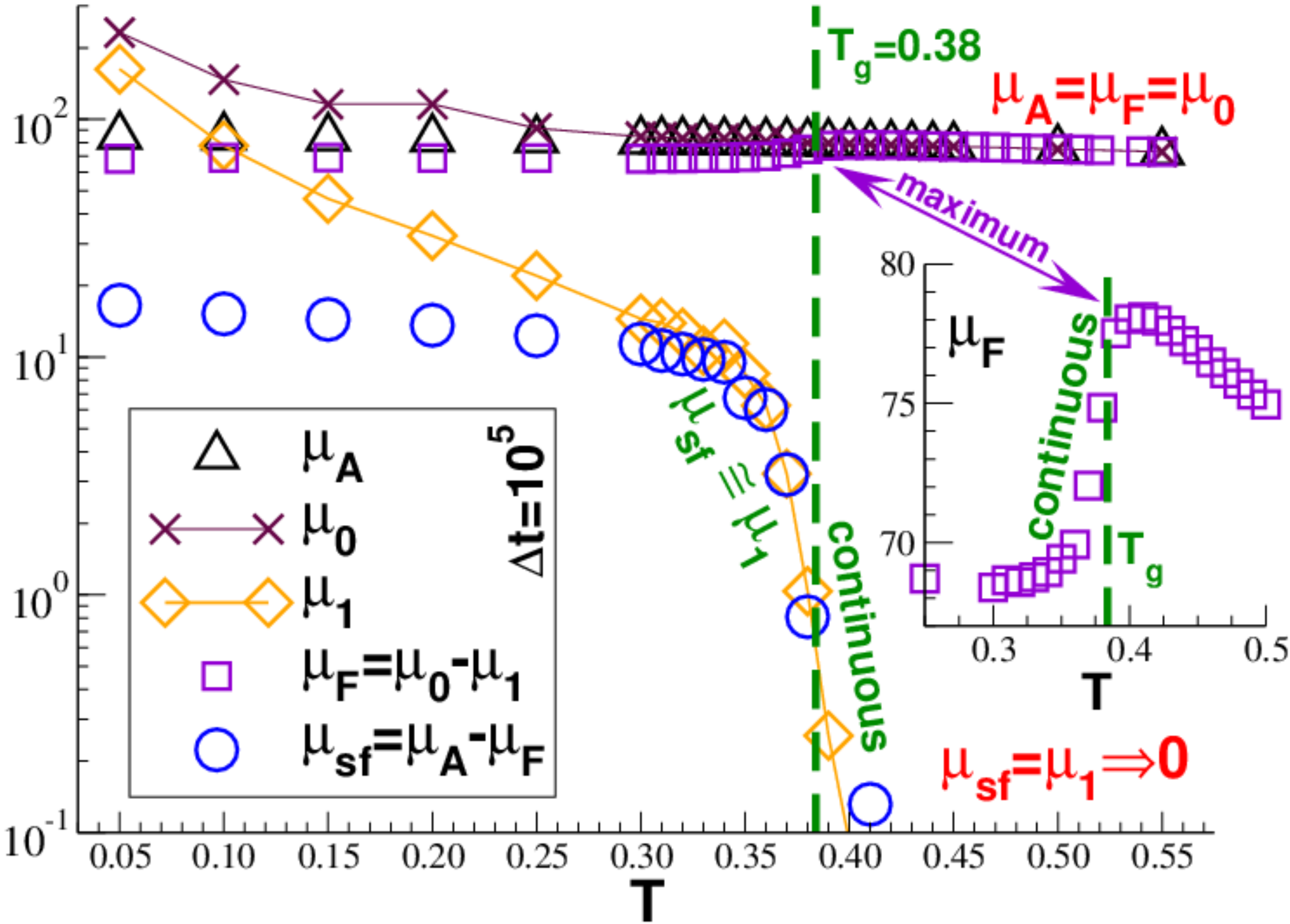}}}
\caption{$\muA$, $\muFtwo$, $\muFone$, $\muF$ and $\GF$ {\em vs.} $T$
using a half-logarithmic representation.
Only data for $\tsamp=\tsampmax=10^5$ are given.
For large temperatures $\GF \approx \muFone$ and $\muF \approx \muFtwo \approx \muA$.
With decreasing temperature $\GF$ increases rapidly around $\Tglass$, but remains continuous.
$\muFtwo$ and $\muFone$ increase rapidly below $T \approx 0.3$ and
$\muFone-\GF$ and $\muFtwo-\muA$ become thus finite. 
Inset: $\muF(T)$ using linear coordinates emphasizing the maximum near $\Tglass$.
\label{fig_mean}
}
\end{figure}

The main panel of Fig.~\ref{fig_mean} presents $\GF(T)$ and its various contributions for 
$\tsamp=\tsampmax=10^5$ using half-logarithmic coordinates. 
As emphasized above, albeit $\GF$ increases rapidly below $\Tglass$,
the data remain continuous in line with findings reported for colloidal glass-formers
\cite{Barrat88,WXP13,LXW16} using also the stress-fluctuation formula.
As one expects, $\GF \approx \muFone \to 0$ in the liquid limit above $\Tglass$.
Using the stress-fluctuation formula $\GF = \muA-\muFtwo+\muFone$, 
this implies $\muF=\muFtwo=\muA$ \cite{WXP13,ivan17a}. 
At variance to this, $\muF < \muA$ below $\Tglass$, i.e. the stress fluctuations do not have 
sufficient time to fully explore the phase space.
In agreement with Lutsko \cite{Lutsko88} and more recent studies \cite{WTBL02,Barrat06,WXP13,LXW16},
$\muF$ does not vanish for $T \to 0$, i.e. $\muA$ is only an upper bound of $\GF=\muA-\muF$ for all $T$.
Between both $T$-limits $\muF(T)$ has a clear maximum near $\Tglass$.
(This can be better seen using the linear representation given in the inset.)
Interestingly, while the difference $\muF=\muFtwo-\muFone$ is more or less constant below $\Tglass$,
its two contributions $\muFtwo$ and $\muFone$ increase rapidly with decreasing $T$. The reason for
this is that strong quenched shear stresses appear which do matter for $\muFtwo$ and $\muFone$,
but nearly cancel out for their difference $\muF$.
One thus expects much stronger fluctuations between different configurations
(and shear planes) for $\muFtwo$ and $\muFone$ than for $\muF$.
We shall verify this in Sec.~\ref{res_fluctu}.
Interestingly, while $\muFtwo$ and $\muA$ are identical at high temperatures, 
they become very different below the glass transition.
We address this finding in the subsequent subsection.

\subsection{$\tsamp$-independent static properties}
\label{res_static}

\begin{figure}[t]
\centerline{\resizebox{1.0\columnwidth}{!}{\includegraphics*{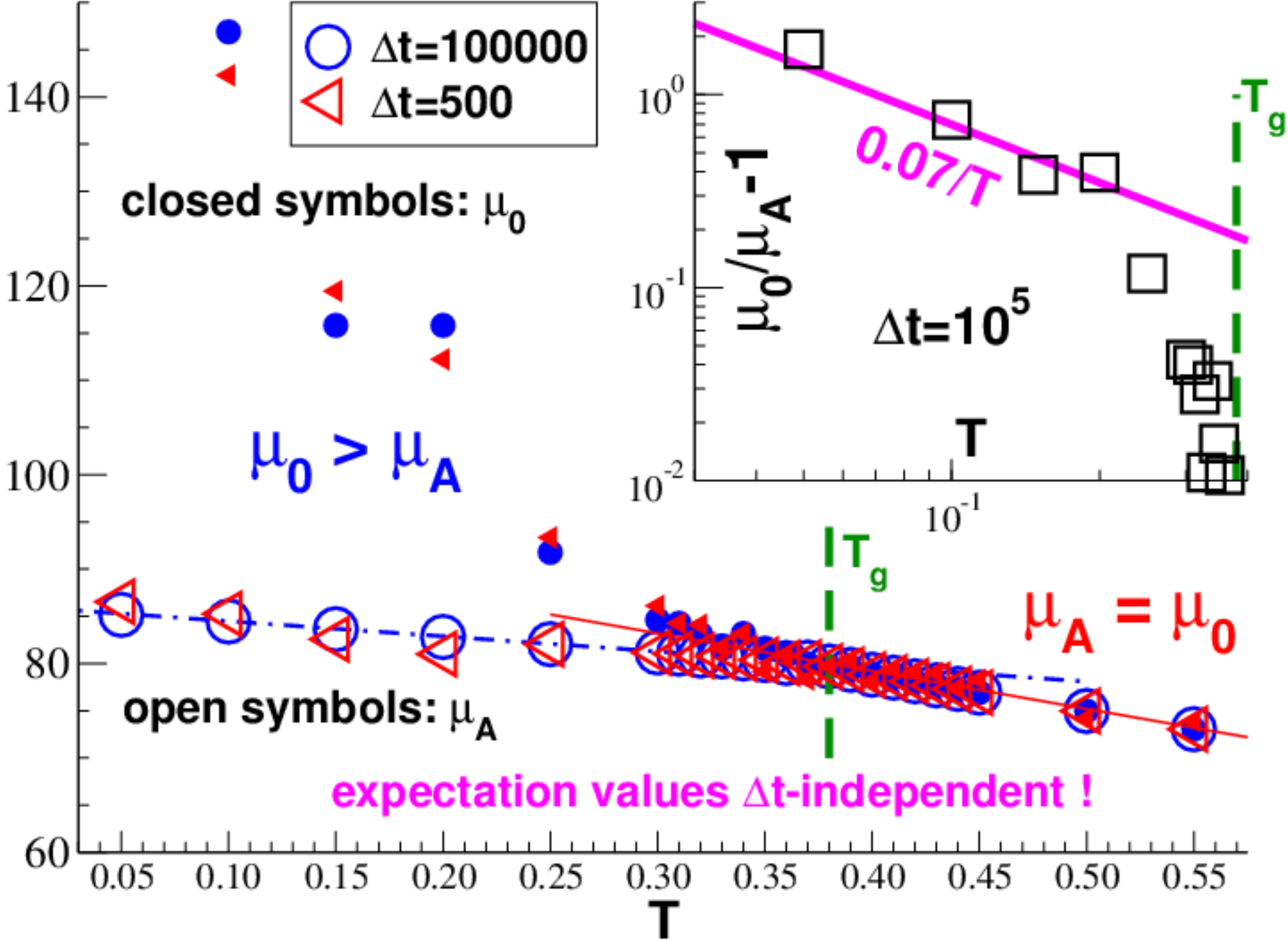}}}
\caption{Temperature dependence of $\muA$ and $\muFtwo$.
Main panel: $\muA$ (open symbols) and $\muFtwo$ (closed symbols) for two sampling times 
illustrating the $\tsamp$-independence expected from the commutation of time and ensemble averages, Eq.~(\ref{eq_commute}).
While $\muA(T)$ becomes (more or less) constant below $\Tglass$, $\muFtwo(T)$ is seen to increase strongly.
The dash-dotted and the solid line indicate two linear fits with 
$\muA(T) = \muA(\Tglass) \ [1 - c \ (T/\Tglass-1)]$ with
$c=0.076$ and $c=0.19$ for, respectively, the low and high temperature
regime and $\Tglass=0.38$ for both.
Inset: Double-logarithmic representation of $\muFtwo/\muA-1$ {\em vs.} $T$.
The ratio decreases inversely with temperature for $T \ll \Tglass$ (solid line). 
}
\label{fig_static}
\end{figure}

Before we return in Sec.~\ref{res_GF_dt} to the sampling time dependence shown in Fig.~\ref{fig_mean_mu},
we need to emphasize that the expectation values of some properties are in fact $\tsamp$-independent.
As expected from crystalline and amorphous solids \cite{LXW16,WXP13}
and permanent \cite{WXP13,WXB16} and transient \cite{WKC16} elastic networks,
this is the case for $\muA$ and $\muFtwo$.
This is demonstrated in Fig.~\ref{fig_static} for two sampling times.
(We determine first $\muAbar$ and $\muFtwobar$ from the first $\tsamp$-window of a given times series over $\tsampmax$
and ensemble-average then over $3m$ shear planes and configurations.)
The observed $\tsamp$-independence can be traced back to the fact that their time and ensemble averages 
commute \cite{WXB16}, i.e.
\begin{equation}
\la \overline{\ahat(t)} \ra = \overline{\la \ahat(t) \ra} 
\sim \tsamp^0 \mbox{ since } \la \ahat(t) \ra \sim \tsamp^0.
\label{eq_commute}
\end{equation}
$\muA$ and $\muFtwo$ are in this sense perfectly defined static observables.

This is of importance since both static properties are seen in Fig.~\ref{fig_mean} and Fig.~\ref{fig_static} 
to behave strikingly different in both temperature limits. 
We remind that $\muA$ is determined solely by the pair correlations of the system while
$\muFtwo$ also contains (in principle) three- and four-point correlations \cite{WXP13}.
While in the liquid limit these higher correlations can be factorized (which implies $\muA=\muFtwo$), 
they become relevant below $\Tglass$. 
That $\muFtwo-\muA$ becomes finite below $\Tglass$, is a non-trivial finding 
(especially in view of our recent work on transient networks \cite{WKC16})
as shown by the following argument. 

Let us suppose that we could have sampled a configuration below (the cooling-rate dependent) $\Tglass$ 
over a huge sampling time $\tsamphuge$ larger than the largest relaxation time $\tauinf(T)$ of the system.
Using the full time series, this would imply that the system must behave as a liquid,
i.e. $\GF=0$ and $\muFone=0$. Using the stress-fluctuation formula Eq.~(\ref{eq_GF}), 
this implies in turn that $\muFtwo-\muA=0$ if both moments are computed over the full $\tsamphuge$.
However, due to Eq.~(\ref{eq_commute}) this must also hold {\em on average}
for subsets of the complete time series of sampling time $\tsamp \ll \tsamphuge$.
The observation that $\muA$ and $\muFtwo$ systematically deviate below $\Tglass$
(Fig.~\ref{fig_static}), thus implies that the $3m$ time series of length $\tsamp$ obtained
from the independently quenched configurations are {\em not equivalent} to 
random subsets of a production run over $\tsamphuge$. 

Basically, $\muFtwo$ increases much more strongly than $\muA$ below $\Tglass$ due to quenched stresses 
which do not arise from equilibrium stress fluctuations at the investigated current temperature
but at some higher temperature of the quench history. Since $\muFtwo \sim \langle \overline{\tauhat^2} \rangle/T$
by definition and assuming $\langle \overline{\tauhat^2} \rangle$ to be quenched below $T \approx 0.3$,
this suggests that the dimensionless ratio $\muFtwo/\muA-1$ should decay inversely with temperature.
Albeit more data points with better statistics are warranted in this limit, 
this idea is consistent with the inset of Fig.~\ref{fig_static}. 

In summary, due to the quenched shear stresses it is not possible to describe the glassy behavior below $\Tglass$ 
by a purely dynamical theory describing the effects of a finite $\tsamp \ll \tauinf$ under the assumption that 
the finite time series are randomly drawn from an equilibrium time evolution of a liquid \cite{foot_TP}.
We note finally that the finding that $\muFtwo \neq \muA$ below $T \approx 0.3$ has important consequences 
for the numerical determination of the shear stress relaxation modulus $G(t)$ \cite{WXB16,ivan17a}. 
This point is addressed in Appendix~\ref{sm_Gt}.

\subsection{$\tsamp$-dependent quasi-static properties}
\label{res_GF_dt}

\begin{figure}[t]
\centerline{\resizebox{1.0\columnwidth}{!}{\includegraphics*{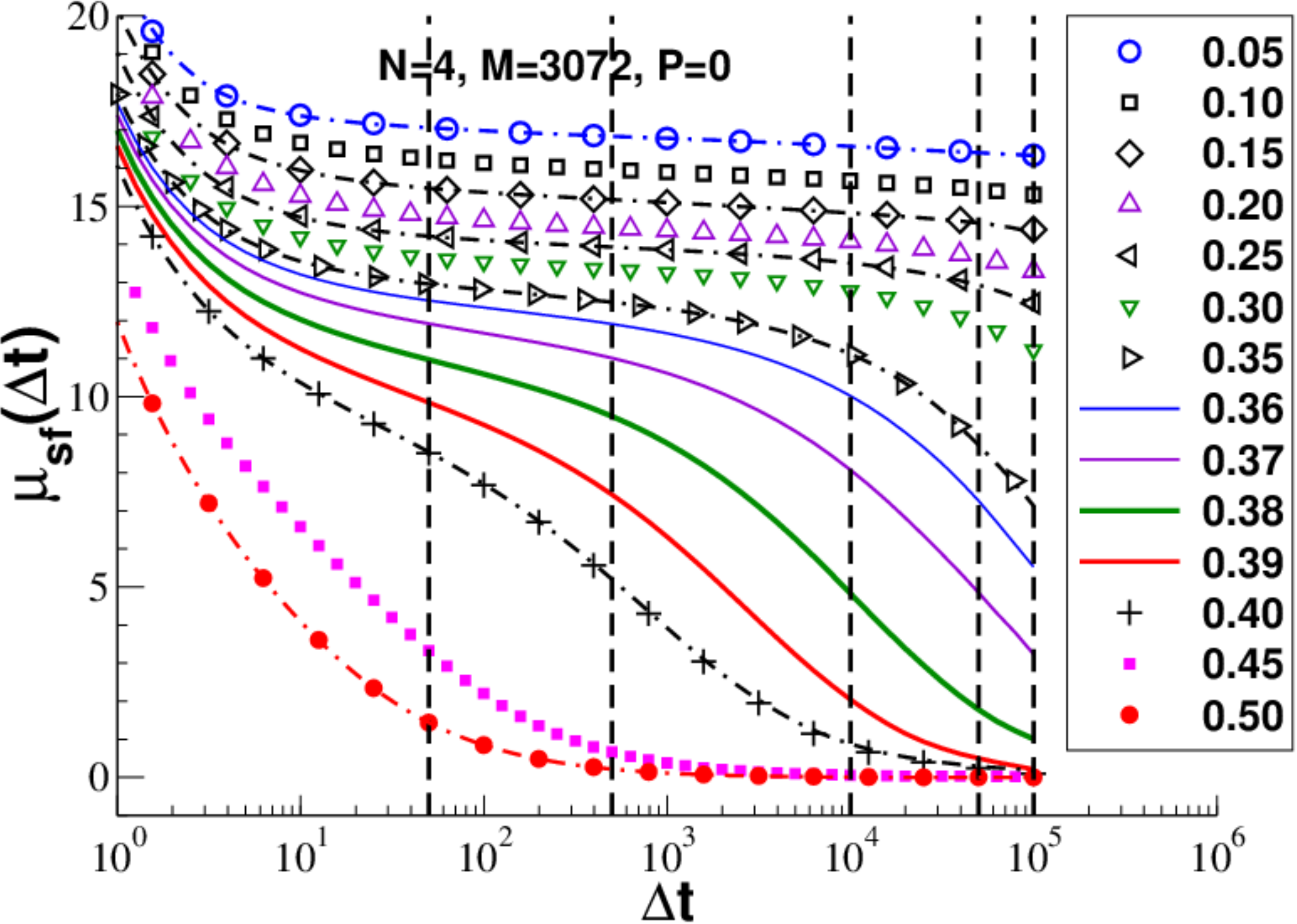}}}
\caption{Shear modulus $\GF$ as a function of sampling time $\tsamp$ for a broad range
of $T$ as indicated in the figure. $\GF(\tsamp)$ decreases continuously with $\tsamp$.
Note that a smaller temperature increment $\Delta T =0.01$ is used around $\Tglass$ (solid lines)
where $\GF(\tsamp;T)$ changes much more rapidly with $T$.
The vertical lines mark the sampling times used in Fig.~\ref{fig_mean_mu}.
The dash-dotted lines are obtained using Eq.~(\ref{eq_mu}) by integrating
the directly measured shear stress relaxation modulus $G(t)$.
}
\label{fig_GF_dt}
\end{figure}

While time and ensemble averages do commute for $\muA$ and $\muFtwo$, 
Eq.~(\ref{eq_commute}) does not hold for $\muFone$, $\muF$ and $\GF$.
We remind that even for permanent elastic networks these observables are
known to depend on $\tsamp$ \cite{WXP13,WXB15,WKC16}.
Since
\begin{equation}
\GF(\tsamp)=\muA - \muF(\tsamp) = (\muA-\muFtwo)+\muFone(\tsamp),
\label{eq_GFdt}
\end{equation}
we can focus here on the $\tsamp$-dependence of $\GF(\tsamp)$ as shown in Fig.~\ref{fig_GF_dt}.
Covering a broad range of temperatures we use subsets of length $\tsamp$ of the total
trajectories of length $\tsampmax$ stored.
It is seen that $\GF(\tsamp)$ decreases {\em both monotonously and continuously with $\tsamp$}.
The figure reveals that $\GF(\tsamp;T)$ decreases also {\em monotonously and continuously with $T$}.
Note that $\GF(\tsamp)$ increases for $T \to 0$ while its $\tsamp$-dependence becomes weaker.
A glance at Fig.~\ref{fig_GF_dt} shows that one expects the transition of $\GF(T;\tsamp)$ to get shifted
to lower $T$ and to become more step-like with increasing $\tsamp$ in agreement with Fig.~\ref{fig_mean_mu}.
It is, however, impossible to reconcile the data with a jump-singularity at a {\em finite} $\tsamp$ and $T$.
As announced in the Introduction, this is the first key result of the present work.
See Appendix~\ref{sm_highT} for the discussion of the technical importance of 
$\GF(\tsamp) \approx \muFone(\tsamp)$ at high temperatures.

\section{Standard deviations, distributions and correlations}
\label{sec_dGF}

\subsection{Standard deviation $\dGF$}
\label{res_fluctu_mu}

\begin{figure}[t]
\centerline{\resizebox{1.0\columnwidth}{!}{\includegraphics*{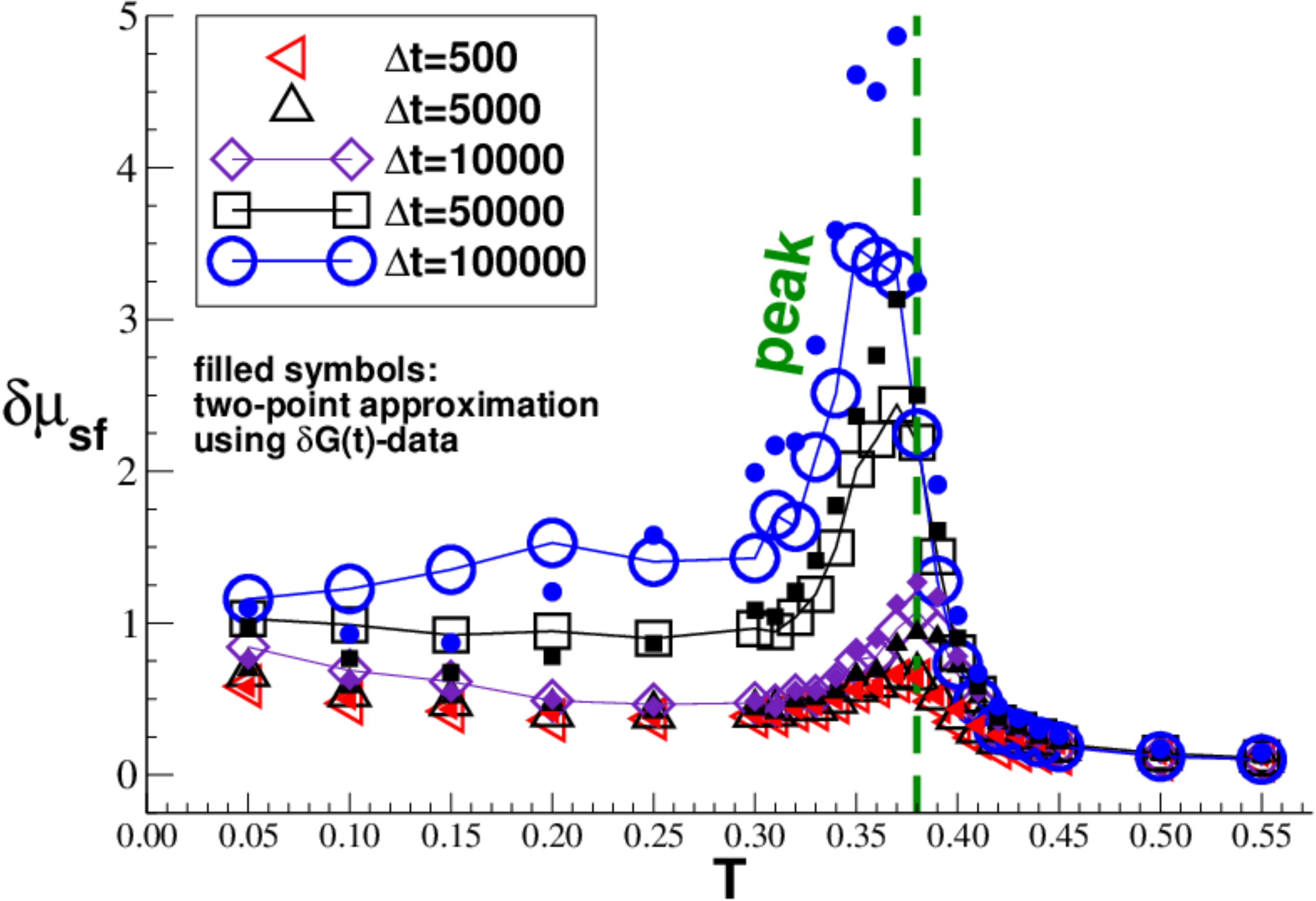}}}
\caption{Standard deviation $\dGF(T)$ for different sampling times $\tsamp$ using a linear representation.
The observed peak slightly below $\Tglass$ becomes sharper with increasing $\tsamp$.
The small filled symbols indicate the values predicted (Sec.~\ref{res_dGt}) 
according to the two-point approximation Eq.~(\ref{eq_dGt2dGF}) from the standard deviation $\delta G(t)$.
While this allows to relate $\GF(\tsamp)$ to $\delta G(t)$ for 
$T \ll \Tglass$ and $T \gg \Tglass$, it fails for large $\tsamp$ around the glass transition.
\label{fig_fluctu_mu}
}
\end{figure}

To characterize the fluctuations between different configurations we take for
various properties the second moment over the ensemble and compute,
e.g., the standard deviation $\dGF$ of the shear modulus, Eq.~(\ref{eq_dGF}).
As seen in Fig.~\ref{fig_fluctu_mu}, 
at variance to the monotonous modulus $\GF(T)$ its standard deviation $\dGF(T)$ is 
non-monotonous with a remarkable peak near $\Tglass$.
(As may be seen from Fig.~8 of Ref.~\cite{WKC16}, similar behavior has been observed
for systems of self-assembled transient networks.)
Note that while $\dGF(T)$ is essentially $\tsamp$-independent above $\Tglass$,
it increases systematically with $\tsamp$ below the transition.
Importantly, the peak of $\dGF(T)$ becomes about a third of the drop of 
the ensemble-averaged shear modulus $\GF(T)$ between $T=0.34$ and $T=0.38$ for $\tsamp=10^5$
(cf. Fig.~\ref{fig_mean_mu}). 
The liquid-solid transition characterized by $\GF(T)$ is thus accompanied 
by strong fluctuations between different quenched configurations.
We note finally that the relative standard deviation $\dGF(\tsamp)/\GF(\tsamp)$ 
is for all temperatures found to increase with $\tsamp$ (not shown). 
In the solid limit this is due to the increase of $\dGF(\tsamp)$,
in the liquid limit due to the $1/\tsamp$-decay of $\GF(\tsamp)$ 
discussed in Appendix~\ref{sm_highT} 
and for temperatures around $\Tglass$ due to a combination of both effects.

\subsection{Distribution of $\GFbar$}
\label{res_muhisto}

\begin{figure}[t]
\centerline{\resizebox{1.0\columnwidth}{!}{\includegraphics*{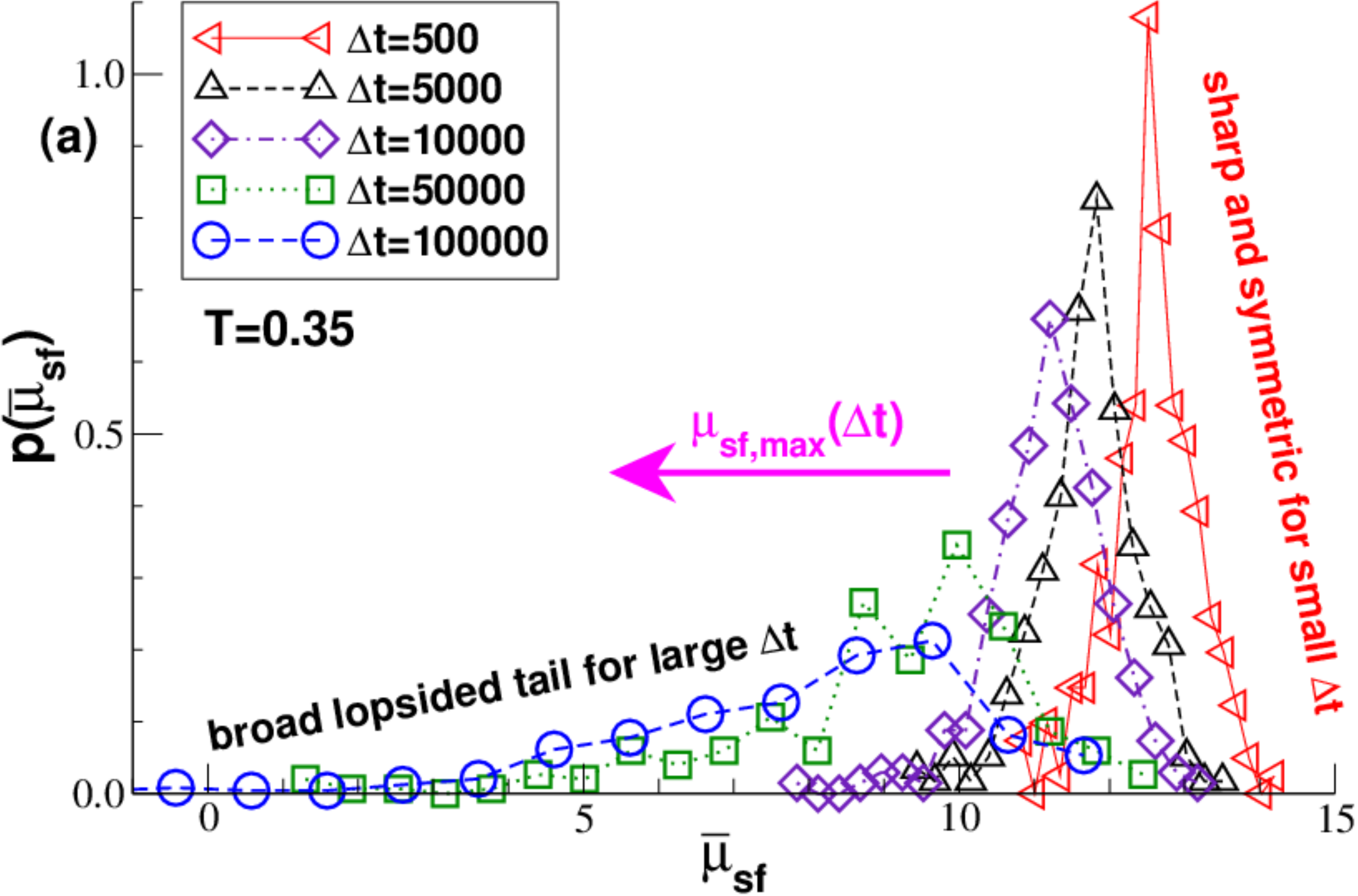}}}
\centerline{\resizebox{1.0\columnwidth}{!}{\includegraphics*{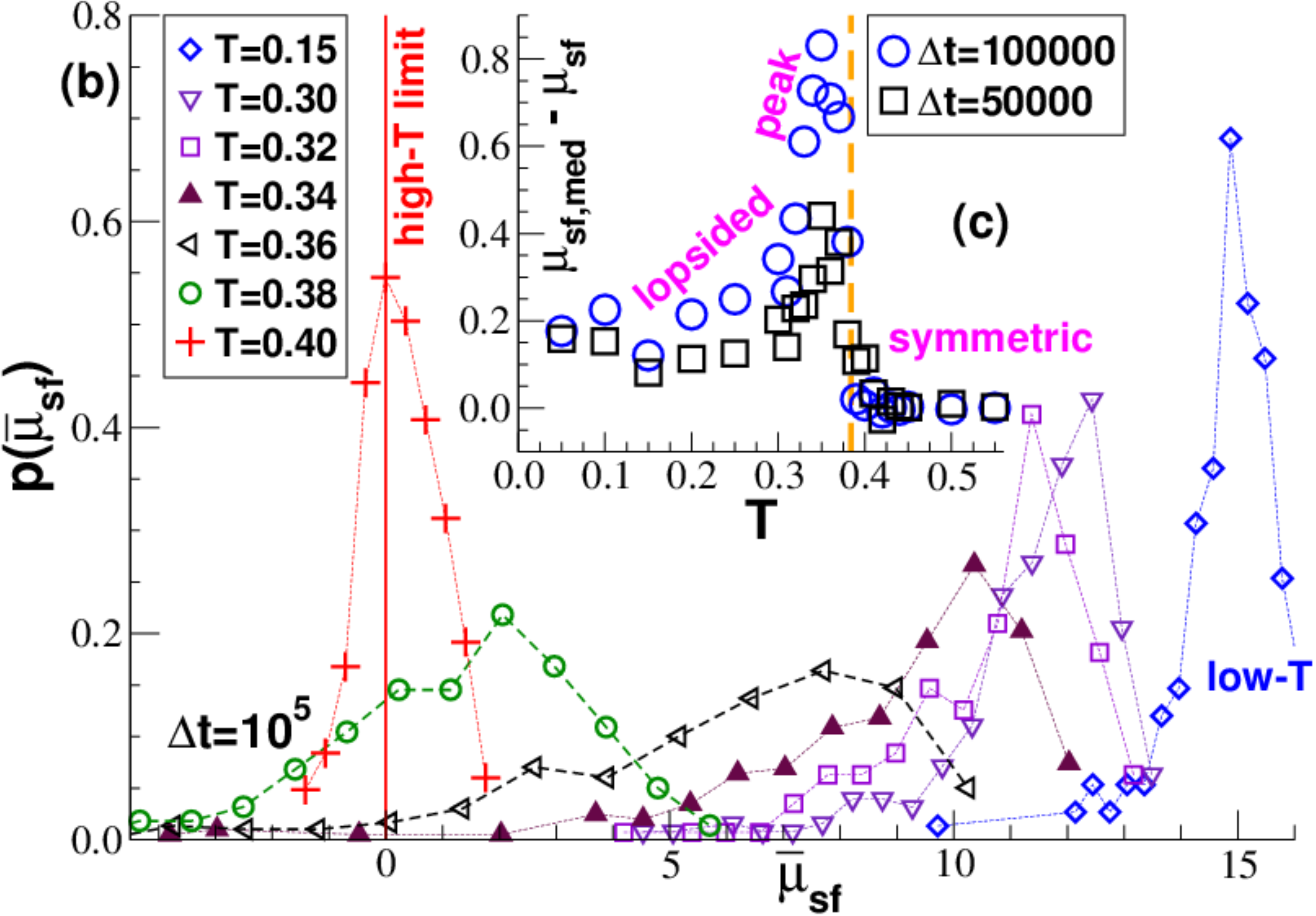}}}
\caption{Distribution $p(\GFbar)$ of the time-averaged modulus $\GFbar$:
{\bf (a)}
$p(\GFbar)$ for $T=0.35$ and several $\tsamp$ as indicated.
The maximum $\GFmax$ shifts to the left with increasing $\tsamp$
and the histogram becomes more lopsided.
{\bf (b)}
$p(\GFbar)$ for $\tsamp=10^5$ and a broad range of $T$.
{\bf (c)}
Difference $\GFmed-\GF$ of the median $\GFmed$ and the ensemble average $\GF$
{\em vs.} $T$ for two sampling times. The difference has a peak slightly below $\Tglass$
corresponding to very lopsided distributions.
}
\label{fig_muhisto}
\end{figure}

The striking peak of $\dGF$ near $\Tglass$ seen in Fig.~\ref{fig_fluctu_mu} begs for a more detailed
characterization of the distribution $p(\GFbar;T,\tsamp)$ of the time-averaged shear modulus $\GFbar$.
Using the available $3 \times m = 300$ independent measurements of $\GFbar$ 
this is presented in Fig.~\ref{fig_muhisto}.
We emphasize first of all that the histograms are {\em unimodal} for all $T$ and $\tsamp$.
The $T$-dependence of $\GF$ and $\dGF$ below $\Tglass$ is thus not due to, e.g., 
the superposition of two configuration populations representing either solid states with finite 
$\GFbar$ and liquid states with $\GFbar \approx 0$.
While the distributions depend only weakly (if at all) on $\tsamp$ in the liquid limit (not shown), 
the distributions become systematically broader and more lopsided with increasing $\tsamp$
below the glass transition temperature $\Tglass$ as seen in panel (a) for $T=0.35$.
This explains the increase of $\dGF$ with $\tsamp$ seen in Fig.~\ref{fig_fluctu_mu}.
Concurrently, the maximum $\GFmax$ and the median $\GFmed$ decrease systematically with increasing $\tsamp$.
Both trends are caused by the higher probability of plastic rearrangements if a configuration 
is probed over a larger time interval. 
Focusing on our largest sampling time $\tsampmax$, panel (b) of Fig.~\ref{fig_muhisto}
presents data for a broad range of temperatures. The maximum $\GFmax$ of the (unimodal) distribution 
systematically shifts to higher values below $\Tglass$
--- in agreement with its first moment $\GF$ (Fig.~\ref{fig_mean_mu}) ---
while the distributions become systematically broader and more lopsided,
i.e. liquid-like configurations with small $\GFbar$ become relevant.
For even smaller temperatures $T \ll \Tglass$, the distributions get again more focused around  
$\GFmax$ and less lopsided in agreement with Fig.~\ref{fig_fluctu_mu}. That the large standard deviations
and the asymmetry of the distributions are related is demonstrated
by comparing the first moment $\GF$ of the distribution, its median $\GFmed$
and its maximum $\GFmax$. One confirms that 
\begin{equation}
0 < \GFmed - \GF < \GFmax -\GF \mbox{ for } T < \Tglass 
\label{eq_GFmed}
\end{equation}
and for all $\tsamp$.
As seen in panel (c) of Fig.~\ref{fig_muhisto}, $\GFmed-\GF$ has a peak similar to $\dGF$
becoming sharper with increasing $\tsamp$.
%

\subsection{Comparison of related standard deviations}
\label{res_fluctu}
\begin{figure}[t]
\centerline{\resizebox{1.0\columnwidth}{!}{\includegraphics*{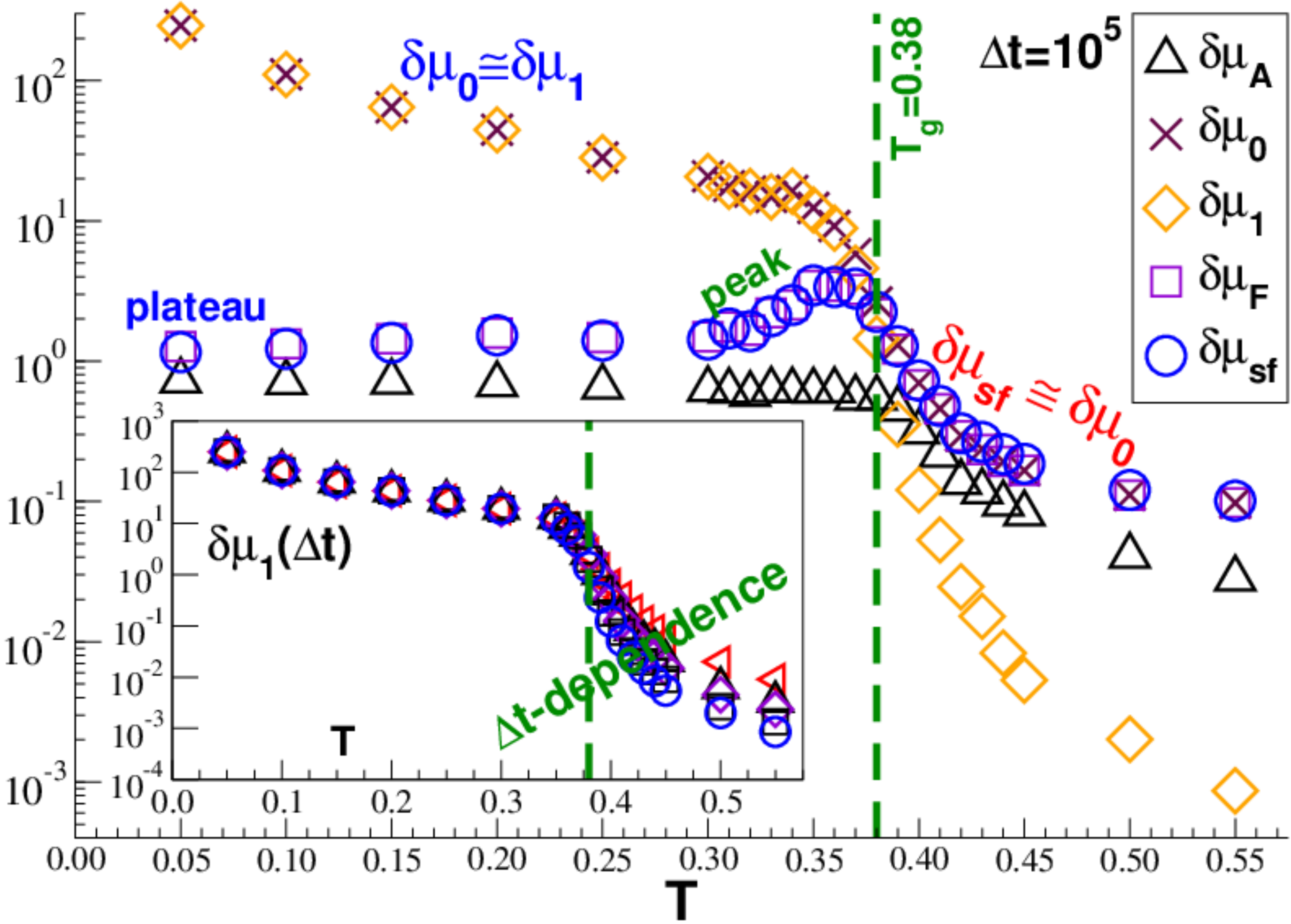}}}
\caption{Standard deviations $\dmuA$, $\dmuFtwo$, $\dmuFone$,
$\dmuF$ and $\dGF$ for $\tsamp=10^5$ as a function of $T$.
$\dmuA$ is found to be small and $\dGF \approx \dmuF$ for all $T$.
$\dmuFtwo$ and $\dmuFone$ become rapidly similar below $\Tglass$ and
orders of magnitude larger than $\dmuF$ confirming the presence of strong frozen shear stresses.
Inset: $\dmuFone(\tsamp)$ for several $\tsamp$ using the same symbols as in Fig.~\ref{fig_fluctu_mu}
for $\dGF(\tsamp)$. While sampling time effects are seen to be irrelevant for $T \ll \Tglass$,
i.e. the frozen stresses cannot relax, they matter for temperatures around and above $\Tglass$.
\label{fig_fluctu}
}
\end{figure}

Using a half-logarithmic representation $\dGF$ is replotted in the main panel of Fig.~\ref{fig_fluctu}
together with the corresponding standard deviations $\dmuA$, $\dmuFtwo$, $\dmuFone$ and $\dmuF$.
Please note that for all these standard deviations ensemble and time averages {\em do not commute},
i.e. these properties depend in principle on the sampling time as we have already seen
for $\dGF$ in Fig.~\ref{fig_fluctu_mu}. Another example is given by $\dmuFone(\tsamp)$ 
in the inset of Fig.~\ref{fig_fluctu} showing that the deviations decrease more rapidly
for larger temperatures with increasing $\tsamp$. The logarithmic scale used for the
vertical axis masks somewhat the effect better visible in linear coordinates.
Returning to the main panel of Fig.~\ref{fig_fluctu} we emphasize first of all
that $\dmuA$ is negligible and $\dGF \approx \dmuF$ for all $T$.
In the high-$T$ regime we find $\dGF \approx \dmuFtwo$ while 
$\dmuFone$ vanishes much more rapidly. 
Interestingly, in the opposite glass-limit $\dGF \approx \dmuF$ becomes 
orders of magnitude smaller than $\dmuFtwo \approx \dmuFone$.
The contributions $\muFtwobar$ and $\muFonebar$ of the difference 
$\muFbar=\muFtwobar-\muFonebar$ thus must be strongly correlated. 

\subsection{Correlations between $\muFtwobar$ and $\muFonebar$}
\label{res_coeffonetwo}

\begin{figure}[t]
\centerline{\resizebox{1.0\columnwidth}{!}{\includegraphics*{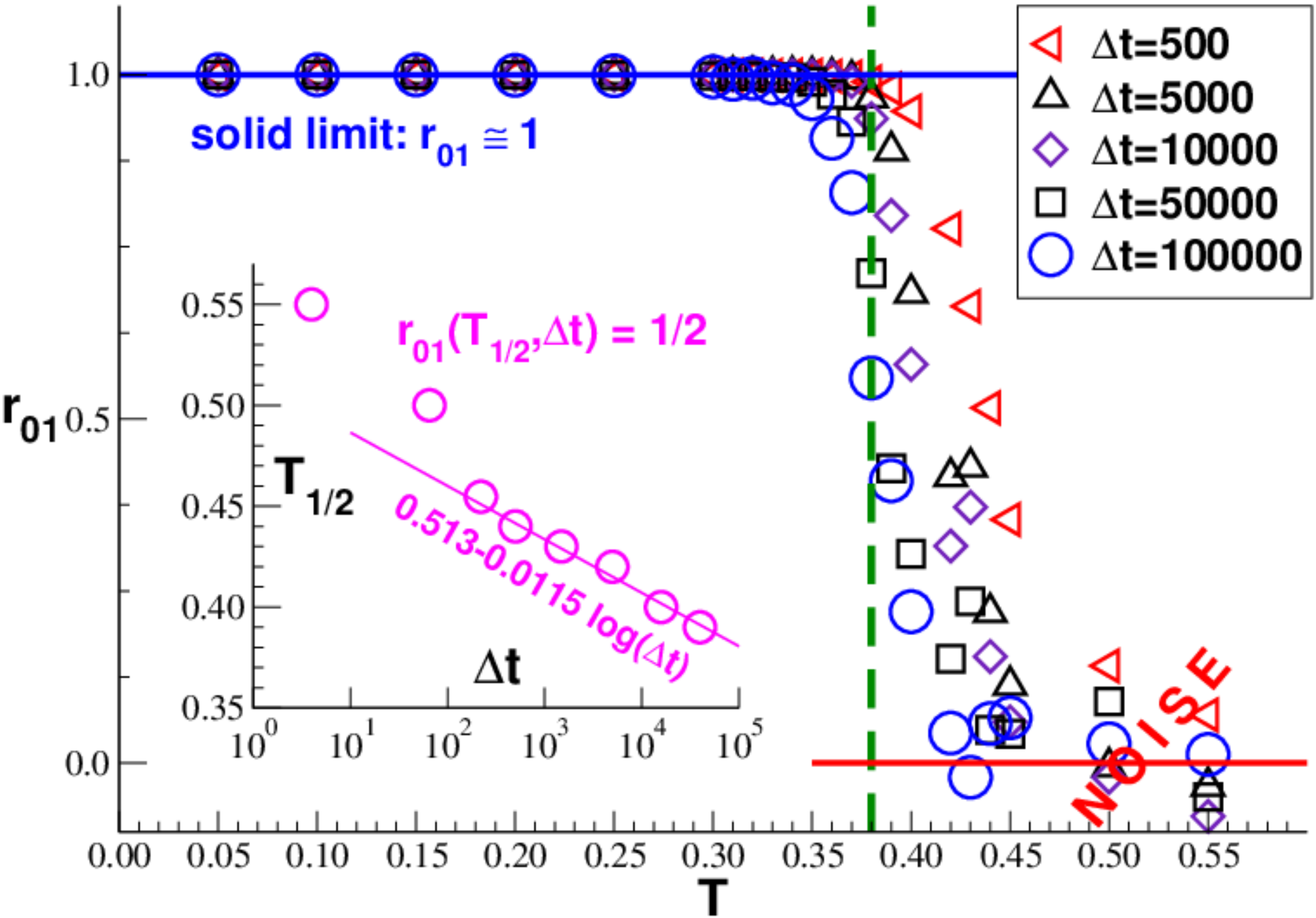}}}
\caption{Correlation coefficient $\coeffonetwo$ as a function of $T$
for different $\tsamp$ showing that $\coeffonetwo \to 0$ for $T \gg \Tglass$
and $\coeffonetwo \approx 1$ for $T \ll \Tglass$. The transition around $T \approx \Tglass$
depends again on $\tsamp$.
Inset: $\Thalf(\tsamp)$ defined by $\coeffonetwo(\Thalf,\tsamp)=1/2$
reveals a logarithmic decay of the correlations with increasing $\tsamp$.
\label{fig_coeffonetwo}
}
\end{figure}
This can be directly verified using the corresponding dimensionless correlation coefficient
\begin{equation}
\coeffonetwo \equiv
\frac{\la (\muFtwobar-\muFtwo) (\muFonebar-\muFone) \ra}{\dmuFtwo \ \dmuFone}.
\label{eq_coeffonetwo}
\end{equation}
As can be seen from Fig.~\ref{fig_coeffonetwo},
$\coeffonetwo(T,\tsamp)$ provides an operationally simple and clear-cut order parameter between 
the liquid limit ($\coeffonetwo \to 0$) and the solid regime ($\coeffonetwo \to 1$).
The latter limit is another manifestation of the frozen shear stresses.
Quite generally, one may write the variance $\dmuF^2$ of $\muF=\muFtwo-\muFone$ as 
\begin{equation}
\dmuF^2 = \dmuFtwo^2 + \dmuFone^2 - 2 \coeffonetwo \dmuFtwo \dmuFone.
\label{eq_crosscorr}
\end{equation}
Since $\coeffonetwo \approx 1^{-}$ for $T \ll \Tglass$, we have $\dmuF \approx \dmuFtwo - \dmuFone \ge 0$.
This explains why
\begin{itemize}
\item
$\dmuF$ becomes very small albeit the fluctuations of its contributions 
$\muFtwo$ and $\muFone$ are large (Fig.~\ref{fig_fluctu}) and
\item
the stress-fluctuation formula, Eq.~(\ref{eq_GF}), for the shear modulus remains a statistically successful 
approach in the solid limit despite the fact that violent stress fluctuations occur between different 
configurations of the ensemble.
\end{itemize}
Please note also that the correlation coefficient $\coeffonetwo(T)$ is again continuous 
and does depend somewhat on the sampling time $\tsamp$. This $\tsamp$-dependence may be characterized
as shown in the inset of Fig.~\ref{fig_coeffonetwo} by means of a temperature $\Thalf$ defined by
$\coeffonetwo(\Thalf,\tsamp) = 1/2$. It is seen that $\Thalf(\tsamp)$ decays logarithmically with $\tsamp$ ---
at least for the $\tsamp$-range we are able to probe. Longer production runs are warranted to clarify
whether $\Thalf(\tsamp)$ continues to decrease as we strongly expect.

\section{Shear stress relaxation modulus}
\label{sec_Gt}

\subsection{Qualitative description of $G(t)$}
\label{res_Gt}

\begin{figure}[t]
\centerline{\resizebox{1.0\columnwidth}{!}{\includegraphics*{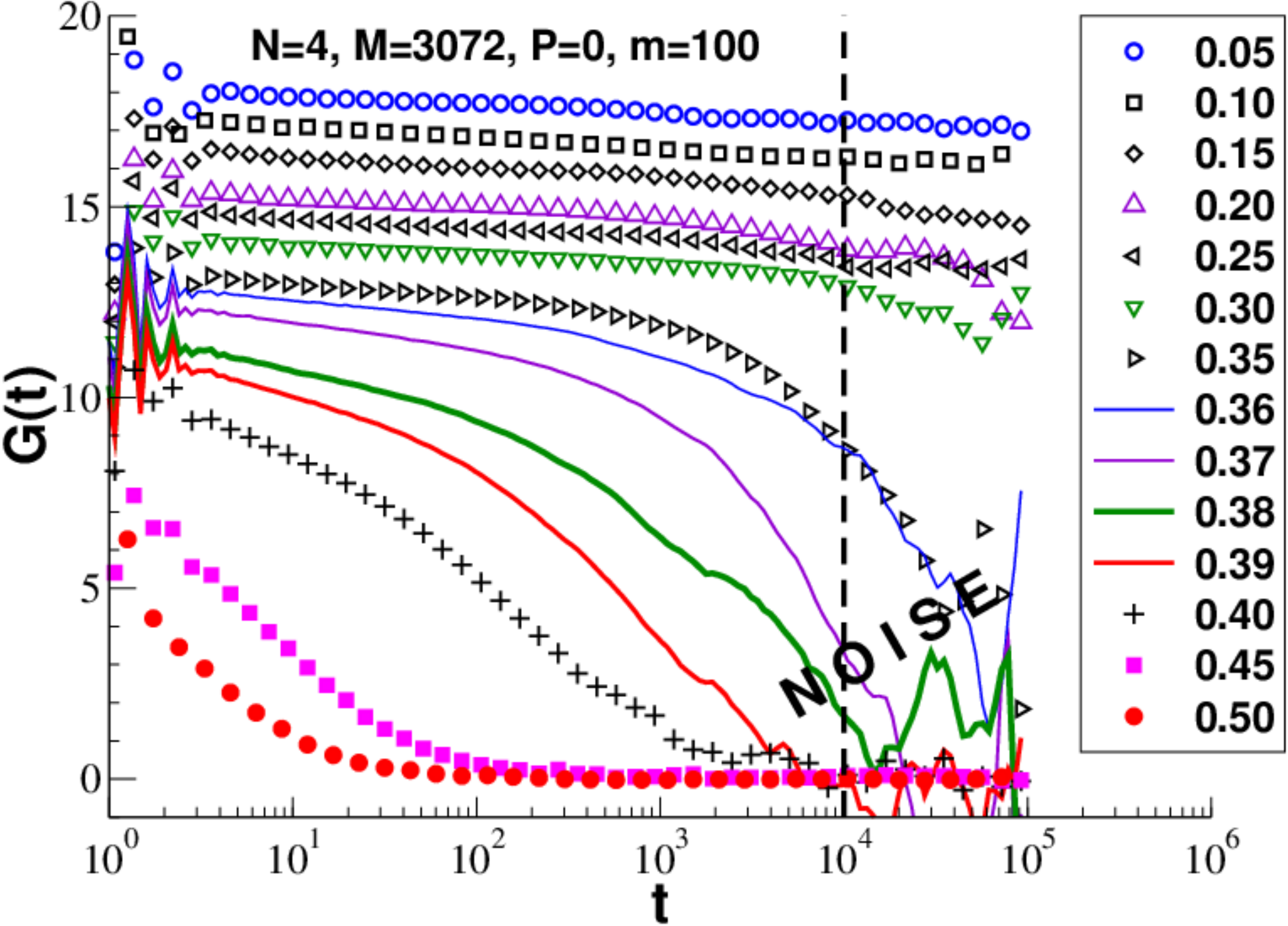}}}
\caption{Stress relaxation modulus $G(t)$ for a broad range of $T$ using half-logarithmic coordinates.
$G(t)$ has been obtained by means of Eq.~(\ref{eq_Gt}) using gliding averages,
i.e. the statistics deteriorates for $t \to \tsamp$ and
the data have been logarithmically averaged for clarity.
The dashed vertical line marks the time used for $G(t)$ in Fig.~\ref{fig_mean_mu}
and Table~\ref{tab_T}.
}
\label{fig_Gt}
\end{figure}

The shear stress relaxation modulus $G(t)$ is an experimentally important observable
\cite{DoiEdwardsBook,WittenPincusBook,RubinsteinBook}.
As shown in Fig.~\ref{fig_Gt}, we have computed $G(t)$ by means of the fluctuation-dissipation 
relation Eq.~(\ref{eq_Gt}) appropriate for canonical ensembles with quenched or sluggish shear 
stresses \cite{WXB16,ivan17a}. This allows us to also sample $G(t)$ below $\Tglass$ where 
$\muFtwo-\muA$ becomes finite as shown in Sec.~\ref{res_static}. 
(See also Appendix~\ref{sm_Gt}.)
Since $G(t)$ is obtained by means of gliding averages along the time series, 
Eq.~(\ref{eq_Gtbar}), the statistics deteriorates for $t \to \tsamp$.
(This will be quantified in Sec.~\ref{res_dGt} where we discuss the standard deviation $\dGt$.)
We have thus logarithmically averaged the data presented in Fig.~\ref{fig_Gt}. 
This suppresses somewhat the oscillations at small times $t \ll 10$, which are, however, irrelevant 
for the present study.
We emphasize the following qualitative properties: 
\begin{itemize}
\item
As it should, $G(t)$ vanishes for large times $t \gg \tauinf(T)$ in the liquid limit above $\Tglass$.
\item
$G(t)$ increases monotonously and continuously with decreasing temperature.
\item
This increase is, however, especially strong around $\Tglass$ where the solid lines indicate,
as in Fig.~\ref{fig_GF_dt}, a smaller temperature increment $\Delta T = 0.01$. 
\item
Albeit we average over $m=100$ configurations and three shear planes, 
$G(t)$ remains rather noisy for $T \approx \Tglass$ and $t > 10^4$.
(See Sec.~\ref{res_dGt} for details.)
\item
$G(t)$ decreases only weakly within the available time window below $T=0.3$.
\end{itemize}
The data presented in Fig.~\ref{fig_Gt} are thus qualitatively very similar to the 
shear modulus $\GF(\tsamp)$ given in Fig.~\ref{fig_GF_dt}.
We describe now $G(t)$ more quantitatively by focusing on the $\tsamp$-dependent moments 
$\mu(\tsamp)$ and $\eta(\tsamp)$ defined in the Introduction.

\subsection{Connection between $\GF(\tsamp)$ and $G(t)$}
\label{res_GtGF}

\begin{figure}[t]
\centerline{\resizebox{1.0\columnwidth}{!}{\includegraphics*{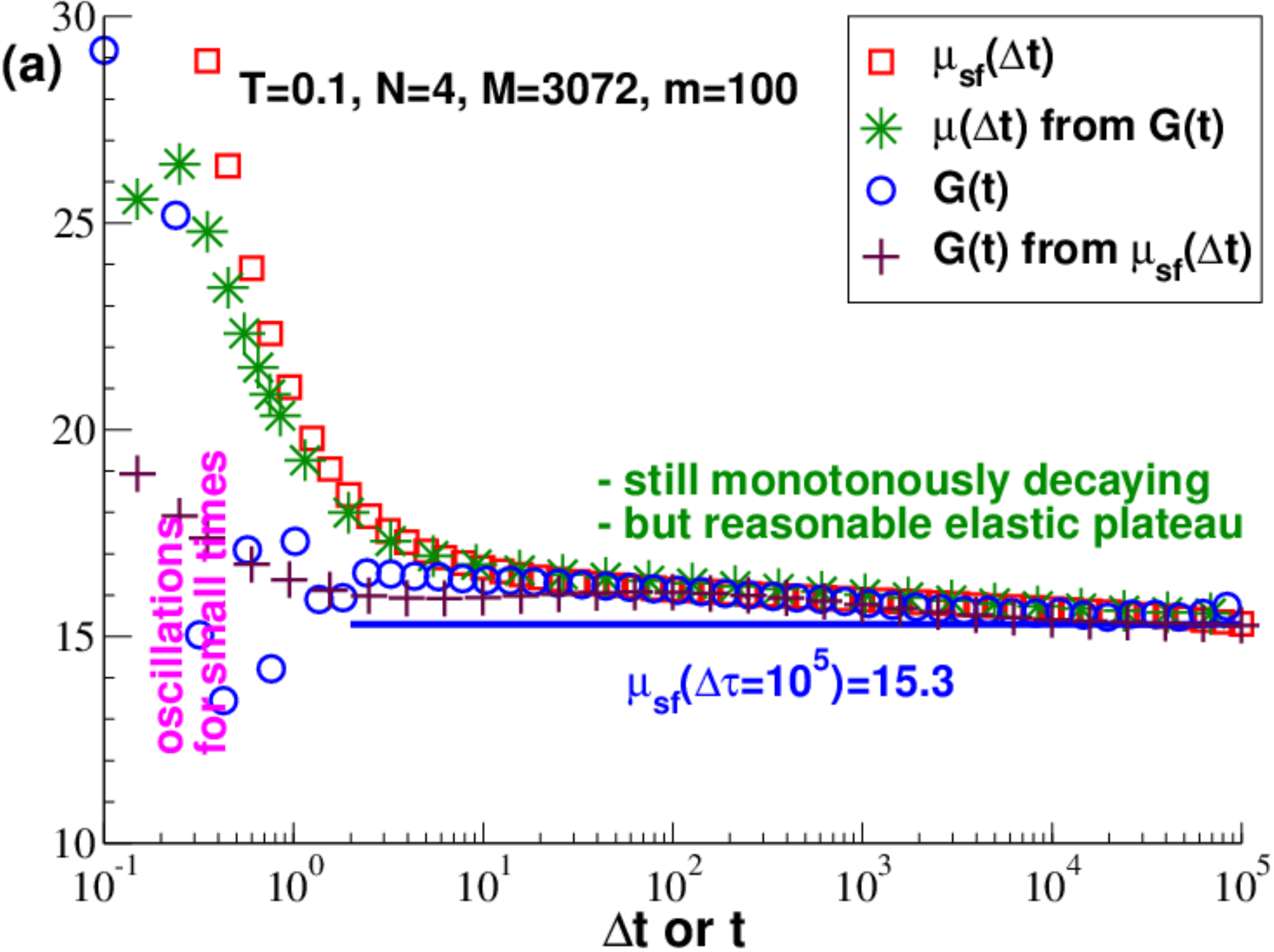}}}
\centerline{\resizebox{1.0\columnwidth}{!}{\includegraphics*{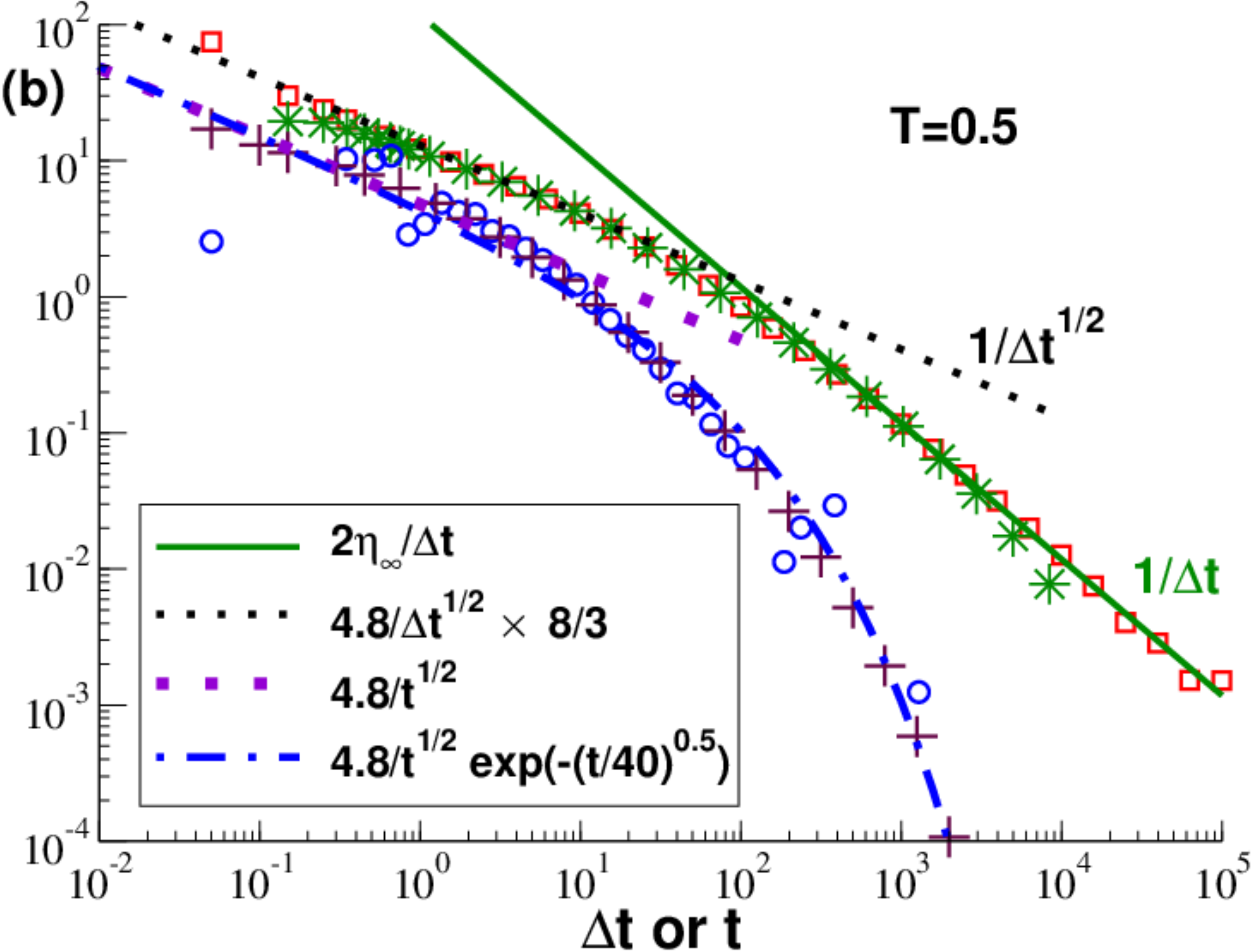}}}
\caption{Comparison of $\GF(\tsamp)$, $\mu(\tsamp)$ and $G(t)$ for $T=0.1$ and $T=0.5$.
$\GF(\tsamp)$ is indicated by squares, $\mu(\tsamp)$ by stars and $G(t)$ by open
cercles if obtained using Eq.~(\ref{eq_Gt}) and by pluses if obtained using 
Eq.~(\ref{eq_mu2Gt}) from the $\GF(\tsamp)$-data.
It is seen that $\GF(\tsamp) \approx \mu(\tsamp)$ and that
also both expressions for $G(t)$ are essentially identical.
Panel {\bf (a)}:
At sufficiently low temperatures $\mu(\tsamp) \approx G(t =\tsamp)$ over a broad plateau.
Panel {\bf (b)}:
At larger temperatures and longer times
$G(t)$ decays faster than its integral $\mu(\tsamp)$.
The dotted lines indicate   
$\mu(\tsamp) \sim 1/\tsamp^{\alpha}$ and $G(t) \sim 1/t^{\alpha}$ with $\alpha=1/2$.
\label{fig_GtGF}
}
\end{figure}

Using the generic $\tsamp$-dependence of time-averaged fluctuations \cite{LandauBinderBook},
we relate now the sampling time dependence of $\GF(\tsamp)$, shown in Fig.~\ref{fig_GF_dt} and 
Fig.~\ref{fig_GtGF}, to the time dependence of the relaxation modulus $G(t)$.
As shown in Appendix~\ref{sm_GtGF}, assuming {\em time-translational invariance}
$\GF(\tsamp)$ should be equivalent to the weighted average $\mu(\tsamp)$
introduced in Eq.~(\ref{eq_mu}) in the Introduction.
Using the directly determined $G(t)$ described in Sec.~\ref{res_Gt}, one readily computes 
$\mu(\tsamp)$ as shown by the thin dash-dotted lines indicated in Fig.~\ref{fig_GF_dt}
and the stars in Fig.~\ref{fig_GtGF} for one very low ($T=0.1$) and one very high ($T=0.5$)
temperature. 
Since $\GF(\tsamp) \approx \mu(\tsamp)$ is found to high accuracy for {\em all} $T$, 
this confirms the assumed time-translational invariance. The $\tsamp$-dependence of $\GF$, 
$\muFone$ and $\muF$ is thus simply due to the upper bound $\tsamp$ used to average 
the relaxation modulus $G(t)$. It is {\em not} due to aging or equilibration problems, 
but to the finite time needed for $G(t)$ to reach its asymptotic limit.
This confirms the third key result of the present work announced in the Introduction.
Since $\GF$ and $\mu$ are identical within numerical accuracy, 
we often drop the notation $\GF$ below. 
%

\subsection{Further consequences}
\label{res_GtGF_continued}

Using integration by parts it is seen that the definition Eq.~(\ref{eq_mu})
is equivalent to
\begin{equation}
\mu(\tsamp) = \frac{2}{\tsamp^2} \int_0^{\tsamp} \ddiff t \int_0^t \ddiff \tp \ G(\tp).
\label{eq_GF_Gt_two}
\end{equation}
This implies that
\begin{eqnarray}
G(t) & = & \left[ t^2 \mu(t)/2 \right]^{\prime\prime}
\label{eq_mu2Gt} \\
& = & \mu(t) + 2 t \mu^{\prime}(t) + t^2 \mu^{\prime\prime}(t)/2
\label{eq_mu2Gt_two}
\end{eqnarray}
where a prime denotes a derivative with respect to the argument.
This suggests that one may use the smooth $\GF(\tsamp)$ presented in Fig.~\ref{fig_GF_dt} to compute $G(t)$. 
This can be done by fitting first to sixth order $y \equiv \ln(\GF(\tsamp)\tsamp^2/2)$ 
as a function of $x \equiv \ln(\tsamp)$ and tracing then 
\begin{equation}
G(t) = \operatorname{e}^{y(x)-2x} \left[ y^{\prime\prime}(x)+y^{\prime}(x) \ (y^{\prime}(x) -1)  \right].
\label{eq_mu2Gt_three}
\end{equation} 
As indicated by the pluses in Fig.~\ref{fig_GtGF},
this yields essentially identical results as the directly computed $G(t)$,
however, with much less noise, especially for large $t$.
Since $\muA=\muFtwo$ for high temperatures (Fig.~\ref{fig_static}), we recommend to replace 
in this limit $\GF(\tsamp)$ by $\muFone(\tsamp)$ to avoid the impulsive corrections discussed 
in Appendix~\ref{sm_highT}.

We emphasize that albeit $\mu(\tsamp)$ and $G(t)$ are similar for all $T$, as may be seen from 
Fig.~\ref{fig_GtGF} and by comparing Fig.~\ref{fig_GF_dt} and Fig.~\ref{fig_Gt}, 
both quantities are only identical if according to Eq.~(\ref{eq_mu2Gt_two})
the second and the third term in Eq.~(\ref{eq_mu2Gt_two}) are negligible compared to the first one.
As may be seen from panel (a) of Fig.~\ref{fig_GtGF}, this is the case in the solid limit 
where $G(t)$ may be well approximated by a constant $\mustar(T)$. Equation~(\ref{eq_mu}) 
thus implies 
\begin{equation}
G(t) \approx \mu(\tsamp) \approx \mustar(T) \mbox{ for } T \le 0.2.
\label{eq_Gt_solidlimit}
\end{equation} 
See the discussion around Eq.~(\ref{eq_constfunc}) in Appendix~\ref{sm_GtGF}.
This becomes different for higher temperatures where relaxation processes become much more important.
As seen for one temperature in panel (b), 
\begin{equation}
\mu(\tsamp) > G(t \approx \tsamp) \mbox{ for } T > 0.2 \mbox{ and } \tsamp \ll \tauinf. 
\label{eq_Gt_inequal}
\end{equation}
The inequality is due to the strong weight of small times $t \ll \tsamp$ in Eq.~(\ref{eq_mu}). 
$\mu(\tsamp)$ converges thus more slowly to any intermediate plateau or the asymptotic
limit ($\tsamp \gg \tauinf$) as the relaxation modulus $G(t)$.
A particular interesting case (not only from the polymer physics point of view) arises 
if the relaxation modulus does not have an intrinsic time scale over a sufficiently broad 
time window and decays as a power law $G(t) \sim 1/t^{\alpha}$ with $1 > \alpha > 0$.
Using Eq.~(\ref{eq_mu2Gt}) it is seen that $\mu(\tsamp)$ decays with the same exponent $\alpha$
and the relative amplitudes are given by
\begin{equation}
\frac{\mu(\tsamp)}{G(t=\tsamp)} = \frac{2}{(1-\alpha)(2-\alpha)} = \frac{8}{3} \mbox{ for } \alpha=1/2.
\label{eq_amplitudes}
\end{equation}
The ratio $8/3$ is consistent with the two dotted power-law slopes indicated in panel (b) 
of Fig.~\ref{fig_GtGF}. 
Note that our chains are too short to reveal a full Rouse dynamics \cite{DoiEdwardsBook}
and the observed $\alpha=1/2$ is due to the superposition of various different effects
(e.g., small-$N$ corrections, crossover from the short-time dynamics to the terminal relaxation).
The point we want to make here is merely that whenever scale-free physics characterized
by an exponent $\alpha < 1$ arises, this implies that $\mu(\tsamp)/G(t \approx \tsamp)$ 
must be constant, however, with a constant larger than unity.

To summarize, $\mu(\tsamp)=\GF(\tsamp)$ only corresponds to a classical thermodynamic
(and thus $\tsamp$-independent) shear (storage) modulus if $G(t)$ becomes essentially constant 
over a broad time window and if the sampling time $\tsamp$ is sufficiently large to probe this window. 
In all other cases $\mu(\tsamp)$ should be seen as a {\em generalized} shear modulus 
or a strong smoothing function over $G(t)$ containing also information from dissipation processes.
That this is indeed the case will be shown in Sec.~\ref{res_eta}.
%

\subsection{Shear viscosity $\eta(\tsamp)$}
\label{res_eta}

\begin{figure}[t]
\centerline{\resizebox{1.0\columnwidth}{!}{\includegraphics*{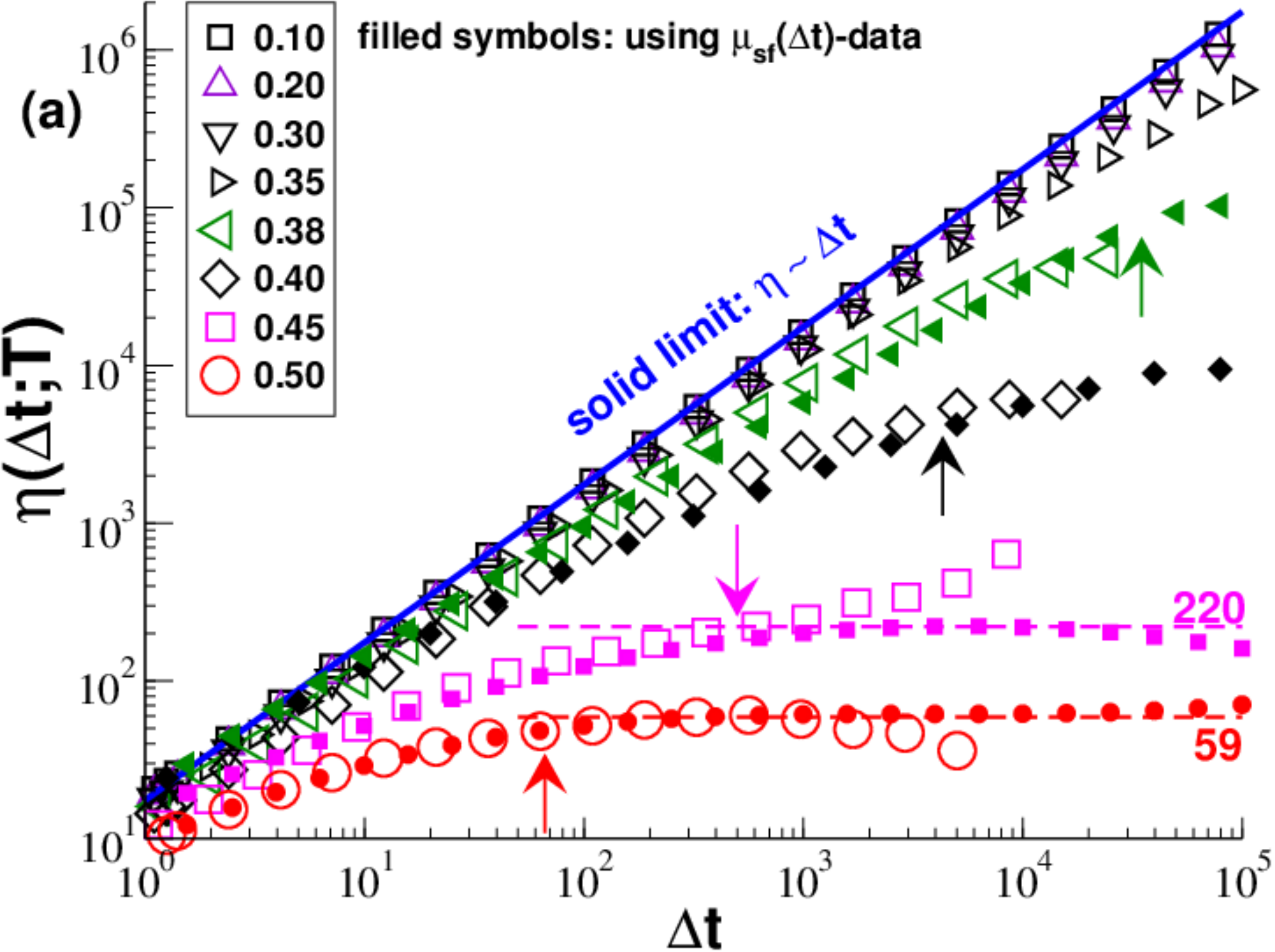}}}
\centerline{\resizebox{1.0\columnwidth}{!}{\includegraphics*{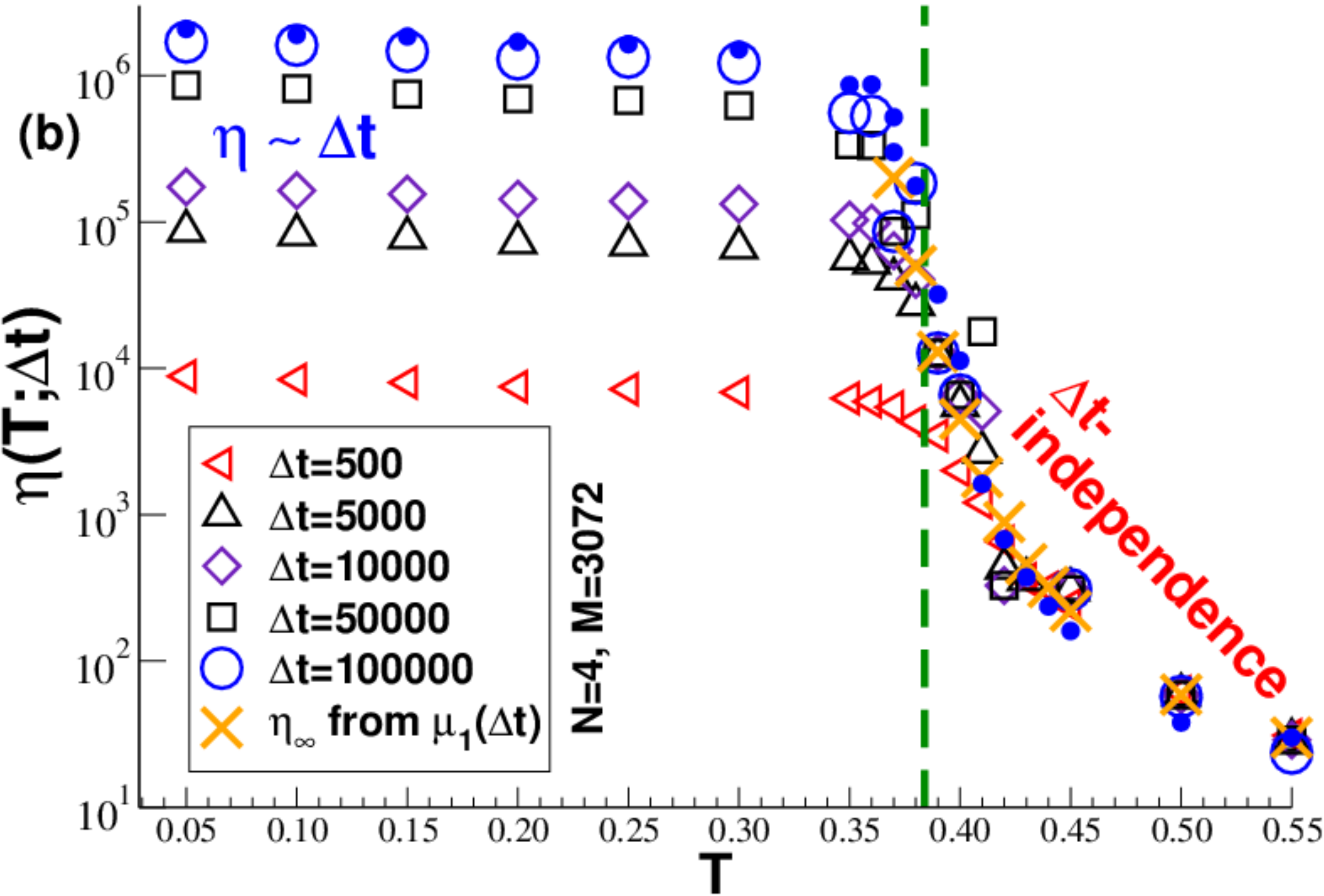}}}
\caption{Generalized shear viscosity $\eta$ with open symbols obtained
using Eq.~(\ref{eq_eta}) and filled symbols using Eq.~(\ref{eq_GF2eta_two}):
{\bf (a)} 
$\eta(\tsamp;T)$ as a function of $\tsamp$ for several $T$.
As shown by the solid line, $\eta(\tsamp)$ increases linearly in the solid limit.
At higher temperatures $\eta(\tsamp)$ increases less strongly and eventually levels off.
The two dashed lines indicate the limits $\eta(\tsamp) \to \etainf(T)$ for $T=0.5$ and $T=0.45$.
The vertical arrows mark for several temperatures the approximative position of the
terminal relaxation time $\tauinf(T)$.
{\bf (b)}
$\eta(T;\tsamp)$ as a function of $T$ for different $\tsamp$.
The crosses represent $\etainf(T)$ from the inset of Fig.~\ref{fig_mu1_dt}. 
}
\label{fig_eta}
\end{figure}

\paragraph*{Standard operational definition.}
We have shown in Sec.~\ref{res_GtGF} that (at least for the glass-forming polymer model investigated) 
the stress-fluctuation formula Eq.~(\ref{eq_GF}) for the shear modulus is equivalent 
to the moment $\mu(\tsamp)$ over $G(t)$, Eq.~(\ref{eq_mu}).
In order to further describe the relaxation modulus $G(t)$,
it is convenient to compute the generalized ($\tsamp$-dependent) shear viscosity
$\eta(\tsamp)$ defined by Eq.~(\ref{eq_eta}) in the Introduction.
As shown by the open symbols in Fig.~\ref{fig_eta}, we have thus computed $\eta$ 
both as a function of $\tsamp$ and as a function of $T$.
As emphasized by the bold solid line in panel (a), $\eta(\tsamp;T) \approx \mustar(T) \tsamp$ 
in the solid limit where $G(t) \approx \mustar(T)$ becomes constant. As one also expects, 
$\eta(\tsamp;T)$ levels off, i.e. becomes $\tsamp$-independent for sufficiently high temperatures 
where $\tsamp$ exceeds the terminal relaxation time $\tauinf$ 
(as characterized using the $\muFone(\tsamp)$-rescaling in Appendix~\ref{sm_highT}).
Please note that the determination of $\eta(\tsamp)$ using Eq.~(\ref{eq_eta})
for the liquid limit is notoriously difficult \cite{AllenTildesleyBook} 
due to the inaccuracy of $G(t) \approx 0$ for large times $t \to \tsamp$. 
Since the noisy $G(t)$ may even become negative, $\eta(\tsamp)$ can get 
non-monotonous as may be seen for $T=0.5$. 
The observed maximum gives in this case a (very rough) guess of $\etainf(T)$.
The open symbols indicated in panel (b) have thus been obtained by terminating
the integration of $G(t)$ as soon as a negative $G(t)$-fluctuation becomes too important.

\paragraph*{Connection between $\mu(\tsamp)$ and $\eta(\tsamp)$.}
Since $\mu(\tsamp)$ and $\eta(\tsamp)$ are moments of the same function $G(t)$,
they must be connected. It follows from Eq.~(\ref{eq_mu2Gt}) that 
\begin{equation}
\eta(\tsamp) = \frac{\ddiff}{\ddiff \tsamp} \left( \frac{\tsamp^2}{2} \ \mu(\tsamp) \right). 
\label{eq_GF2eta}
\end{equation}
Since $\mu(\tsamp) = \GF(\tsamp)$, $\eta(\tsamp)$ may be determined
by numerical differentiation of the smooth $\GF(\tsamp)$-data. 
(In the liquid limit $\GF(\tsamp)$ may be replaced by $\muFone(\tsamp)$.)
Using the same notations as in Eq.~(\ref{eq_mu2Gt_three}) this implies
\begin{equation}
\eta(\tsamp) = \operatorname{e}^{y(x)-x} y^{\prime}(x).
\label{eq_GF2eta_two}
\end{equation}
Results obtained for several temperatures are shown by the filled symbols in Fig.~\ref{fig_eta}.
While both methods yield equivalent results for small temperatures and for small $\tsamp$,
the second method allows a slightly improved characterization of the plateau value $\etainf$
in the liquid limit.
Even more importantly, it follows directly from Eq.~(\ref{eq_GF2eta}) that our forth key result, 
Eq.~(\ref{eq_etainf}), must hold for $\tsamp \gg \tauinf$.
Using in addition that $\GF(\tsamp) \approx \muFone(\tsamp)$ for high temperatures, we have 
used Eq.~(\ref{eq_etainf}) in Appendix~\ref{sm_highT} to determine $\etainf(T)$ from $\muFone(\tsamp)$.
The corresponding data are indicated by the two horizontal dashed lines for $T=0.5$ and $T=0.45$ 
in panel (a) of Fig.~\ref{fig_eta} and by crosses in panel (b). As one expects, it is seen that 
$\eta(\tsamp) \to \etainf(T)$ in the large-$\tsamp$ limit.

\paragraph*{Continuous behavior.}
As may be seen from panel (b) of Fig.~\ref{fig_eta}, $\eta(T;\tsamp)$ increases
monotonously with decreasing temperature around $\Tglass$ for all $\tsamp$.
Albeit this increase becomes for larger $\tsamp$ more dramatic and numerically
more difficult to describe, it remains {\em continuous} for all $\tsamp$,
especially for the asymptotic limit $\etainf(T)$.
This is expected from the $\GF$-data presented in Fig.~\ref{fig_mean_mu} and Fig.~\ref{fig_GF_dt}.
Interestingly, the last argument can be turned around.
If one accepts that $\eta(\tsamp;T)$ is monotonous and continuous for both $\tsamp$ and $T$,
this implies that
\begin{equation}
\GF(\tsamp;T) = \mu(\tsamp;T) = \frac{2}{\tsamp^2} \int_0^{\tsamp} \ddiff t \ \eta(t;T) 
\label{eq_eta2GF}
\end{equation}
must have the same properties \cite{foot_intconst}.
%

\subsection{Standard deviation $\dGt$}
\label{res_dGt}

\begin{figure}[t]
\centerline{\resizebox{1.0\columnwidth}{!}{\includegraphics*{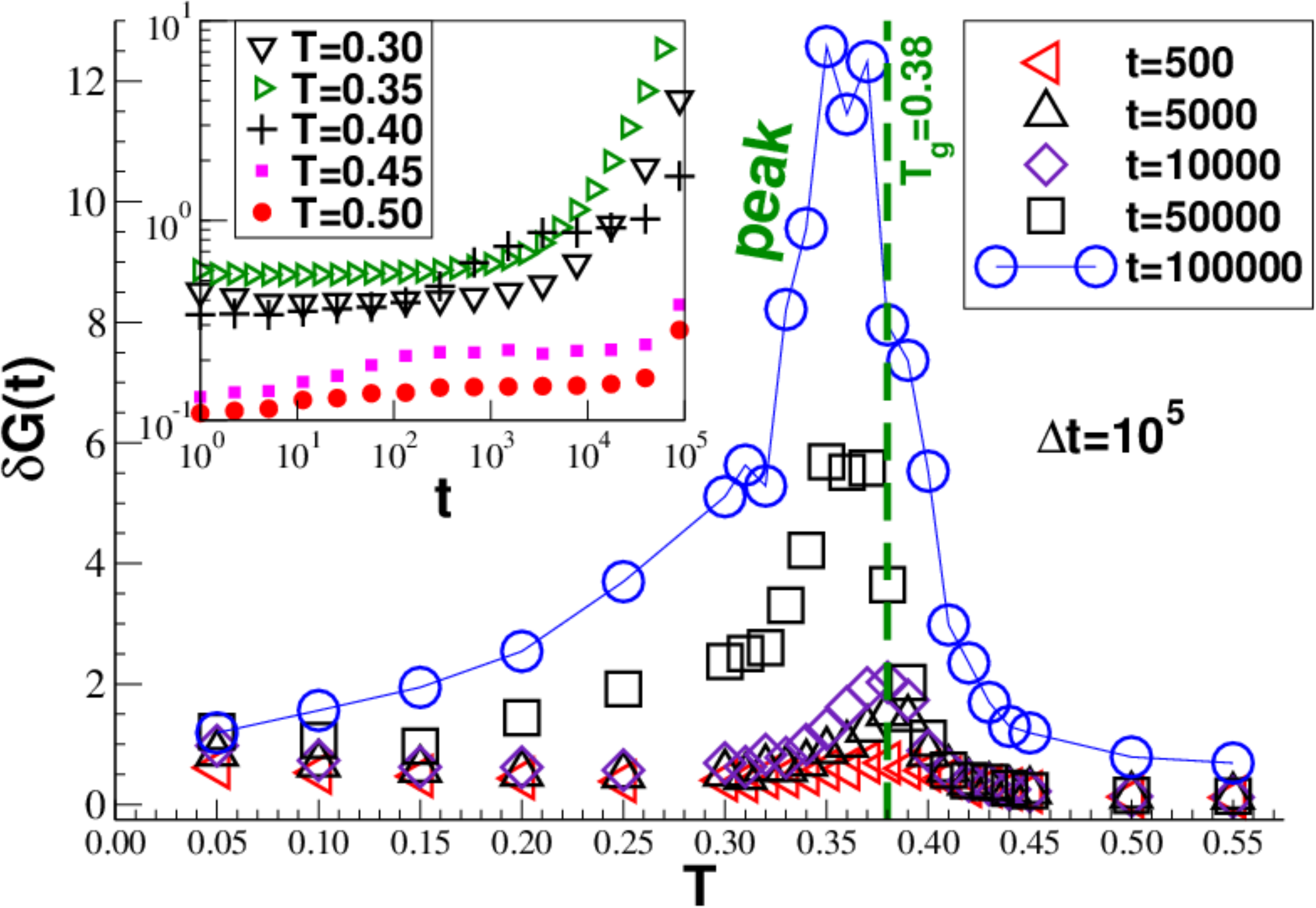}}}
\caption{Standard deviation $\dGt$ of the relaxation modulus $G(t)$ presented in Fig.~\ref{fig_Gt}.
All data have been obtained with gliding averages over time series of length $\tsamp=10^5$.
Inset: Double-logarithmic representation of $\dGt$ {\em vs.} time $t$ for several temperatures $T$.
Main panel: $\delta G$ as a function of temperature for several times $t$ as indicated.
$\delta G$ is non-monotonous with a strong maximum slightly below $\Tglass$.
}
\label{fig_dGt}
\end{figure}

As we have already pointed out in Sec.~\ref{res_Gt}, the statistics of $G(t)$ deteriorates for 
$t \to \tsamp$ since we have naturally used gliding averages along the time series,
Eq.~(\ref{eq_Gtbar}), to compute this property. More importantly, it can be seen from Fig.~\ref{fig_Gt}
that fluctuations of $G(t)$ become stronger around $\Tglass$.
Both observations are confirmed quantitatively in Fig.~\ref{fig_dGt}
where we present the standard deviation $\dGt$.
Using double-logarithmic coordinates $\dGt$ is shown in the inset for several 
temperatures. It is seen that $\dGt$ increases strongly for $t \to \tsamp$.
This effect is particularily pronounced for $T=0.35$.
Using a similar representation as in Fig.~\ref{fig_fluctu_mu} for $\dGF(T)$,
the main panel of Fig.~\ref{fig_dGt} presents $\delta G(T)$ as a function of temperature
for several times $t$. As one expects, one observes a strong peak near $\Tglass$.
Interestingly, this peak is more pronounced as the one for $\dGF(T)$ and $\tsamp=10^5$.
As we have shown in Sec.~\ref{res_GtGF},
$\GF(\tsamp)$ is given by the integral Eq.~(\ref{eq_mu}) over $G(t)$.
It is natural to attempt similarily to describe the variance $\dGF^2(\tsamp)$
in terms of the variance $\dGt^2$ using the weighted average
\begin{equation}
\dGF^2(\tsamp) \approx \frac{2}{\tsamp^2} \int_0^{\tsamp} \ddiff t \ (\tsamp -t) \ \dGt^2.
\label{eq_dGt2dGF}
\end{equation}
This relation is a two-point approximation assuming that the standard deviations at different
times are decorrelated. As shown by the filled symbols in Fig.~\ref{fig_fluctu_mu},
we have used Eq.~(\ref{eq_dGt2dGF}) to predict $\dGF(\tsamp)$ using the data for $\dGt$. 
The simple two-point approximation works quite reasonably, especially in the solid limit and the liquid limit. 
However, it overpredicts $\GF(\tsamp)$ around $\Tglass$. (Note that a much better fit of the measured
standard deviation $\dGF$ is obtained if we replace in Eq.~(\ref{eq_dGt2dGF}) the variances by the 
corresponding standard deviations. Unfortunately, this is more difficult to justify.)
A closer inspection of higher order (three and four point) correlations is thus needed. 
This is beyond the scope of the present paper. We only note that much longer time 
series are warranted to clarify this issue.

\section{System-size effects}
\label{res_V}

\begin{figure}[t]
\centerline{\resizebox{1.0\columnwidth}{!}{\includegraphics*{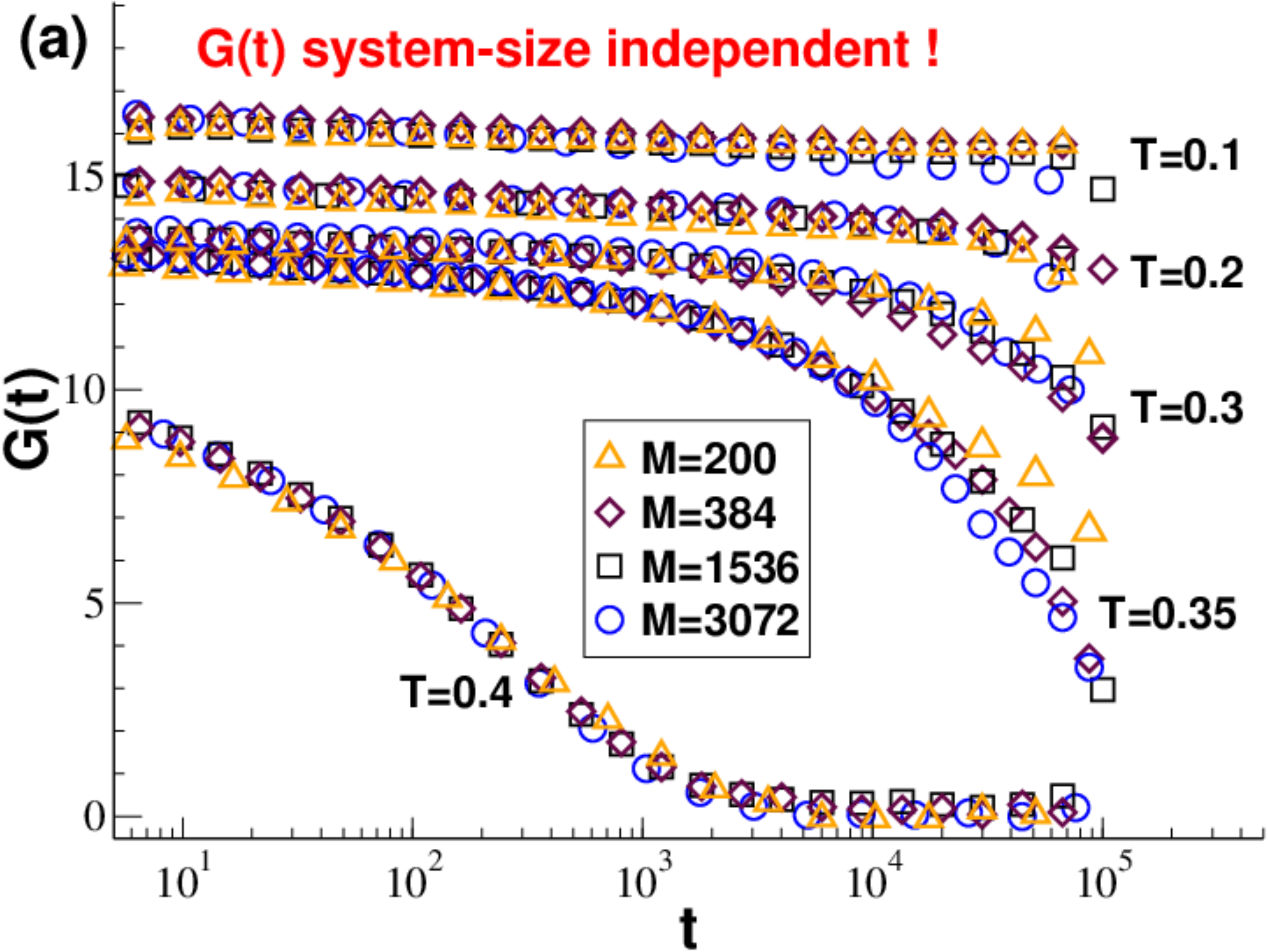}}}
\centerline{\resizebox{1.0\columnwidth}{!}{\includegraphics*{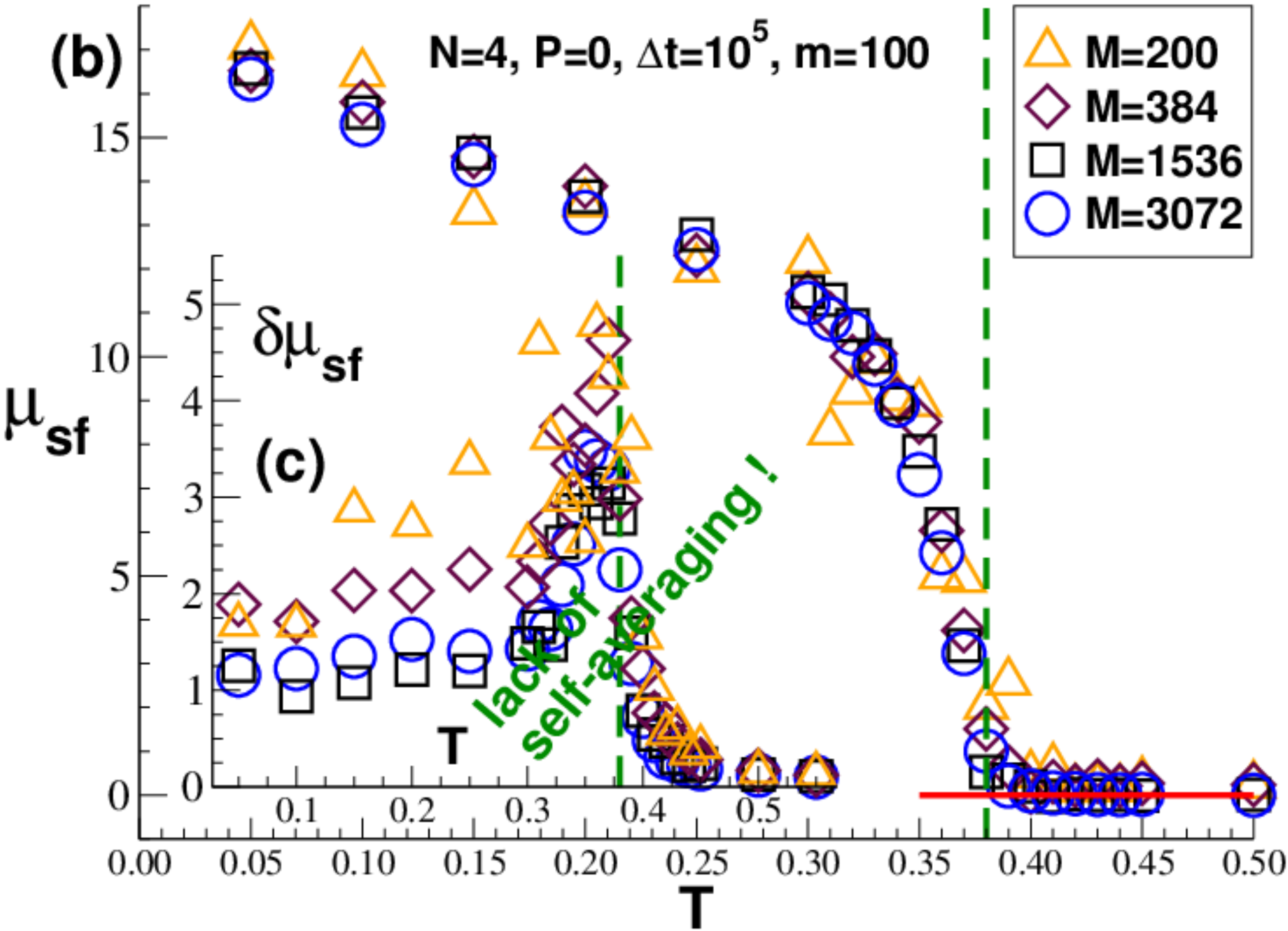}}}
\caption{
Test of system-size effects using the four chain numbers $M=200$, $384$, $1536$ and $3072$:
{\bf (a)} 
Shear-stress relaxation modulus $G(t)$ for five temperatures $T$.
{\bf (b)}
$\GF(T;\tsamp=10^5)$ for several $M$ demonstrating the expected
system-size independence of the continuous transition of the shear modulus
for finite sampling times $\tsamp$.
{\bf (c)}
Standard deviation $\dGF(T)$ as a function of temperature $T$ for $\tsamp=10^5$.
\label{fig_V}
}
\end{figure}

We comment finally on on-going work on system-size effects presenting data 
obtained for four chain numbers $M=200$, $384$, $1536$ and $3072$.
The largest number $M=3072$, corresponding to a total mass $n = N M = 12288$, is assumed everywhere else in the paper.
As before, all data have been obtained for $N=4$ by averaging over $m=100$ independently quenched configurations and 
three shear planes.

Panel (a) of Fig.~\ref{fig_V} shows the shear-stress relaxation modulus $G(t)$ for five selected 
temperatures $T$. All data are obtained using the relation Eq.~(\ref{eq_Gt}) for systems 
with quenched (sluggish) shear stresses \cite{ivan17a}. The data have been logarithmically averaged for clarity.
As can be seen, $G(t)$ is essentially $M$-independent for all $T$. 
(Our smallest system with $M=200$ differs slightly for $T=0.3$ and $T=0.35$ for long times.
It relaxes more slowly. This observation is similar to the results reported in Ref.~\cite{Berthier12},
but longer simulation runs are needed to scrutinze this effect.)
Since $G(t)$ is $M$-independent and since the shear modulus $\GF$ obtained by Eq.~(\ref{eq_GF}) is 
equivalent to the integral $\mu(\tsamp)$ over $G(t)$, Eq.~(\ref{eq_mu}), one also expects 
$\GF(T)$ to be $M$-independent.
As shown in panel (b) of Fig.~\ref{fig_V} for $\tsamp=10^5$ this holds for all our data.
This demonstrates that the continuous transition
of $\GF(T;\tsamp)$ observed for finite sampling times $\tsamp$ is not
due to finite-size effects --- as may happen for standard phase transitions \cite{LandauBinderBook} ---
but is expected to hold for asymptotically large configurations.
Naturally, the same system-size independence follows also (not shown) 
for the generalized shear-stress viscocity $\eta(\tsamp)$, Eq.~(\ref{eq_eta}),
and the relaxation time $\tau(\tsamp)$, Eq.~(\ref{eq_tau}).

Focusing again on the sampling time $\tsamp=10^5$, the standard deviation $\dGF(T)$ of the
shear modulus $\GF(T)$ is given in panel (c). 
The central point for the current study is that $\dGF(T)$ 
becomes system-size independent for temperatures $T$ around the glass transition temperature 
$\Tglass$ and beyond, i.e. $\dGF/\GF \approx 1$ holds irrespective of the system size for 
$T \approx \Tglass$.
One expects from Procaccia {\em et al.} \cite{Procaccia16} and Wittmer {\em et al.} \cite{WKC16} 
that the standard deviations $\delta G(t)$ and $\dGF$ should decrease with $M$ in the solid limit 
($T \ll \Tglass$) where plastic rearrangements become rare and uncorrelated and, hence, irrelevant.
This is qualitatively confirmed by our data, but larger configuration ensembles (with $m \gg 10^2$) 
and system sizes (with chain numbers $M \gg 10^4$) are warranted to verify this expectation more precisely. 

\section{Conclusion}
\label{sec_conc}

\subsection{Summary}
\label{conc_summary}

We have investigated by means of MD simulations a coarse-grained model of polymer glasses (Sec.~\ref{sec_algo}).
The linear shear mechanical response of this model has been characterized 
from the (ensemble-averaged) expectation values of the (time-averaged) 
contributions to the stress-fluctuation relation $\GF = \muA- \muF = \muA -\muFtwo + \muFone$ 
for the shear modulus (Sec.~\ref{sec_GF}),
their corresponding standard deviations and cross-correlations (Sec.~\ref{sec_dGF})
and using the shear-stress relaxation modulus $G(t)$ and its in general
$\tsamp$-dependent moments $\mu$ and $\eta$ (Sec.~\ref{sec_Gt}). 
The relaxation modulus has been directly determined by means of the recently proposed 
general fluctuation-dissipation relation, Eq.~(\ref{eq_Gt}), which can be also used
for systems where $\muA=\muFtwo$ does not hold (Appendix~\ref{sm_Gt}).
We emphasize the following central results:
\begin{itemize}
\item
Key result I:
The liquid-solid transition characterized by the shear modulus $\GF(T)$ depends on the sampling 
time $\tsamp$ (Fig.~\ref{fig_GF_dt}). It is continuous with respect to the temperature $T$
for all $\tsamp$ becoming, however, sharper with increasing $\tsamp$ (Fig.~\ref{fig_mean_mu}). 
\item
Strong quenched shear stresses arise naturally at and below the glass transition and 
$\muFtwo$ and $\muFone$ increase thus dramatically with decreasing $T$ (Figs.~\ref{fig_mean} and \ref{fig_static}).
\item
Since time and ensemble averages commute for $\muA$, $\muFtwo$ and $G(t)$, 
their expectation values do not depend on the sampling time $\tsamp$
while all other properties studied here, especially all standard deviations, do in principle.
\item
Together with the observation that $\muFtwo \gg \muA $ below $\Tglass$, 
the $\tsamp$-independence of $\muA$ and $\muFtwo$ implies that
our low-$T$ configurations are not compatible with 
the assumption that the finite-$\tsamp$ time series are randomly drawn from an equilibrium 
time evolution of a liquid (Sec.~\ref{res_static}).
This falling out of equilibrium is not an artefact of the particular 
preparation (quench) history of our configurations, but a central generic feature of 
the glass transition.
\item
Key result II:
While the glass transition characterized by $\GF$ becomes continuously sharper 
{\em on average} with increasing $\tsamp$, increasingly strong fluctuations $\dGF$ 
between different configurations underly the transition (Fig.~\ref{fig_fluctu_mu}). 
The broad and lopsided distribution $p(\GFbar)$ below $\Tglass$ makes the prediction of $\GFbar$
for a single configuration elusive (Fig.~\ref{fig_muhisto}). 
It is thus insufficient to discuss only the average shear modulus at the glass transition,
fluctuations need also to be considered theoretically.
\item
A clear-cut order parameter of the glass transition is given
by the dimensionless correlation coefficient $\coeffonetwo$
of the time-averaged moments $\muFtwobar$ and $\muFonebar$
showing that the transition is logarithmically shifted to
lower $T$ with increasing $\tsamp$ (Fig.~\ref{fig_coeffonetwo}). 
\item
Key result III:
The observed $\tsamp$-dependence of $\GF$ (Figs.~\ref{fig_mean_mu}, \ref{fig_GF_dt} and \ref{fig_GtGF}) 
and its contributions $\muFone$ (Fig.~\ref{fig_mu1_dt}) and $\muF$ can be traced back 
to the finite time (time-averaged) stress fluctuations need to explore 
the phase space. This effect is perfectly described by the weighted integral
$\mu(\tsamp)$ over the shear-stress relaxation modulus $G(t)$, Eq.~(\ref{eq_mu}), 
shown to be identical to $\GF(\tsamp)$.
\item
Albeit some ageing must occur in our systems, this shows that this must happen on much larger
time scales and that within the $\tsampmax$-window available, time translational invariance
holds to high accuracy for the macroscopic properties of interest here.
\item
The stress-fluctuation formula $\GF$ corresponds to an equilibrium storage modulus only
if $G(t)$ becomes constant over a sufficiently large time window. This is of relevance
in the solid limit well below $\Tglass$ and in the liquid limit for $\tsamp \gg \tauinf$.
In all other cases one should see $\GF$ as a ``generalized shear modulus" or a useful 
smoothing function of $G(t)$, Eq.~(\ref{eq_mu2Gt}), which also contains information 
related to dissipative processes (Sec.~\ref{res_GtGF_continued}). 
\item
Since $G(t,T)$ is monotonous and continuous with respect to $T$ and $\tsamp$ (Fig.~\ref{fig_Gt}), 
this implies the same behavior for $\GF(\tsamp,T)=\mu(\tsamp,T)$, $\eta(\tsamp,T)$ 
and $\tau(\tsamp,T)$ as seen in 
Figs.~\ref{fig_mean_mu}, \ref{fig_GF_dt}, \ref{fig_eta} and \ref{fig_tau}.
\item 
Since $\mu=\GF$ in general and $\muFone=\GF$ for higher temperatures where $\muA=\muFtwo$, 
this allows to express $G(t)$, $\mu(\tsamp)$, $\eta(\tsamp)$ and $\tau(\tsamp)$
above $\Tglass$ in terms of the numerically better behaved $\muFone(\tsamp,T)$,
Eqs.~(\ref{eq_mu2Gt_three},\ref{eq_GF2eta_two},\ref{eq_GF2tau}).
Using these effectively low-pass filters one avoids some of the problems related to the precision 
of the tail of $G(t)$.
\item
Key result IV:
The $1/\tsamp$-decay of $\GF \approx \muFone$ at high temperatures, Eq.~(\ref{eq_etainf}),
allows a simple and precise determination of the shear viscosity $\etainf$ 
(Fig.~\ref{fig_mu1_dt}) \cite{foot_Gloss2eta}. 
\item
We have verified by varying the number of chains that our numerical results
(especially our key findings) are not due to system-size effects (Fig.~\ref{fig_V}).
\end{itemize}

\subsection{Outlook}
\label{conc_outlook}

While $\GF$ and its contributions $\muA$, $\muF$, $\muFtwo$ and $\muFone$ 
do not depend on the system size, this is more intricate for the corresponding 
standard deviations and must be addressed in the future following recent work
on colloidal glasses \cite{Procaccia16} and on self-assembled networks \cite{WKC16}.
The latter study suggests a strong self-averaging for the affine shear modulus $\muA$, 
i.e. $\dmuA \sim 1/M^{1/2}$, and a complete lack of self-averaging for $\muFtwo$ and $\muFone$,
i.e. $\dmuFtwo \sim \dmuFone \sim M^0$, for all temperatures.
As already seen in panel (c) of Fig.~\ref{fig_V}, we expect that further simulations
with larger configurations will confirm that 
\begin{equation}
\delta G(t) \sim \dmuF \sim \dmu \sim \deta \sim M^0
\label{eq_aroundTg}
\end{equation}
around and above the glass transition (lack of self-averaging).  
In this limit long-ranged viscoelastically interacting activated events
should dominate the plastic reorganizations of the particle contacts \cite{Barrat14b}.
At much lower temperatures some self-averaging must become relevant,
i.e. one expects 
\begin{equation}
\delta G(t) \sim \dGF \approx \dmu \sim 1/M^{\alpha}
\mbox{ with } 0 < \alpha \le 1/2.
\label{eq_greatexpectations}
\end{equation}
Our work on self-assembled transient networks \cite{WKC16} suggests strong self-averaging, 
i.e. $\alpha=1/2$, while the study by Procaccia {\em et al.} \cite{Procaccia16} points to 
a somewhat smaller exponent.
Since the two temperature regimes must match, such a scaling would imply the existence of
a characteristic length scale $\xi(T)$. Our guess is that such a length scale
must be related to (and perhaps even be set by) the distance over which the sound waves generated by a given
plastic particle rearrangement are able to tricker subsequent plastic events. 
Qualitatively different behavior is thus to be expected if three-dimensional
polymer melts are compared to effectively two-dimensional melts confined to thin films.

We emphasize finally that albeit the presented work has focused on the shear-stress 
fluctuations, similar $\tsamp$-dependent results are expected for shear-strain fluctuations, 
mixed stress-strain fluctations and trajectory analysis in reciprocal space following
Klix {\em et al.} \cite{Klix12,Klix15}.
It should be possible, e.g., to trace back the observed $\tsamp$-dependence in the latter case
to a weighted integral over a wavevector-dependent creep compliance. 

\vspace*{0.2cm} 
\begin{acknowledgments}
I.K. thanks the IRTG Soft Matter for financial support.
We are indebted to A.N. Semenov (ICS, Strasbourg) and 
H.~Xu (Univ. Lorraine, Metz) for helpful discussions.
We thank the University of Strasbourg for CPU time through GENCI/EQUIP${@}$MESO.
\end{acknowledgments}

\appendix
\section{Canonical affine shear strains}
\label{sm_affine}

\paragraph*{Canonical affine transform.}
Let us consider a small shear strain increment $\gamma$ in the $xy$-plane as it would be used to 
determine the shear relaxation modulus $G(t)$ by means of a direct out-of-equilibrium simulation
\cite{AllenTildesleyBook,WXB15,WXBB15,WKB15,WXB16,WKC16}.
For simplicity all particles are in the principal simulation box \cite{AllenTildesleyBook}.
It is assumed that all particle positions $\rvec$ and particle momenta $\pvec$
follow the imposed ``macroscopic" strain in a {\em canonical affine} manner according to \cite{WXBB15}
\begin{equation}
\rx \to \rx + \gamma \ \ry \mbox{ and } 
\py \to \py - \gamma \ \px
\label{eq_cantrans}
\end{equation}
where the negative sign in the second transform assures that Liouville's theorem \cite{Goldstein}
is satisfied. 

\paragraph*{General definitions.}
The (instantaneous) Hamiltonian $\Hhat$ of the given configuration will thus change as
\begin{equation}
(\Hhat(\gamma)-\Hhat(\gamma=0))/V \approx \tauAhat \gamma + \frac{1}{2} \muAhat \gamma^2
\mbox{ for } |\gamma| \ll 1.
\label{eq_Hhatexpand}
\end{equation}
We thus define the instantaneous {\em affine shear stress} $\tauAhat$ and the instantaneous 
{\em affine shear modulus} $\muAhat$ by
\begin{eqnarray}
\tauAhat & \equiv & \Hhat^{\prime}(\gamma)/V|_{\gamma=0} \label{eq_tauAhatdef} \mbox{ and } \\
\muAhat & \equiv & 
\Hhat^{\prime\prime}(\gamma)/V|_{\gamma=0} =
\tauAhat^{\prime}(\gamma)|_{\gamma=0} 
\label{eq_muAhatdef}
\end{eqnarray}
where a prime denotes a functional derivative with respect to the 
imposed canonical affine transformation \cite{WXBB15}.
It follows from the last equality in Eq.~(\ref{eq_muAhatdef}) that
\begin{equation}
\hat{G}(t=0)=\muAhat
\label{eq_Gtzero_muAhat}
\end{equation}
for the shear relaxation modulus of one configuration.
Assuming the Hamiltonian $\Hhat = \Hidhat + \Hexhat$ to be the sum of an ideal and 
an excess contribution $\Hidhat$ and $\Hexhat$, similar relations apply for the corresponding 
contributions $\tauAidhat$ and $\tauAexhat$ to  $\tauAhat =\tauAidhat + \tauAexhat$ and 
for the contributions $\muAidhat$ and $\muAexhat$ to $\muAhat = \muAidhat + \muAexhat$.
By explicitly applying Eq.~(\ref{eq_cantrans}) to a given configuration
using a broad range of shear strains $\gamma$, all four expansion coefficients $\tauAidhat$, 
$\tauAexhat$, $\muAidhat$ and $\muAexhat$ are in principle directly measurable observables 
irrespective of the specific Hamiltonian used \cite{WXBB15}.

\paragraph*{Some useful formulae.}
As shown elsewhere \cite{WXBB15}, the ideal contributions become
\begin{eqnarray}
\tauAidhat & = & - \frac{1}{V} \sum_{i=1}^n \pix \piy / m_i \label{eq_tauAidhat} \mbox{ and } \\
\muAidhat & = & \frac{1}{V} \sum_{i=1}^n \pix^2 /m_i \label{eq_muAidhat} 
\end{eqnarray}
where the sums run over all $n$ particles of mass $m_i$.
Note that the minus sign for the ideal shear stress follows from the minus sign in Eq.~(\ref{eq_cantrans})
required for a canonical transformation.
Assuming a pairwise central conservative potential $\Hexhat = \sum_l u(\rl)$
with $l$ labeling the interactions and $\rl$ the distance between the pair of monomers,
one obtains the excess contributions \cite{WXBB15}
\begin{eqnarray}
\tauAexhat & = & \frac{1}{V} \sum_l \rl u^{\prime}(\rl) \ \nlx \nly   \label{eq_tauAexhat} \ \mbox{ and } \\
\muAexhat & = & \frac{1}{V} \sum_l  \left( \rl^2 u^{\prime\prime}(\rl)
- \rl u^{\prime}(\rl) \right) \nlx^2 \nly^2 \nonumber \\
& + & \frac{1}{V} \sum_l \rl u^{\prime}(\rl) \ \nly^2  \label{eq_muAexhat}
\end{eqnarray}
with $\nvecl = \rvecl/\rl$ being the normalized distance vector.
As one expects, Eq.~(\ref{eq_tauAexhat}) is strictly identical to the
corresponding off-diagonal term of the Irving-Kirkwood stress tensor  \cite{AllenTildesleyBook}.
The index ``A" for the shear stress has thus been dropped in other parts of this paper.
The last term in Eq.~(\ref{eq_muAexhat}) takes into account the excess contribution of the average normal 
pressure \cite{XWP12}.

\paragraph*{Comments.}
Similar relations are obtained for the $xz$- and the $yz$-plane.
For an isotropic system the averages of all three affine shear moduli are finite and equal.
We keep the index ``A" to remind that the (time and ensemble averaged) $\muA$ assumes a strictly {\em affine} 
strain without relaxation. It thus provides only an upper bound to the shear modulus of the configuration
\cite{Lutsko88,WTBL02,Barrat06,WXP13,WXBB15,LXW16}. 
Please note that $\muAexhat$ depends on the second derivative $u^{\prime\prime}(r)$ of the pair potential. 
As emphasized in Appendix~\ref{sm_highT} (Fig.~\ref{fig_GF_dtlarge}), 
impulsive corrections need to be taken into account due to this 
term if the first derivative $u^{\prime}(r)$ of the potential is not continuous \cite{XWP12}. 
Unfortunately, this is the case at the cut-off of the LJ potential used in the current study
(Sec.~\ref{algo_hamiltonian}).


\begin{figure}[t]
\centerline{\resizebox{1.0\columnwidth}{!}{\includegraphics*{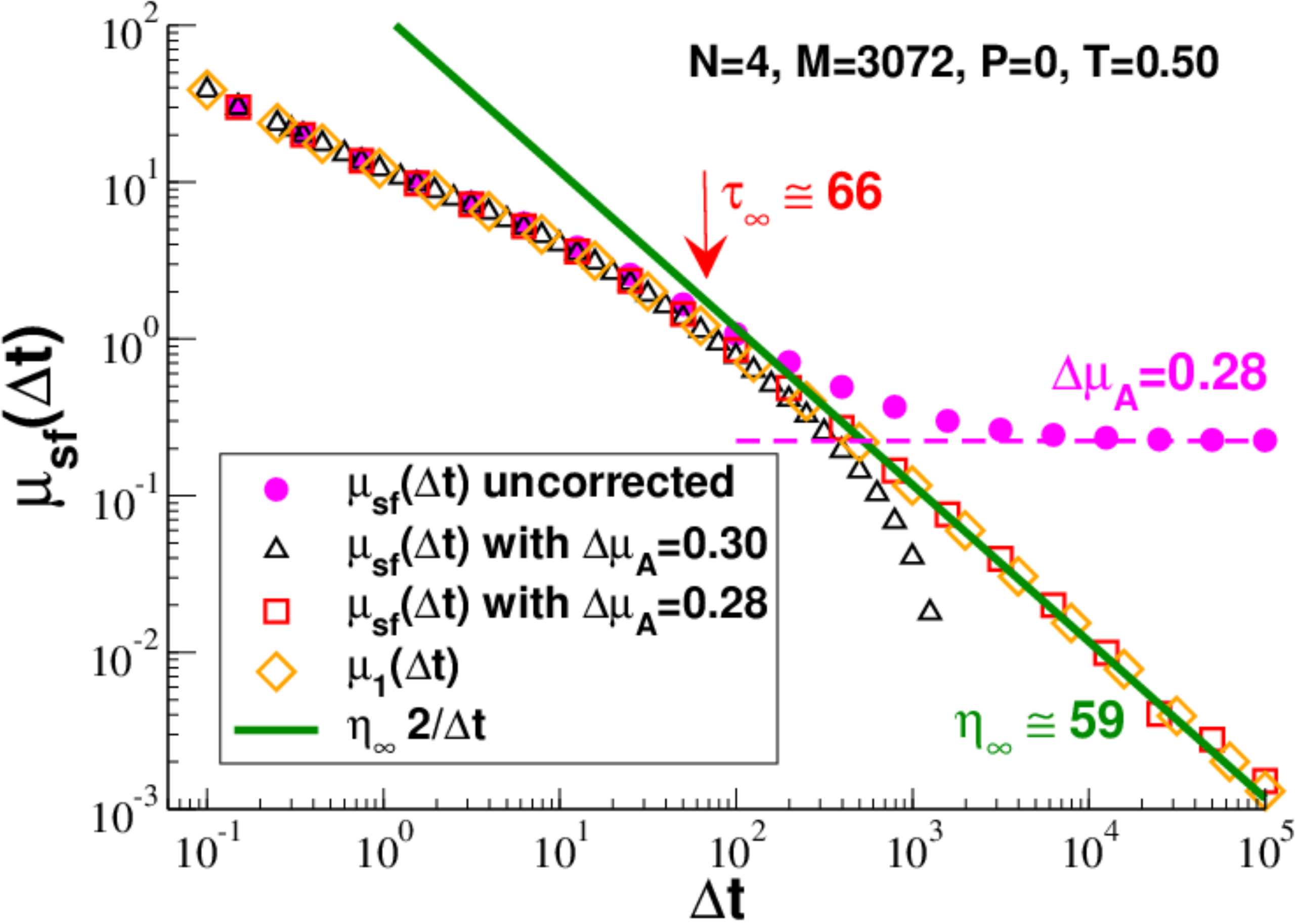}}}
\caption{$\GF(\tsamp)$ and its contribution $\muFone(\tsamp)$ for the high temperature $T=0.5$ 
using logarithmic coordinates.
The small filled cercles represent uncorrected $\GF(\tsamp)$-data where we do not take into account
the impulsive truncation corrections for $\muA$ \cite{XWP12}. The uncorrected data
saturate at a small, but finite value $\Delta \muA \approx 0.28$ (dashed horizontal line). 
If correctly shifted (open squares), $\GF(\tsamp) = \muFone(\tsamp)$ for all $\tsamp$
and $\GF(\tsamp) = \etainf 2/\tsamp$ for $\tsamp \gg \tauinf$ (solid line).
The vertical arrow marks the terminal relaxation time $\tauinf=66$ for $T=0.5$
set as reference for the rescaling of $\muFone(\tsamp)$ in Fig.~\ref{fig_mu1_dt}.
}
\label{fig_GF_dtlarge}
\end{figure}

\section{High-$T$ limit of $\GF(\tsamp)$ and $\muFone(\tsamp)$}
\label{sm_highT}

\paragraph*{Introduction.}
We address now three points, which, albeit valid in principle for all $T$, 
are in practice only relevant at sufficiently high temperatures where $\tsampmax=10^5 \gg \tauinf(T)$.
(An estimate of the terminal relaxation time $\tauinf(T)$ will be given below.)
These points are illustrated for one temperature ($T=0.5$) in Fig.~\ref{fig_GF_dtlarge} 
where we present $\GF(\tsamp)$ and $\muFone(\tsamp)$ using double-logarithmic coordinates.
This allows to pay attention to the large-$\tsamp$ behavior where both observables become very small.
We remind that according to Eq.~(\ref{eq_GFdt}) one expects $\GF(\tsamp)=\muFone(\tsamp)$ 
for all $\tsamp$ in the liquid limit where $\muA=\muFtwo$.

\paragraph*{Truncation effect.}
The first, more technical point we want to make concerns the truncation corrections due to the contribution 
of the second derivative of the potential to the affine shear modulus $\muA$ discussed in Ref.~\cite{XWP12}. 
As shown by the filled disks, the bare, uncorrected values of $\GF(\tsamp)$ saturate at some finite, 
positive value $\Delta \muA$ as indicated by the dashed horizontal line. Albeit the effect is small, 
this finding is clearly unphysical since the true thermodynamic shear modulus of a liquid must rigorously 
vanish for large $\tsamp$ \cite{RubinsteinBook}.
This is essentially the case if the data for the affine shear modulus is shifted,
$\muA \rightarrow \muA - \Delta \muA$, using the constant $\Delta \muA\approx 0.3$ 
suggested by the histogram method described in Ref.~\cite{XWP12}. 
Unfortunately, as can be seen from the open triangles, using this value for $\Delta \mu$,
$\GF(\tsamp)$ is yet not identical to $\muFone(\tsamp)$ for large $\tsamp$. 
Since it is currently not possible using the histogram method to obtain a more precise 
value for $\Delta \muA$ \cite{foot_truncprob}, we have fine-tuned $\Delta \muA$ 
by insisting on $\GF(\tsamp) \approx \muFone(\tsamp)$ for all $\tsamp$. 
This yields the refined shift constant $\Delta \muA\approx 0.28$ used for the open squares.
Similar values are obtained for other temperatures above $\Tglass$. 
Using these more precise $\Delta \muA$-values, one confirms the $1/\tsamp$-asymptote (bold solid line) 
for $\GF(\tsamp)$ and $\tsamp \gg \tauinf$ expected on general grounds for finite-sampling-time 
corrections of time-preaveraged fluctuations 
\cite{LandauBinderBook,WXP13,WXB15,WXBB15,WKB15,WXB16,WKC16}.

\begin{figure}[t]
\centerline{\resizebox{1.0\columnwidth}{!}{\includegraphics*{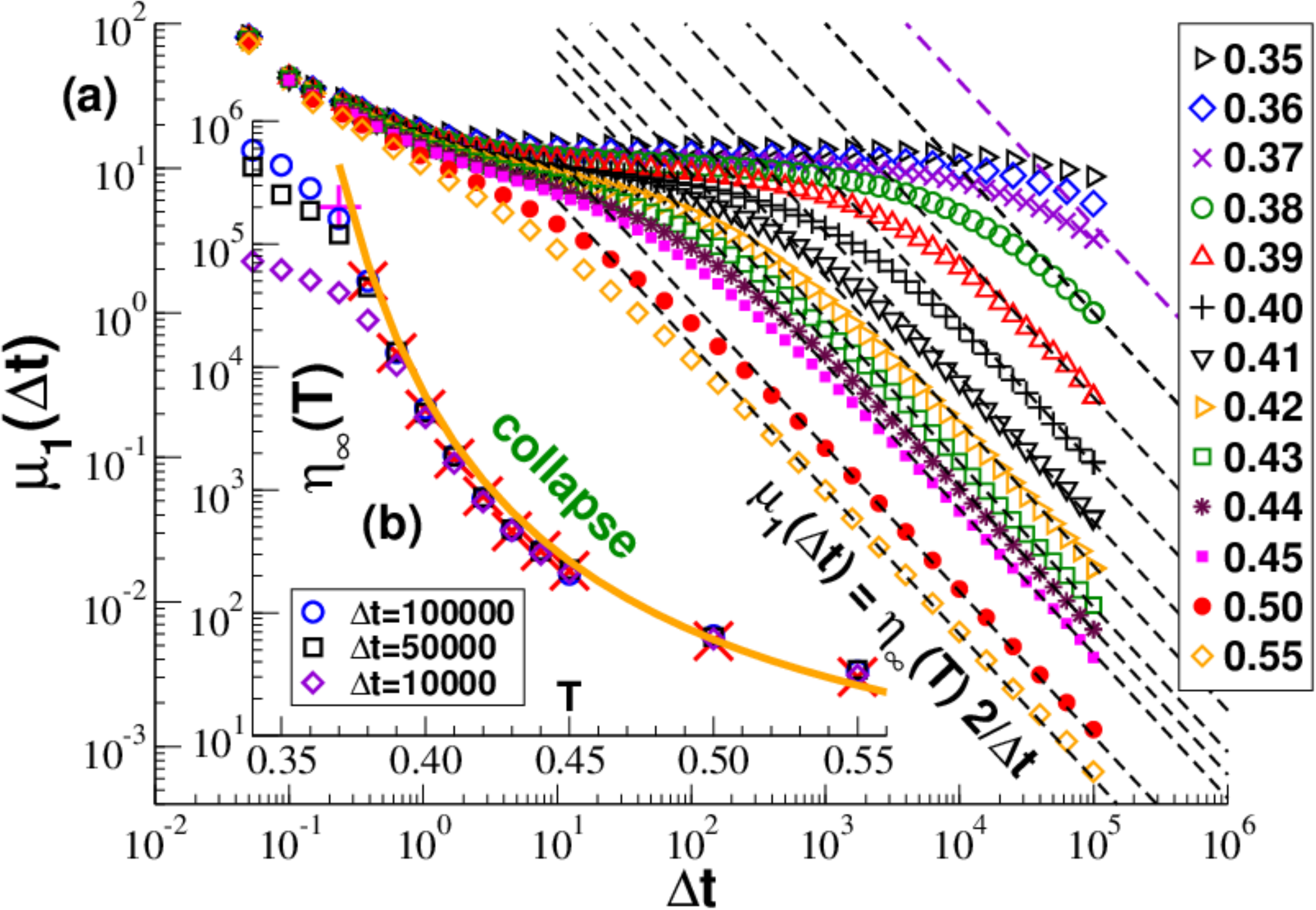}}}
\centerline{\resizebox{1.0\columnwidth}{!}{\includegraphics*{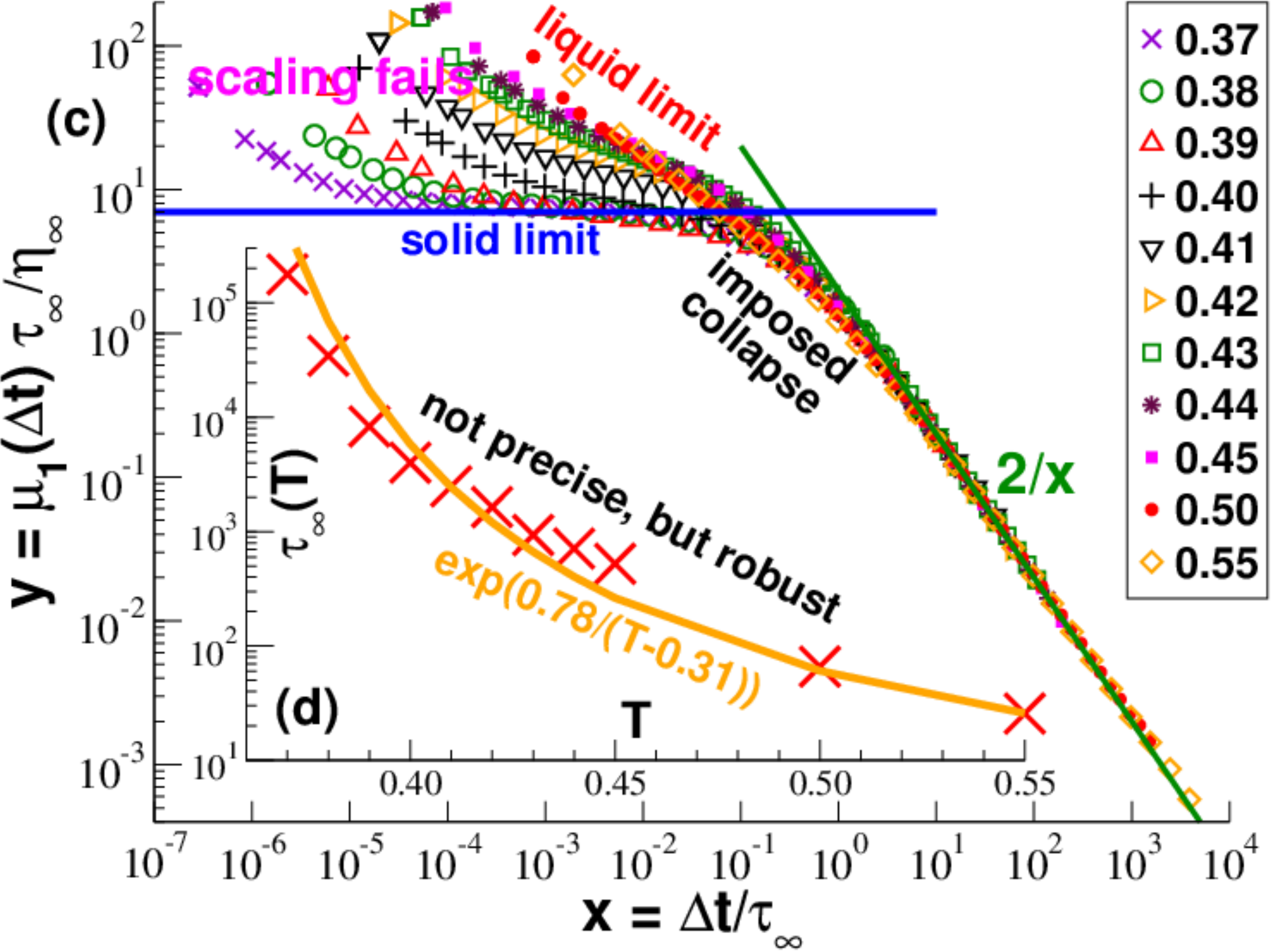}}}
\caption{$\muFone(\tsamp)$ for higher temperatures where $\muA=\muFtwo$ holds:
{\bf (a)} 
Using double-logarithmic coordinates the shear viscosity $\etainf(T)$ can be estimated 
for $T \gtrsim 0.38$ by fitting $\muFone(\tsamp) = \etainf 2/\tsamp$ as indicated by the dashed lines.
{\bf (b)}
$\etainf(T)$ increases over four orders of magnitude between $T=0.55$ and $T=0.38$.
The plus is a fair guess for $T=0.37$. 
For comparison we indicate $\muFone(T;\tsamp) \tsamp/2$ for $\tsamp=100000$, $50000$ and $10000$ 
(open symbols). The bold line represents Eq.~(\ref{eq_VFT}).
{\bf (c)}
Characterization of $\tauinf(T)$ by tracing $y = \muFone(\tsamp) \tauinf/\etainf$ 
{\em vs.} $x = \tsamp/\tauinf$. 
{\bf (d)}
Terminal relaxation time $\tauinf(T)$ used for the rescaling of $\muFone(\tsamp)$.
}
\label{fig_mu1_dt}
\end{figure}

\paragraph*{Shear viscosity.}
This leads us to the second point we want to make. 
By fitting the amplitude of Eq.~(\ref{eq_etainf}) given in the Introduction,
one obtains in fact a rather precise estimate of the shear viscosity $\etainf(T)$.
This relation, being a direct consequence of the key result Eq.~(\ref{eq_mu}), 
is justified in Sec.~\ref{res_eta}. 
Due to the tricky determination of the truncation shift constant $\Delta \muA$ 
it is even more convenient to use directly $\muFone(\tsamp)$ instead of $\GF(\tsamp)$
in the liquid limit as illustrated in panel (a) of Fig.~\ref{fig_mu1_dt}.
As shown by the thin dashed lines, it is thus possible to estimate $\etainf(T)$ 
for temperatures down to $T\approx 0.38$ \cite{foot_fitetainf}. 
As seen from panel (b) of Fig.~\ref{fig_mu1_dt}, the obtained values (crosses) are quite reasonable: 
$\etainf(T)$ increases both monotonously and continuously over four orders of magnitude 
between $T=0.55$ and $T=0.38$. 
The data can be well fitted (bold line) using a Vogel-Fulcher-Tammann (VFT) law 
\begin{equation}
\etainf(T) \approx 1.0 \exp[0.78/(T-0.31)] \mbox{ for } T \ge 0.37.
\label{eq_VFT}
\end{equation}
For slightly lower temperatures Eq.~(\ref{eq_etainf}) allows at least to guess 
the power-law amplitude. A possible value is indicated for $T=0.37$.
By tracing $\muFone(T;\tsamp) \tsamp/2$ for several $\tsamp$ (open symbols)
one obtains an additional simple test of the observed values and lower bounds for
$\etainf(T)$ at even smaller temperatures. While a perfect data collapse is observed
for large $T$ or $\tsamp$, the scaling naturally fails in the opposite limit. 

\paragraph*{Terminal relaxation time.}
Being the third point we want to make here, this failure allows an estimation 
of the terminal relaxation time $\tauinf(T)$ of the system and the sampling time 
$\tsamp \gg \tauinf(T)$ needed. For instance,  $\tsamp=10000$ (diamonds) becomes 
insufficient for $T \approx 0.39$ and $\tauinf(T)$ should be of this order.
This estimation of $\tauinf(T)$ can be made more quantitative by attempting a scaling
plot for $\muFone(\tsamp;T)$ as shown in panel (c) of Fig.~\ref{fig_mu1_dt}.
We trace here the dimensionless $y = \muFone(\tsamp;T) \tauinf(T)/\etainf(T)$
as a function of the reduced sampling time $x =\tsamp/\tauinf(T)$ using
the $\etainf(T)$-values obtained from Eq.~(\ref{eq_etainf}). The terminal
relaxation times $\tauinf(T)$ are fixed by imposing a data collapse around 
$x \approx 1$ and by setting $\tauinf(T=0.5) \approx 66$ as a reference.
This reference is indicated in Fig.~\ref{fig_GF_dtlarge} by a vertical arrow.
It is motivated by the more systematic but numerically more difficult
operational definition Eq.~(\ref{eq_tau}) discussed in Sec.~\ref{sm_tau}.
We thus obtain, e.g., $\tauinf(T=0.39) \approx 8300$ consistently with the 
failure of the scaling observed for the $\tsamp=10000$-data at $T=0.39$ in panel (c).
The terminal relaxation times determined using the $\muFone(\tsamp)$-rescaling
are indicated in panel (d) of Fig.~\ref{fig_mu1_dt} (crosses).
A dramatic monotonous increase over (again) four orders of magnitude is 
observed between $T=0.55$ and $T=0.37$.  
The bold line compares the $\tauinf$-data with the same VFT law Eq.~(\ref{eq_VFT}) used for the viscosity.

\paragraph*{Summary.}
It is possible to obtain reasonable values for $\etainf(T)$ and $\tauinf(T)$
from the asymptotic $1/\tsamp$-decay and the scaling of $\muFone(\tsamp;T)$ 
for not too low temperatures. These values are given in Table~\ref{tab_T}.
Note that $\etainf(T)$ has been determined with a much higher precision than $\tauinf(T)$.
We emphasize that both $\etainf(T)$ and $\tauinf(T)$ are completely continuous 
and no jump-singularity is observed.
Judging from the available data it appears to be reasonable that by extending in the near
future the production runs up to $\tsampmax=10^6$ using the same system size or 
$\tsampmax=10^7$ using smaller systems, reliable values for both quantities 
down to $T=0.35$ are feasible.


\begin{figure}[t]
\centerline{\resizebox{1.0\columnwidth}{!}{\includegraphics*{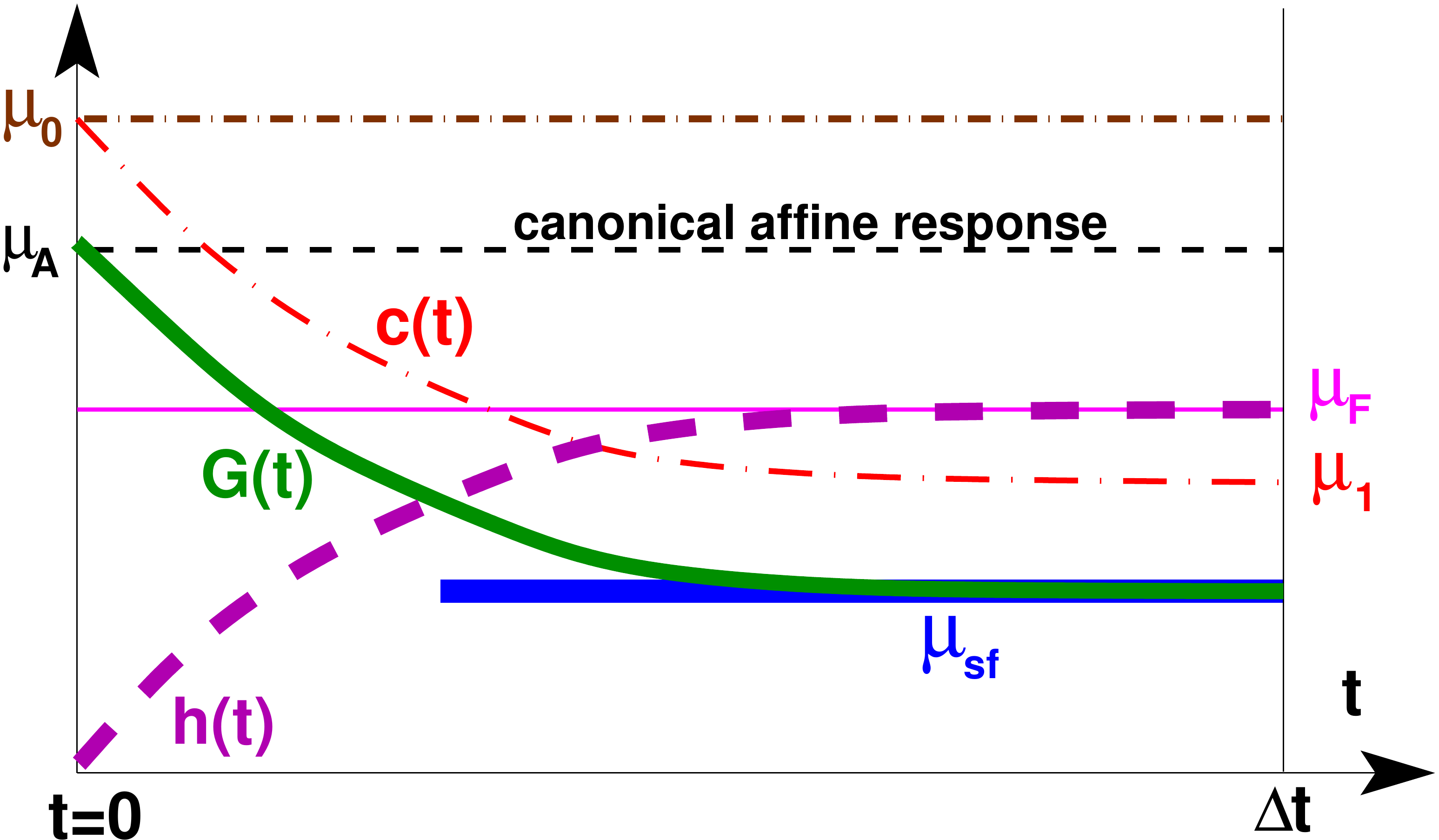}}}
\caption{Sketch of properties of interest for the determination of 
the shear-stress relaxation modulus $G(t)$ focusing on solids.
Static properties are indicated by horizontal lines,
$G(t)$ by the bold solid line,
$c(t)$ by the thin dash-dotted line and
$h(t)=c(0)-c(t)$ by the bold dashed line.
A canonical affine shear transformation at $t=0$
implies $G(t=0) = \muA$ while for large times $G(t) \to \GF$.
At variance to this $c(t)$ decays from $c(t=0)=\muFtwo$ to $c(t=\tsamp)=\muFone$.
In general $\muA \ne \muFtwo$, hence $G(t) \ne c(t)$.
Depending on the type of dynamics and the thermostat used, the dynamical behavior
may be more complex, especially at short times where oscillations may appear, and the representation
is thus strongly simplified.
\label{fig_Gtsketch}
}
\end{figure}

\section{Determination of $G(t)$}
\label{sm_Gt}

\paragraph*{Definitions and motivation.}
A central rheological property characterizing both liquids and solid elastic bodies 
is the shear relaxation modulus $G(t)$ \cite{RubinsteinBook,DoiEdwardsBook,HansenBook,AllenTildesleyBook}.
Assuming for simplicity an isotropic system, $G(t) \equiv \delta \tau(t)/\gamma$
may be obtained from the stress increment $\delta \tau(t) = \la \tauhat(t) - \tauhat(0^{-}) \ra$ 
after a small step strain with $|\gamma| \ll 1$ has been imposed at time $t=0$.
A schematic representation of $G(t)$ is given in Fig.~\ref{fig_Gtsketch}.
The direct numerical computation of $G(t)$ by means of an out-of-equilibrium simulation,
using the response to an imposed strain increment, is for technical reasons in general tedious 
\cite{AllenTildesleyBook,WXB15,WXBB15,WKB15,WXB16,WKC16}.
It is thus of high importance to compute $G(t)$ correctly and efficiently ``on the fly" 
by means of the appropriate linear-response fluctuation-dissipation relation for the 
standard canonical ensemble at imposed particle number $n$, 
volume $V$, shear strain $\gamma$ and temperature $T$ \cite{HansenBook,AllenTildesleyBook}.

\begin{figure}[t]
\centerline{\resizebox{1.0\columnwidth}{!}{\includegraphics*{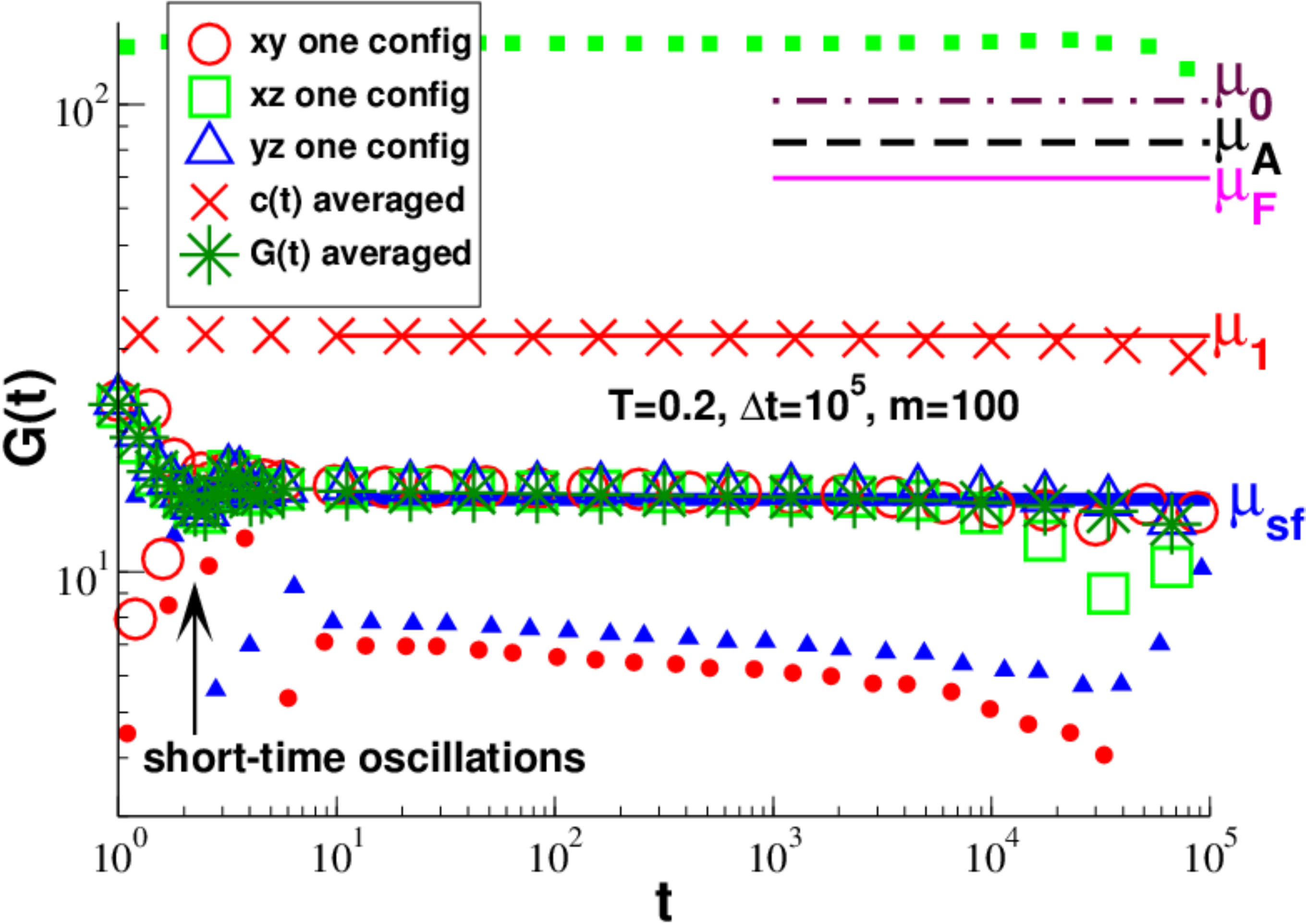}}}
\caption{Determination of $G(t)$ at $T=0.2$ following Ref.~\cite{ivan17a}.
The values of the ensemble-averaged static properties $\muFtwo$, $\muA$, $\muF$, $\muFone$ and $\GF$
(from top to bottom) are represented by horizontal lines.
The filled symbols indicate $\ctbar$ for the three shear planes of one configuration.
The ensemble-averaged ACF $c(t)$ (crosses) is similar to $\muFone$.
As indicated by the open symbols, Eq.~(\ref{eq_Gtbar}) yields essentially for each shear plane
the {\em same} behavior as the ensemble-averaged relation (stars).
\label{fig_Gtmethod}
}
\end{figure}

\begin{figure}[t]
\centerline{\resizebox{1.0\columnwidth}{!}{\includegraphics*{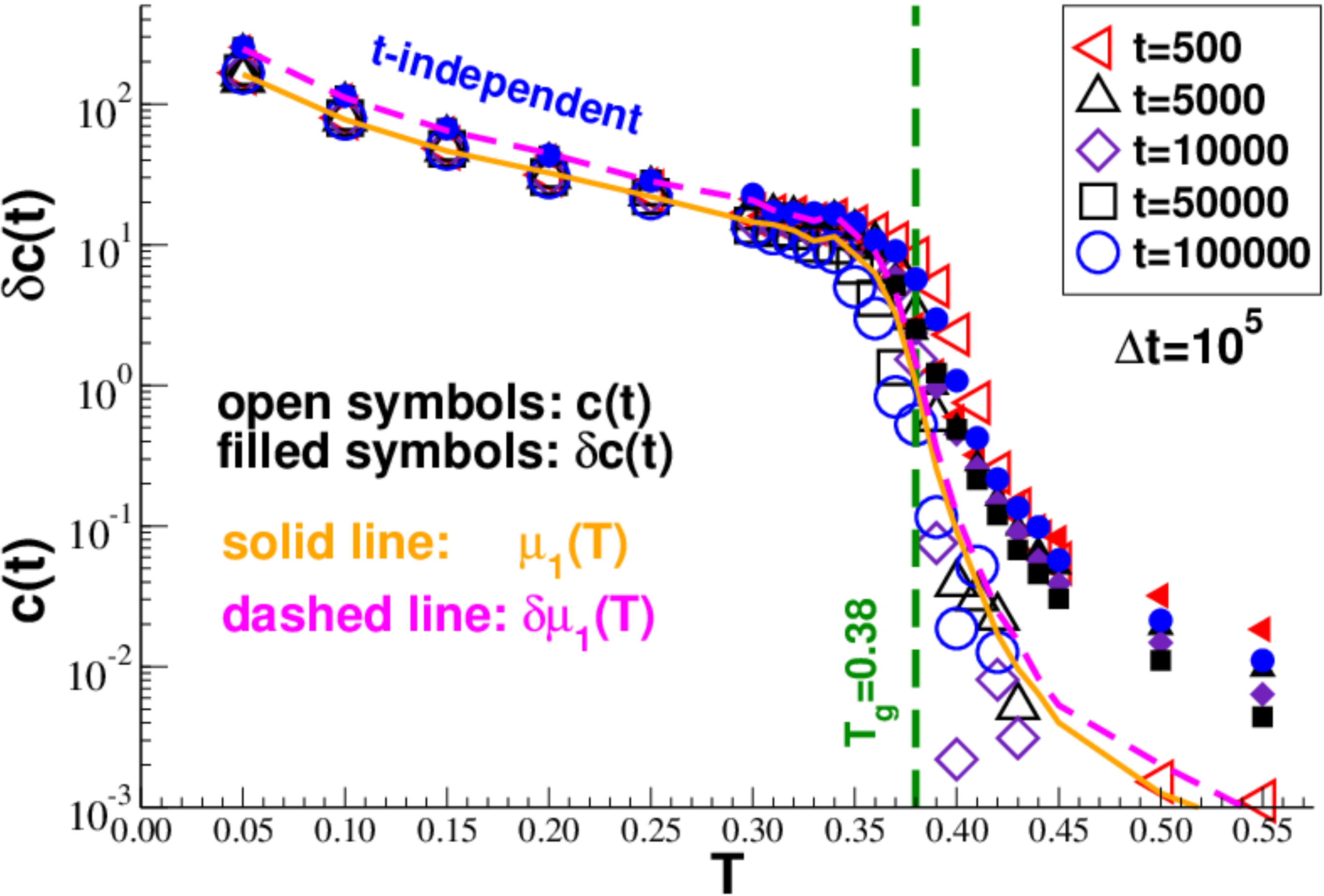}}}
\caption{Ensemble-averaged ACF $c(t)$ (open symbols) and associated standard deviation $\dct$ (filled symbols)
as a function of temperature $T$ for several times $t$. Note that $\dct$ is always larger $c(t)$ for all $T$ and $t$.
The bold solid line indicates $\muFone(T)$ for $\tsamp=10^5$,
the dashed line the corresponding standard deviation $\dmuFone(T)$. 
Below $\Tglass$ one observes $c(t) \approx \muFone(T)$ and $\dct \approx \dmuFone(T)$.
\label{fig_ct_T}
}
\end{figure}

\paragraph*{Shear-stress autocorrelation function.}
It is widely assumed \cite{AllenTildesleyBook,Klix12} that quite generally $G(t)=c(t)$
must hold with the shear-stress ACF $c(t)$ as defined in Eq.~(\ref{eq_ct}).
A schematic representation of $c(t)$ is given in Fig.~\ref{fig_Gtsketch},
data for a polymer-glass at a low temperature $T=0.2$ in Fig.~\ref{fig_Gtmethod}
and data for $c(t)$ as a function of temperature $T$ for several times $t$ in Fig.~\ref{fig_ct_T}.
As indicated in Fig.~\ref{fig_Gtsketch}, $c(t=0)=\muFtwo$
with $\muFtwo$ as defined by Eq.~(\ref{eq_muFtwo}) in the main text. 
This extreme short-time limit is not visible in the representations used in Fig.~\ref{fig_Gtmethod}
and Fig.~\ref{fig_ct_T}.
The opposite long-time limit $c(t=\tsamp)$ is slightly more intricate. It is of conceptional importance 
that for solid bodies, such as permanent elastic networks \cite{WXP13,WXB15,WXBB15,WXB16}, 
\begin{equation}
c(t=\tsamp) = \muFone = \la \muFonebar \ra
\mbox{ with } \muFonebar \equiv \beta V \overline{ \tauhat}^2.
\label{eqsm_muFone}
\end{equation}
As shown in Sec.~\ref{sm_GtGF}, this relation may also be justified 
as the specific limit of a more general relation Eq.~(\ref{eq_ct2muFone}) 
if plastic rearrangements can be neglected on the time-scale probed. 
The relevance of Eq.~(\ref{eqsm_muFone}) for our polymer glasses at low temperatures
is demonstrated in Fig.~\ref{fig_Gtmethod} where we indicate the time-averaged $\ctbar$ 
for the three shear planes of one arbitrary configuration (filled symbols). 
The observed strong fluctuation of the latter data and the plateau values are 
due to the different quenched stresses $\overline{\tauhat}$ of each shear plane. 
Note also that the ensemble-averaged $c(t)$ indicated by the crosses becomes rapidly identical to 
$\muFone \approx 32$ (thin horizontal line). 
As shown in Fig.~\ref{fig_ct_T}, Eq.~(\ref{eqsm_muFone}) holds quite generally for temperatures
below $\Tglass$ and the fluctuations of the ACF are perfectly described by $\dct \approx \dmuFone(T)$
for a broad range of times $t$.  
As emphasized in Refs.~\cite{WXB15,WXBB15}, an important consequence of $c(t) \to \muFone$ 
is now that $G(t)=c(t)$ is inconsistent with $G(t) \to \GF$ and the stress-fluctuation 
formula $\GF = (\muA-\muFtwo)+\muFone$
which must hold quite generally for solid bodies 
\cite{Hoover69,Lutsko88,WTBL02,Barrat06,WXP13,WXBB15,LXW16}.

\paragraph*{More general relation.}
This problem is resolved by means of the slightly more general fluctuation-dissipation relation 
\cite{WKB15,WXB16,WKC16,ivan17a,ivan17c} stated by Eq.~(\ref{eq_Gtbar}) and Eq.~(\ref{eq_Gt}) in the main text.
This relation has been called a ``simple-average expression" in \cite{WKB15,WXB16},
since both terms $\muAbar$ and $\htbar$ transform as simple averages \cite{AllenTildesleyBook}
between the conjugated ensembles at constant shear strain and constant shear stress.
For elastic solids this is one possibility to derive Eq.~(\ref{eq_Gtbar}) within a few lines \cite{WKB15}.
Please note that the ACF $c(t)$ and the MSD $h(t)$ are related by \cite{DoiEdwardsBook}
\begin{equation}
h(t) = c(0) - c(t) = \muFtwo - c(t).
\label{eq_htct}
\end{equation}
Using Eq.~(\ref{eqsm_muFone}) and $h(t) \to \muF$ for large times this implies that
Eq.~(\ref{eq_Gt}) is consistent with 
\begin{equation}
\lim_{t\to \tsamp} G(t) = \GF(\tsamp) =\muA-\muF(\tsamp)
\label{eq_asitshould}
\end{equation}
as it should. 
As one sees from Fig.~\ref{fig_Gtmethod} (open symbols), Eq.~(\ref{eq_Gt}) remains valid below $\Tglass$ 
and is, moreover, statistically well-behaved despite the strong fluctuations of the frozen shear stresses
in the different shear planes. The reason for this is simply that the frozen stresses directly 
drop out of the stress difference computed in the MSD $h(t)$. 
See Sec.~\ref{res_dGt} for the low-temperature behavior of the standard deviation $\dGt$.
We emphasize finally that since $\muA=\muFtwo$ holds in the liquid limit (Sec.~\ref{res_static}),
Eq.~(\ref{eq_Gt}) simplifies to $G(t)=c(t)$ as one expects.
%


\section{Relation between $G(t)$ and $\GF(\tsamp)$}
\label{sm_GtGF}

\paragraph*{Some useful properties of a functional.}
With $y(t)$ being an arbitrary well-behaved function of $t$,
let us consider the linear functional
\begin{equation}
\Tver[y(t)] \equiv \frac{2}{\tsamp^2} \int_0^{\tsamp} \ddiff t \ (\tsamp-t) \ y(t).
\label{eq_Tver_def}
\end{equation}
Interestingly, for a constant function
\begin{equation}
y(t)=c_0 \mbox{ we have } \Tver[c_0] = c_0,
\label{eq_constfunc}
\end{equation}
i.e. the $\tsamp$-dependence drops out.
This even holds to leading order if $y(t) \approx c_0$ only for large $t$ or 
for a finite $t$-window if this window becomes sufficiently large.
Note that contributions at the lower boundary of the integral have a strong weight 
due to the factor $(\tsamp-t)$. If $y(t)$ is a strictly monotonously decreasing function,
we have 
\begin{equation}
y(t=\tsamp) < \Tver[y(t)].
\label{eq_inequal}
\end{equation}
This inequality also holds if $y(t)$ is only 
in a finite, but broad intermediate time window a monotonously decreasing function.

\paragraph*{Time-translational invariance.}
Let us consider a time series $(x_1,\ldots,x_n, \ldots x_N)$  with entries $x_n$
sampled at equidistant time intervals $\ddiff t$.
The time averaged variance of this time series may be written without approximation as
\begin{eqnarray}
\overline{x^2}-\overline{x}^2 & = & \overline{(x_n-\overline{x})^2}=
\frac{1}{2N^2} \sum_{n,m=1} (x_n-x_m)^2 \nonumber\\
&=& \frac{2}{N^2} \sum_{s=0}^{N-1} (N-s) \ \overline{h}(s,N).
\label{eq_var_hs} 
\end{eqnarray}
We have defined here the gliding average
\begin{equation}
\overline{h}(s,N) \equiv \frac{1}{2} \ \frac{1}{N-s} \sum_{n=1}^{N-s} \ (x_{n+s}-x_n)^2
\label{eq_hs_def}
\end{equation}
which depends {\em a priori} on both $s$ and $N$. 
The latter representation is useful if the time series is stationary, 
i.e. time-translational invariance can be assumed on average.
Taking the expectation value $\langle \ldots \rangle$ over an ensemble of such time series yields
\begin{eqnarray}
\la \overline{x^2}-\overline{x}^2 \ra & = & \frac{2}{N^2} \sum_{s=0}^{N-1} (N-s) \ h(s) 
\label{eq_var_hsav} \\
\mbox{ with } h(s) & \equiv & \la \overline{h}(s,N) \ra = \frac{1}{2} \la \overline{(x_s-x_0)^2} \ra
\label{eq_hs_defav}
\end{eqnarray}
being now a proper MSD depending only on the time-increment as the one introduced in Eq.~(\ref{eq_ht}).
In the continuum limit for $N \gg 1$ the latter result becomes
\begin{equation}
\la \overline{x^2}-\overline{x}^2 \ra = \Tver[h(t)]
\label{eq_var_hscont}
\end{equation}
where we use that the time series have been sampled with equidistant time steps,
i.e. $t \approx s \ddiff t$ and $\tsamp \approx N \ddiff t$.

\paragraph*{Back to current problem.}
Setting $x(t) \equiv \sqrt{\beta V} \tauhat(t)$ and assuming time translational invariance for
the sampled instantaneous shear stresses $\tauhat$, Eq.~(\ref{eq_var_hscont}) thus implies
\begin{equation}
\muF(\tsamp) \equiv \muFtwo - \muFone(\tsamp) = \Tver[h(t)]
\label{eq_muFtsamp}
\end{equation}
for the $\tsamp$-dependence of the shear-stress fluctuations in agreement with the more direct
demonstration given in Ref.~\cite{WXB15}.
Since $\muFtwo$ does not depend explicitly on $\tsamp$, 
it follows using Eq.~(\ref{eq_htct}) that
\begin{equation}
\muFone(\tsamp) = \muFtwo - \Tver[h(t)] = \Tver[c(t)].
\label{eq_ct2muFone}
\end{equation}
Importantly, this reduces to the relation $c(t=\tsamp)=\muFone$ for solids
if the ACF $c(t)$ becomes constant for large times as in the example given in Fig.~\ref{fig_Gtmethod}.
We have thus obtained a generalization of Eq.~(\ref{eq_muFone}) being also valid for general
viscoelastic fluids.
Since $\muA$  is constant, Eq.~(\ref{eq_constfunc}) and Eq.~(\ref{eq_Gt}) imply
\begin{eqnarray}
\GF(\tsamp) & \equiv & \muA - \muF(\tsamp) = \Tver[G(t)] \nonumber \\
            & = & \frac{2}{\tsamp^2} \int_0^{\tsamp} (\tsamp -t) \ G(t) \ \ddiff t 
\label{eq_GFtsamp}
\end{eqnarray}
in agreement with Eq.~(\ref{eq_mu}) stated in the Introduction.
%


\begin{figure}[t]
\centerline{\resizebox{1.0\columnwidth}{!}{\includegraphics*{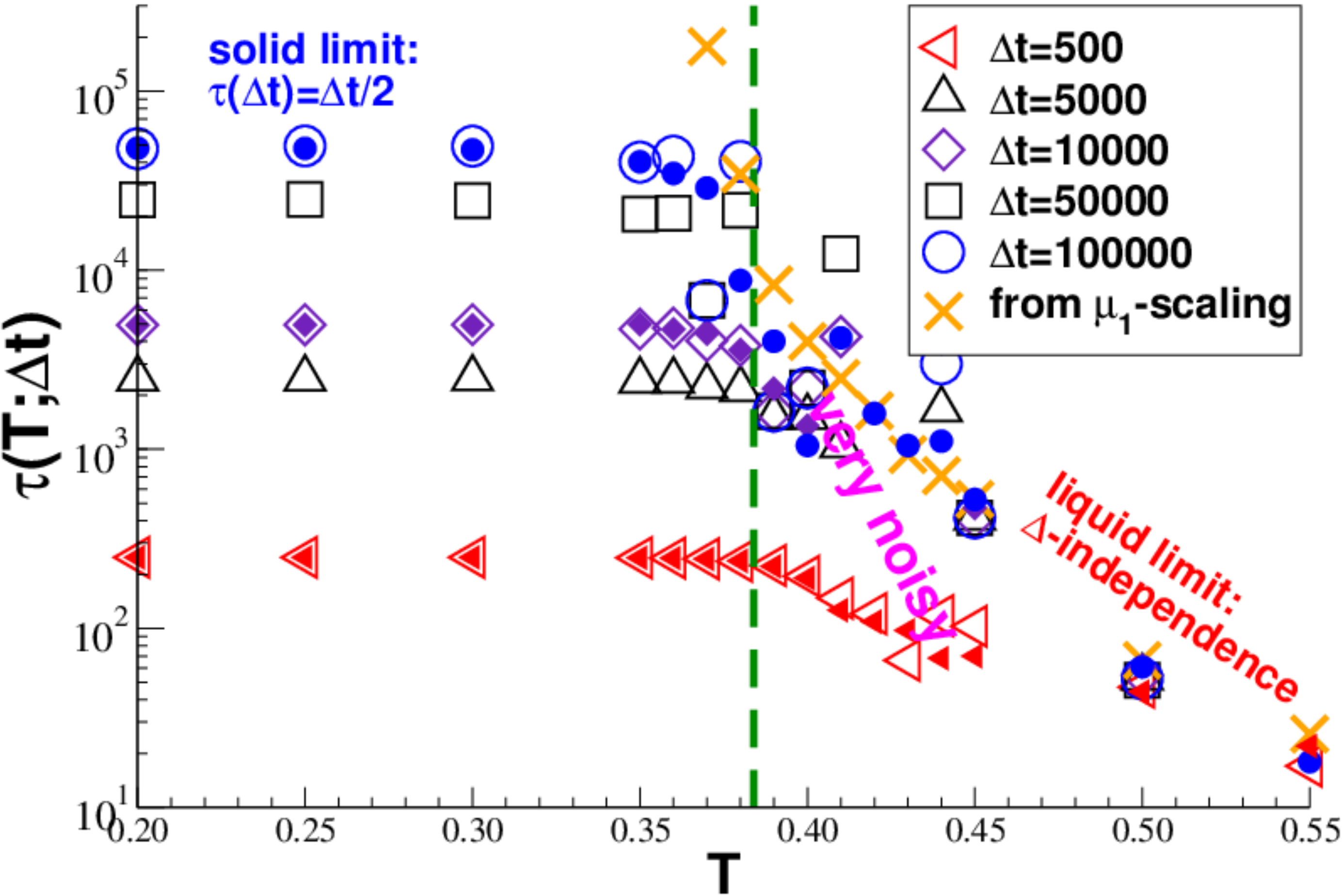}}}
\caption{Generalized shear-stress relaxation time $\tau(T;\tsamp)$ for several $\tsamp$
with open symbols obtained using Eq.~(\ref{eq_tau}) and filled symbols
using Eq.~(\ref{eq_GF2tau}) taking advantage of the $\GF(\tsamp)$-data. 
For both methods the data is noisy and unreliable around $\Tglass$.
This is at variance to the smooth $\tauinf(T)$-values (crosses) obtained in Fig.~\ref{fig_mu1_dt}
from $\muFone(\tsamp)$. 
}
\label{fig_tau}
\end{figure}

\section{Relaxation time $\tau(\tsamp)$}
\label{sm_tau}
It is tempting to consider as an additional moment over $G(t)$ the
generalized $\tsamp$-dependent shear-stress relaxation time 
\begin{equation}
\tau(\tsamp) \equiv \frac{1}{\eta(\tsamp)} \ \int_{t=0}^{\tsamp} \ddiff t \ t \ G(t) 
\label{eq_tau}
\end{equation}
with $\tauinf \equiv \lim_{\tsamp \to \infty} \tau(\tsamp)$
being the experimentally relevant terminal relaxation time of the system.
Note that $\tau(\tsamp)=\tauinf$ for all $\tsamp$ for a Maxwell fluid with 
$G(t) \sim \exp(-t/\tauinf)$ as observed, e.g., for equilibrium polymers
\cite{CC90}, vitrimers \cite{Leibler11,Leibler13} or self-assembled transient networks \cite{WKC16}.
Unfortunately, Eq.~(\ref{eq_tau}) must be dominated even more strongly by the upper bound
of the integral over $G(t)$ than the generalized shear viscosity $\eta(\tsamp)$.
Since $G(t)$ is strongly fluctuating for $t \to \tsampmax=10^5$, especially around $\Tglass$
(Fig.~\ref{fig_Gt}), this leads to an unreliable estimation of $\tau(\tsamp)$ for $\tsamp \to \tsampmax$. 
This suggests to reexpress Eq.~(\ref{eq_tau}) in terms of $\GF(\tsamp)$ with $\tsamp \le \tsampmax$.
As seen from Eqs.~(\ref{eq_mu}-\ref{eq_tau}), the three moments $\mu(\tsamp)$, $\eta(\tsamp)$ and 
$\tau(\tsamp)$ are related by
\begin{equation}
\tau(\tsamp) = \tsamp - \left[\mu(\tsamp) \tsamp^2/2\right]/\eta(\tsamp).
\label{eq_muetatau}
\end{equation}
One may thus obtain the generalized relaxation time from
\begin{equation}
\tau(\tsamp) = \operatorname{e}^x \ \left[1 - 1/y^{\prime}(x) \right]
\label{eq_GF2tau}
\end{equation}
with again $x \equiv \ln(\tsamp)$ and $y \equiv \ln(\GF(\tsamp)\tsamp^2/2)$
fitted to sixth order and $\muFone(\tsamp)$ replacing $\GF(\tsamp)$ for large temperatures.

Figure~\ref{fig_tau} presents $\tau(T;\tsamp)$ as a function of temperature
using half-logarithmic coordinates. 
The open symbols indicate values obtained using the direct integral Eq.~(\ref{eq_tau}), 
filled symbols data using Eq.~(\ref{eq_GF2tau}) and the crosses the terminal
relaxation times estimated using the rescaling of $\muFone(\tsamp)$ presented
in panel (c) of Fig.~\ref{fig_mu1_dt}.
The first two methods yield identical results for small $\tsamp$ and in the solid limit.
As before for $\eta(\tsamp)$ one observes that $\tau(\tsamp)$ increases linearly 
in the solid limit where $\tau(\tsamp) = \tsamp/2$ for $\tsamp \ll \tauinf(T)$.
For the more interesting higher temperatures the first method
yields slightly erratic results although the integration is terminated at the
first occurance of a strong negative $G(t)$-fluctuation. The observed $\tsamp$-independence
for large temperatures is thus trivially imposed and not the confirmation of an expected result.
Unfortunately, similar rather erratic data are obtained using Eq.~(\ref{eq_GF2tau}) as shown for $\tsamp=10^5$.
However, albeit very noisy, both data sets approach with increasing $\tsamp$ the terminal relaxation time
estimated using the $\muFone(\tsamp)$-rescaling (crosses).
Being certainly not very precise (not even on the used logarithmic scale) our $\tau(\tsamp;T)$-data 
are thus at least consistent with the $\muFone(\tsamp)$-data and give reasonable lower bounds.


\begin{thebibliography}{52}
\bibitem[{foo({\natexlab{a}})}]{foot_however}
\bibinfo{note}{The ensemble-averaged prediction is thus irrelevant for a single
  configuration. This does, however, not imply that a single configuration may
  not have a finite well-defined shear modulus.}
\bibitem[{foo({\natexlab{b}})}]{foot_Gloss2eta}
\bibinfo{note}{We remind that the shear viscosity is obtained experimentally
  from the low-frequency limit of the loss modulus $\Gloss$ according to
  $\Gloss/\omega \to \etainf$ for $\omega \tauinf \ll 1$. We shall show
  elsewhere that this is equivalent to Eq.~(\ref{eq_etainf}).}
\bibitem[{foo({\natexlab{c}})}]{foot_otherdata}
\bibinfo{note}{In addition to the presented data we have sampled systematically
  a broad range of chain lengths $N$ and cooling rates $\Gamma$. As one expects
  \cite{SBM11,Frey15,BaschRev16}, the glass transition temperature $\Tglass$ is
  found to depend somewhat on both parameters without, however, changing the
  qualitative behavior.}
\bibitem[{foo({\natexlab{d}})}]{foot_Vbar}
\bibinfo{note}{The second half of the tempering step at constant pressure is
  used to determine the average volume $\Vbar$ which is then used for the
  subsequent constant volume simulations.}
\bibitem[{foo({\natexlab{e}})}]{foot_TP}
\bibinfo{note}{This is, e.g., the case for self-assembled transient networks
  formed by reversibly bridging and breaking bonds between colloids
  \cite{WKC16}. It can be shown that in this case $\muA=\muFtwo$ holds not only
  in the liquid regime, but also through the glass transition into the quenched
  armorphous solid limit.}
\bibitem[{foo({\natexlab{f}})}]{foot_intconst}
\bibinfo{note}{Equation~(\ref{eq_eta2GF}) may be obtained by integration by
  parts from Eq.~(\ref{eq_mu}) or by integration of Eq.~(\ref{eq_GF2eta}). The
  integration constant vanishes as can be seen from the fact that of sides of
  Eq.~(\ref{eq_eta2GF}) becomes $\muA$ for $\tsamp \to 0$.}
\bibitem[{foo({\natexlab{g}})}]{foot_truncprob}
\bibinfo{note}{Surprisingly, it appears that the value of $\Delta \muA$ depends
  somewhat on the time-increment $\delta t$ of the velocity-Verlet integration
  step. A complete understanding and description of this technical issue is
  still missing. This observation is presumably due to small detailed balance
  violations of the MD integration scheme which lead to inconsistencies with
  the rigorous thermodynamic relations put forward in Ref.~\cite{XWP12}.}
\bibitem[{foo({\natexlab{h}})}]{foot_fitetainf}
\bibinfo{note}{$\etainf(T)$ is most readily determined from the plateau at
  large $\tsamp$ of $\muFone(\tsamp) \tsamp/2$ traced using a half-logarithmic
  representation.}
\end{thebibliography}

\end{document}